\documentclass[longauth]{aa}

\usepackage{graphicx}
\usepackage{natbib}
\usepackage{scalerel}

\usepackage[table]{xcolor}

\usepackage{txfonts}
\usepackage[pdfencoding=auto,psdextra]{hyperref}
\hypersetup{
    colorlinks=true,
    linkcolor=blue,
    filecolor=magenta,      
    urlcolor=blue,
    citecolor=blue
}
\makeatletter
\renewcommand*\aa@pageof{, page \thepage{} of \pageref*{LastPage}}
\makeatother

\usepackage[utf8]{inputenc}

\usepackage[switch, modulo]{lineno}
\renewcommand{\linenumbers}[0]{}

\usepackage{euclid}
\usepackage{multirow}
\usepackage{bm}	
\usepackage[pdfencoding=auto,psdextra]{hyperref}
\usepackage{mathtools}
\usepackage{glossaries}
\glsdisablehyper

\newcommand{\ev}[1]{\left\langle #1 \right\rangle}

\newcommand{\orcid}[1]{} 

\newcommand{\xv}{\bm{x}}

\newcommand{\kv}{\bm{k}}

\newcommand{\rv}{\bm{r}}

\newcommand{\qv}{\bm{q}}
\newcommand{\sv}{\bm{s}}
\newcommand{\uv}{\bm{u}}

\newcommand{\kmax}{k_\mathrm{max}}

\newcommand{\deltainit}{\delta_\mathrm{L}}
\newcommand{\deltag}{\delta_\mathrm{g}}
\newcommand{\dirac}{\delta_\mathrm{D}}
\newcommand{\Plin}{P_\mathrm{L}}

\newcommand{\Pgg}{P_\mathrm{gg}}
\newcommand{\Pgt}{P_{\mathrm{g}\theta}}
\newcommand{\Ptt}{P_{\theta\theta}}

\newcommand{\bGtwo}{b_{\mathcal{G}_2}}
\newcommand{\bGthree}{b_{\Gamma_3}}

\newcommand{\czero}{c_0}
\newcommand{\ctwo}{c_2}
\newcommand{\cfour}{c_4}

\newcommand{\avir}{a_{\rm vir}}

\newcommand{\Gtwo}{\mathcal{G}_2}
\newcommand{\Gthree}{\Gamma_3}
\newcommand{\Ps}{P^{\,s}}

\newcommand{\PsSPT}{\Ps_{\rm gg,\,SPT}}
\newcommand{\Pslin}{\Ps_{\rm gg,\,L}}
\newcommand{\Pstwotwo}{\Ps_{\rm gg,22}}
\newcommand{\Psonethree}{\Ps_{\rm gg,13}}
\newcommand{\PsTNS}{\Ps_{\rm gg,\,TNS}}
\newcommand{\sigv}{\sigma_v}
\newcommand{\PsEFT}{\Ps_{\rm gg,\,EFT}}
\newcommand{\Psnlo}{P^{\,s,\,{\rm IR-NLO}}_{\rm gg}}
\newcommand{\Psctr}{P^{\,s,\,{\rm ctr}}_{\rm gg}}
\newcommand{\Psstoch}{P^{\,s,\,{\rm stoch}}_{\rm gg}}
\newcommand{\Pell}{P_\ell}
\newcommand{\xiell}{\xi_\ell}
\newcommand{\xis}{\xi^{\,s}}
\newcommand{\Psgg}{\Ps_{\rm gg}}
\newcommand{\sigeight}{\sigma_8}
\newcommand{\sigtwelve}{\sigma_{12}}
\newcommand{\qpar}{q_\parallel}
\newcommand{\qperp}{q_\perp}

\newcommand{\bdm}{\begin{displaymath}}
\newcommand{\edm}{\end{displaymath}}

\usepackage[normalem]{ulem}
\usepackage{color}
\definecolor{darkgreen}{RGB}{0,128,0}
\definecolor{darkorange}{RGB}{213,113,0}

\newcommand{\eg}{{e.g.}\xspace}
\newcommand{\Ms}{\, h^{-1} \, M_\odot}
\newcommand{\Mpc}{\, h^{-1} \, {\rm Mpc}}
\newcommand{\cMpc}{\, h^{-3} \, {\rm Mpc}^3}

\newcommand{\kMpc}{\, h \, {\rm Mpc}^{-1}}
\newcommand{\kcMpc}{\, h^3 \, {\rm Mpc}^{-3}}

\newcommand{\As}{A_\mathrm{s}}
\newcommand{\ns}{n_\mathrm{s}}

\newcommand{\omegac}{\omega_\text{c}}
\newcommand{\omegab}{\omega_\text{b}}

\newcommand{\flagship}{Flagship 1\xspace}
\newcommand{\flagshiptwo}{Flagship 2\xspace}
\newcommand{\pkdgrav}{\texttt{PKDGRAV3}\xspace}

\newcommand{\euclidemulatortwo}{\texttt{EuclidEmulator2}\xspace}
\newcommand{\comet}{\texttt{COMET}\xspace}

\newcommand{\smin}{s_{\mathrm{min}}}
\newcommand{\chired}{\chi^2_{\rm red}}
\newcommand{\tm}[1]{\textrm{#1}}

\newacronym{2pcf}{2PCF}{2-point correlation function}
\newacronym{3pcf}{3PCF}{3-point correlation function}
\newacronym{spt}{SPT}{standard perturbation theory}
\newacronym{rpt}{RPT}{renormalised perturbation theory}
\newacronym{eft}{EFT}{effective field theory of large-scale structure}
\newacronym{vdg}{VDG$_\infty$}{velocity difference generating function}
\newacronym{clpt}{CLPT}{convolutional Lagrangian perturbation theory}
\newacronym{cleft}{CLEFT}{convolutional Lagrangian effective field theory}
\newacronym{bao}{BAO}{baryon acoustic oscillation}
\newacronym{rsd}{RSD}{redshift-space distortions}
\newacronym{hod}{HOD}{halo occupation distribution}
\newacronym{los}{LOS}{line of sight}
\newacronym{fob}{FoB}{figure of bias}
\newacronym{fom}{FoM}{figure of merit}
\newacronym{mgf}{MGF}{moment generating function}
\newacronym{left}{LEFT}{Lagrangian effective field theory}
\newacronym{cdm}{CDM}{cold dark matter}
\newacronym{lpt}{LPT}{Lagrangian perturbation theory}
\newacronym{llapprox}{LL}{local Lagrangian}
\newacronym{gprocess}{GP}{Gaussian process}
\newacronym{dof}{dof}{degrees of freedom}
\newacronym{pdfabbrev}{PDF}{probability distribution function}
\newacronym{DRone}{DR1}{Data Release 1}
\newacronym{DRthree}{DR3}{Data Release 3}

\newcommand{\tpcf}{\gls{2pcf}\xspace}
\newcommand{\threepcf}{\gls{3pcf}\xspace}
\newcommand{\spt}{\gls{spt}\xspace}

\newcommand{\eft}{\gls{eft}\xspace}
\newcommand{\vdg}{\gls{vdg}\xspace}
\newcommand{\clpt}{\gls{clpt}\xspace}
\newcommand{\cleft}{\gls{cleft}\xspace}
\newcommand{\bao}{\gls{bao}\xspace}
\newcommand{\rsd}{\gls{rsd}\xspace}
\newcommand{\lcdm}{$\Lambda$CDM\xspace}
\newcommand{\hod}{\gls{hod}\xspace}
\newcommand{\los}{\gls{los}\xspace}
\newcommand{\fob}{\gls{fob}\xspace}
\newcommand{\fom}{\gls{fom}\xspace}
\newcommand{\wcdm}{$w$CDM\xspace}

\newcommand{\mgf}{\gls{mgf}\xspace}
\newcommand{\lefttheo}{\gls{left}\xspace}
\newcommand{\cdm}{\gls{cdm}\xspace}
\newcommand{\lpt}{\gls{lpt}\xspace}
\newcommand{\LL}{\gls{llapprox}\xspace}
\newcommand{\GP}{\gls{gprocess}\xspace}
\newcommand{\dof}{\gls{dof}\xspace}
\newcommand{\pdfabbrev}{\gls{pdfabbrev}\xspace}
\newcommand{\DRone}{\gls{DRone}\xspace}
\newcommand{\DRthree}{\gls{DRthree}\xspace}

\begin{document}
%
%


\title{\Euclid preparation}
\subtitle{Galaxy 2-point correlation function modelling in redshift space}    



\author{Euclid Collaboration: M.~K{\"a}rcher\orcid{0000-0001-5868-647X}\thanks{\email{martin.karcher@unimi.it}}\inst{\ref{aff1},\ref{aff2},\ref{aff3}}
\and M.-A.~Breton\inst{\ref{aff4}}
\and S.~de~la~Torre\inst{\ref{aff1}}
\and A.~Veropalumbo\orcid{0000-0003-2387-1194}\inst{\ref{aff5},\ref{aff6},\ref{aff7}}
\and A.~Eggemeier\orcid{0000-0002-1841-8910}\inst{\ref{aff8}}
\and M.~Crocce\orcid{0000-0002-9745-6228}\inst{\ref{aff9},\ref{aff10}}
\and E.~Sefusatti\orcid{0000-0003-0473-1567}\inst{\ref{aff11},\ref{aff12},\ref{aff13}}
\and E.~Sarpa\orcid{0000-0002-1256-655X}\inst{\ref{aff14},\ref{aff15},\ref{aff13}}
\and R.~E.~Angulo\orcid{0000-0003-2953-3970}\inst{\ref{aff16},\ref{aff17}}
\and B.~Camacho~Quevedo\orcid{0000-0002-8789-4232}\inst{\ref{aff12},\ref{aff14},\ref{aff11}}
\and L.~Castiblanco\orcid{0000-0002-2324-7335}\inst{\ref{aff18},\ref{aff19}}
\and E.~Castorina\inst{\ref{aff3},\ref{aff20}}
\and A.~Chudaykin\inst{\ref{aff21}}
\and V.~Desjacques\orcid{0000-0003-2062-8172}\inst{\ref{aff22}}
\and A.~Farina\orcid{0009-0000-3420-929X}\inst{\ref{aff7},\ref{aff5},\ref{aff6}}
\and G.~Gambardella\orcid{0009-0001-1281-1746}\inst{\ref{aff9},\ref{aff10}}
\and M.~Guidi\orcid{0000-0001-9408-1101}\inst{\ref{aff23},\ref{aff24}}
\and D.~Linde\orcid{0000-0001-7192-1067}\inst{\ref{aff25}}
\and F.~Marulli\orcid{0000-0002-8850-0303}\inst{\ref{aff26},\ref{aff24},\ref{aff27}}
\and A.~Moradinezhad~Dizgah\orcid{0000-0001-8841-9989}\inst{\ref{aff28}}
\and M.~Moresco\orcid{0000-0002-7616-7136}\inst{\ref{aff26},\ref{aff24}}
\and C.~Moretti\orcid{0000-0003-3314-8936}\inst{\ref{aff11},\ref{aff12},\ref{aff13}}
\and K.~Pardede\orcid{0000-0002-7728-8220}\inst{\ref{aff25}}
\and A.~Pezzotta\orcid{0000-0003-0726-2268}\inst{\ref{aff5}}
\and M.~Pellejero~Iba\~nez\orcid{0000-0003-4680-7275}\inst{\ref{aff29}}
\and C.~Porciani\orcid{0000-0002-7797-2508}\inst{\ref{aff8}}
\and A.~Pugno\inst{\ref{aff8}}
\and M.~Zennaro\orcid{0000-0002-4458-1754}\inst{\ref{aff30}}
\and N.~Aghanim\orcid{0000-0002-6688-8992}\inst{\ref{aff31}}
\and B.~Altieri\orcid{0000-0003-3936-0284}\inst{\ref{aff32}}
\and L.~Amendola\orcid{0000-0002-0835-233X}\inst{\ref{aff33}}
\and S.~Andreon\orcid{0000-0002-2041-8784}\inst{\ref{aff5}}
\and N.~Auricchio\orcid{0000-0003-4444-8651}\inst{\ref{aff24}}
\and C.~Baccigalupi\orcid{0000-0002-8211-1630}\inst{\ref{aff12},\ref{aff11},\ref{aff13},\ref{aff14}}
\and M.~Baldi\orcid{0000-0003-4145-1943}\inst{\ref{aff23},\ref{aff24},\ref{aff27}}
\and S.~Bardelli\orcid{0000-0002-8900-0298}\inst{\ref{aff24}}
\and A.~Biviano\orcid{0000-0002-0857-0732}\inst{\ref{aff11},\ref{aff12}}
\and E.~Branchini\orcid{0000-0002-0808-6908}\inst{\ref{aff7},\ref{aff6},\ref{aff5}}
\and M.~Brescia\orcid{0000-0001-9506-5680}\inst{\ref{aff34},\ref{aff35}}
\and S.~Camera\orcid{0000-0003-3399-3574}\inst{\ref{aff36},\ref{aff37},\ref{aff38}}
\and G.~Ca\~nas-Herrera\orcid{0000-0003-2796-2149}\inst{\ref{aff29},\ref{aff39}}
\and V.~Capobianco\orcid{0000-0002-3309-7692}\inst{\ref{aff38}}
\and C.~Carbone\orcid{0000-0003-0125-3563}\inst{\ref{aff40}}
\and V.~F.~Cardone\inst{\ref{aff41},\ref{aff42}}
\and J.~Carretero\orcid{0000-0002-3130-0204}\inst{\ref{aff43},\ref{aff44}}
\and M.~Castellano\orcid{0000-0001-9875-8263}\inst{\ref{aff41}}
\and G.~Castignani\orcid{0000-0001-6831-0687}\inst{\ref{aff24}}
\and S.~Cavuoti\orcid{0000-0002-3787-4196}\inst{\ref{aff35},\ref{aff45}}
\and K.~C.~Chambers\orcid{0000-0001-6965-7789}\inst{\ref{aff46}}
\and A.~Cimatti\inst{\ref{aff47}}
\and C.~Colodro-Conde\inst{\ref{aff48}}
\and G.~Congedo\orcid{0000-0003-2508-0046}\inst{\ref{aff29}}
\and L.~Conversi\orcid{0000-0002-6710-8476}\inst{\ref{aff49},\ref{aff32}}
\and Y.~Copin\orcid{0000-0002-5317-7518}\inst{\ref{aff50}}
\and F.~Courbin\orcid{0000-0003-0758-6510}\inst{\ref{aff51},\ref{aff52},\ref{aff53}}
\and H.~M.~Courtois\orcid{0000-0003-0509-1776}\inst{\ref{aff54}}
\and H.~Degaudenzi\orcid{0000-0002-5887-6799}\inst{\ref{aff55}}
\and G.~De~Lucia\orcid{0000-0002-6220-9104}\inst{\ref{aff11}}
\and H.~Dole\orcid{0000-0002-9767-3839}\inst{\ref{aff31}}
\and F.~Dubath\orcid{0000-0002-6533-2810}\inst{\ref{aff55}}
\and X.~Dupac\inst{\ref{aff32}}
\and S.~Dusini\orcid{0000-0002-1128-0664}\inst{\ref{aff56}}
\and A.~Ealet\orcid{0000-0003-3070-014X}\inst{\ref{aff50}}
\and S.~Escoffier\orcid{0000-0002-2847-7498}\inst{\ref{aff57}}
\and M.~Farina\orcid{0000-0002-3089-7846}\inst{\ref{aff58}}
\and R.~Farinelli\inst{\ref{aff24}}
\and F.~Faustini\orcid{0000-0001-6274-5145}\inst{\ref{aff41},\ref{aff59}}
\and S.~Ferriol\inst{\ref{aff50}}
\and F.~Finelli\orcid{0000-0002-6694-3269}\inst{\ref{aff24},\ref{aff60}}
\and P.~Fosalba\orcid{0000-0002-1510-5214}\inst{\ref{aff10},\ref{aff9}}
\and N.~Fourmanoit\orcid{0009-0005-6816-6925}\inst{\ref{aff57}}
\and M.~Frailis\orcid{0000-0002-7400-2135}\inst{\ref{aff11}}
\and E.~Franceschi\orcid{0000-0002-0585-6591}\inst{\ref{aff24}}
\and M.~Fumana\orcid{0000-0001-6787-5950}\inst{\ref{aff40}}
\and S.~Galeotta\orcid{0000-0002-3748-5115}\inst{\ref{aff11}}
\and K.~George\orcid{0000-0002-1734-8455}\inst{\ref{aff61}}
\and W.~Gillard\orcid{0000-0003-4744-9748}\inst{\ref{aff57}}
\and B.~Gillis\orcid{0000-0002-4478-1270}\inst{\ref{aff29}}
\and C.~Giocoli\orcid{0000-0002-9590-7961}\inst{\ref{aff24},\ref{aff27}}
\and J.~Gracia-Carpio\inst{\ref{aff62}}
\and A.~Grazian\orcid{0000-0002-5688-0663}\inst{\ref{aff63}}
\and F.~Grupp\inst{\ref{aff62},\ref{aff64}}
\and L.~Guzzo\orcid{0000-0001-8264-5192}\inst{\ref{aff3},\ref{aff5},\ref{aff20}}
\and S.~V.~H.~Haugan\orcid{0000-0001-9648-7260}\inst{\ref{aff65}}
\and W.~Holmes\inst{\ref{aff66}}
\and F.~Hormuth\inst{\ref{aff67}}
\and A.~Hornstrup\orcid{0000-0002-3363-0936}\inst{\ref{aff68},\ref{aff69}}
\and K.~Jahnke\orcid{0000-0003-3804-2137}\inst{\ref{aff70}}
\and M.~Jhabvala\inst{\ref{aff71}}
\and B.~Joachimi\orcid{0000-0001-7494-1303}\inst{\ref{aff72}}
\and E.~Keih\"anen\orcid{0000-0003-1804-7715}\inst{\ref{aff73}}
\and S.~Kermiche\orcid{0000-0002-0302-5735}\inst{\ref{aff57}}
\and A.~Kiessling\orcid{0000-0002-2590-1273}\inst{\ref{aff66}}
\and B.~Kubik\orcid{0009-0006-5823-4880}\inst{\ref{aff50}}
\and M.~K\"ummel\orcid{0000-0003-2791-2117}\inst{\ref{aff64}}
\and M.~Kunz\orcid{0000-0002-3052-7394}\inst{\ref{aff21}}
\and H.~Kurki-Suonio\orcid{0000-0002-4618-3063}\inst{\ref{aff74},\ref{aff75}}
\and A.~M.~C.~Le~Brun\orcid{0000-0002-0936-4594}\inst{\ref{aff76}}
\and S.~Ligori\orcid{0000-0003-4172-4606}\inst{\ref{aff38}}
\and P.~B.~Lilje\orcid{0000-0003-4324-7794}\inst{\ref{aff65}}
\and V.~Lindholm\orcid{0000-0003-2317-5471}\inst{\ref{aff74},\ref{aff75}}
\and I.~Lloro\orcid{0000-0001-5966-1434}\inst{\ref{aff77}}
\and G.~Mainetti\orcid{0000-0003-2384-2377}\inst{\ref{aff78}}
\and D.~Maino\inst{\ref{aff3},\ref{aff40},\ref{aff20}}
\and E.~Maiorano\orcid{0000-0003-2593-4355}\inst{\ref{aff24}}
\and O.~Mansutti\orcid{0000-0001-5758-4658}\inst{\ref{aff11}}
\and S.~Marcin\inst{\ref{aff79}}
\and O.~Marggraf\orcid{0000-0001-7242-3852}\inst{\ref{aff8}}
\and M.~Martinelli\orcid{0000-0002-6943-7732}\inst{\ref{aff41},\ref{aff42}}
\and N.~Martinet\orcid{0000-0003-2786-7790}\inst{\ref{aff1}}
\and R.~J.~Massey\orcid{0000-0002-6085-3780}\inst{\ref{aff80}}
\and E.~Medinaceli\orcid{0000-0002-4040-7783}\inst{\ref{aff24}}
\and S.~Mei\orcid{0000-0002-2849-559X}\inst{\ref{aff81},\ref{aff82}}
\and M.~Melchior\inst{\ref{aff83}}
\and Y.~Mellier\inst{\ref{aff84},\ref{aff85}}
\and M.~Meneghetti\orcid{0000-0003-1225-7084}\inst{\ref{aff24},\ref{aff27}}
\and E.~Merlin\orcid{0000-0001-6870-8900}\inst{\ref{aff41}}
\and G.~Meylan\inst{\ref{aff86}}
\and A.~Mora\orcid{0000-0002-1922-8529}\inst{\ref{aff87}}
\and L.~Moscardini\orcid{0000-0002-3473-6716}\inst{\ref{aff26},\ref{aff24},\ref{aff27}}
\and C.~Neissner\orcid{0000-0001-8524-4968}\inst{\ref{aff88},\ref{aff44}}
\and S.-M.~Niemi\orcid{0009-0005-0247-0086}\inst{\ref{aff89}}
\and C.~Padilla\orcid{0000-0001-7951-0166}\inst{\ref{aff88}}
\and F.~Pasian\orcid{0000-0002-4869-3227}\inst{\ref{aff11}}
\and J.~A.~Peacock\orcid{0000-0002-1168-8299}\inst{\ref{aff29}}
\and K.~Pedersen\inst{\ref{aff90}}
\and W.~J.~Percival\orcid{0000-0002-0644-5727}\inst{\ref{aff91},\ref{aff92},\ref{aff93}}
\and V.~Pettorino\orcid{0000-0002-4203-9320}\inst{\ref{aff89}}
\and S.~Pires\orcid{0000-0002-0249-2104}\inst{\ref{aff4}}
\and G.~Polenta\orcid{0000-0003-4067-9196}\inst{\ref{aff59}}
\and M.~Poncet\inst{\ref{aff94}}
\and L.~A.~Popa\inst{\ref{aff95}}
\and F.~Raison\orcid{0000-0002-7819-6918}\inst{\ref{aff62}}
\and G.~Riccio\inst{\ref{aff35}}
\and E.~Romelli\orcid{0000-0003-3069-9222}\inst{\ref{aff11}}
\and M.~Roncarelli\orcid{0000-0001-9587-7822}\inst{\ref{aff24}}
\and C.~Rosset\orcid{0000-0003-0286-2192}\inst{\ref{aff81}}
\and R.~Saglia\orcid{0000-0003-0378-7032}\inst{\ref{aff64},\ref{aff62}}
\and Z.~Sakr\orcid{0000-0002-4823-3757}\inst{\ref{aff33},\ref{aff96},\ref{aff97}}
\and A.~G.~S\'anchez\orcid{0000-0003-1198-831X}\inst{\ref{aff62}}
\and D.~Sapone\orcid{0000-0001-7089-4503}\inst{\ref{aff98}}
\and B.~Sartoris\orcid{0000-0003-1337-5269}\inst{\ref{aff64},\ref{aff11}}
\and P.~Schneider\orcid{0000-0001-8561-2679}\inst{\ref{aff8}}
\and T.~Schrabback\orcid{0000-0002-6987-7834}\inst{\ref{aff99}}
\and A.~Secroun\orcid{0000-0003-0505-3710}\inst{\ref{aff57}}
\and G.~Seidel\orcid{0000-0003-2907-353X}\inst{\ref{aff70}}
\and S.~Serrano\orcid{0000-0002-0211-2861}\inst{\ref{aff10},\ref{aff100},\ref{aff9}}
\and P.~Simon\inst{\ref{aff8}}
\and C.~Sirignano\orcid{0000-0002-0995-7146}\inst{\ref{aff101},\ref{aff56}}
\and G.~Sirri\orcid{0000-0003-2626-2853}\inst{\ref{aff27}}
\and L.~Stanco\orcid{0000-0002-9706-5104}\inst{\ref{aff56}}
\and J.~Steinwagner\orcid{0000-0001-7443-1047}\inst{\ref{aff62}}
\and P.~Tallada-Cresp\'{i}\orcid{0000-0002-1336-8328}\inst{\ref{aff43},\ref{aff44}}
\and A.~N.~Taylor\inst{\ref{aff29}}
\and I.~Tereno\orcid{0000-0002-4537-6218}\inst{\ref{aff102},\ref{aff103}}
\and N.~Tessore\orcid{0000-0002-9696-7931}\inst{\ref{aff104}}
\and S.~Toft\orcid{0000-0003-3631-7176}\inst{\ref{aff105},\ref{aff106}}
\and R.~Toledo-Moreo\orcid{0000-0002-2997-4859}\inst{\ref{aff107}}
\and F.~Torradeflot\orcid{0000-0003-1160-1517}\inst{\ref{aff44},\ref{aff43}}
\and I.~Tutusaus\orcid{0000-0002-3199-0399}\inst{\ref{aff9},\ref{aff10},\ref{aff96}}
\and J.~Valiviita\orcid{0000-0001-6225-3693}\inst{\ref{aff74},\ref{aff75}}
\and T.~Vassallo\orcid{0000-0001-6512-6358}\inst{\ref{aff11}}
\and Y.~Wang\orcid{0000-0002-4749-2984}\inst{\ref{aff108}}
\and J.~Weller\orcid{0000-0002-8282-2010}\inst{\ref{aff64},\ref{aff62}}
\and G.~Zamorani\orcid{0000-0002-2318-301X}\inst{\ref{aff24}}
\and F.~M.~Zerbi\inst{\ref{aff5}}
\and E.~Zucca\orcid{0000-0002-5845-8132}\inst{\ref{aff24}}
\and V.~Allevato\orcid{0000-0001-7232-5152}\inst{\ref{aff35}}
\and M.~Ballardini\orcid{0000-0003-4481-3559}\inst{\ref{aff109},\ref{aff110},\ref{aff24}}
\and M.~Bolzonella\orcid{0000-0003-3278-4607}\inst{\ref{aff24}}
\and A.~Boucaud\orcid{0000-0001-7387-2633}\inst{\ref{aff81}}
\and E.~Bozzo\orcid{0000-0002-8201-1525}\inst{\ref{aff55}}
\and C.~Burigana\orcid{0000-0002-3005-5796}\inst{\ref{aff111},\ref{aff60}}
\and R.~Cabanac\orcid{0000-0001-6679-2600}\inst{\ref{aff96}}
\and M.~Calabrese\orcid{0000-0002-2637-2422}\inst{\ref{aff112},\ref{aff40}}
\and A.~Cappi\inst{\ref{aff113},\ref{aff24}}
\and T.~Castro\orcid{0000-0002-6292-3228}\inst{\ref{aff11},\ref{aff13},\ref{aff12},\ref{aff15}}
\and J.~A.~Escartin~Vigo\inst{\ref{aff62}}
\and L.~Gabarra\orcid{0000-0002-8486-8856}\inst{\ref{aff30}}
\and J.~Garc\'ia-Bellido\orcid{0000-0002-9370-8360}\inst{\ref{aff114}}
\and V.~Gautard\inst{\ref{aff115}}
\and J.~Macias-Perez\orcid{0000-0002-5385-2763}\inst{\ref{aff116}}
\and R.~Maoli\orcid{0000-0002-6065-3025}\inst{\ref{aff117},\ref{aff41}}
\and J.~Mart\'{i}n-Fleitas\orcid{0000-0002-8594-569X}\inst{\ref{aff118}}
\and M.~Maturi\orcid{0000-0002-3517-2422}\inst{\ref{aff33},\ref{aff119}}
\and N.~Mauri\orcid{0000-0001-8196-1548}\inst{\ref{aff47},\ref{aff27}}
\and R.~B.~Metcalf\orcid{0000-0003-3167-2574}\inst{\ref{aff26},\ref{aff24}}
\and P.~Monaco\orcid{0000-0003-2083-7564}\inst{\ref{aff120},\ref{aff11},\ref{aff13},\ref{aff12}}
\and M.~P\"ontinen\orcid{0000-0001-5442-2530}\inst{\ref{aff74}}
\and I.~Risso\orcid{0000-0003-2525-7761}\inst{\ref{aff5},\ref{aff6}}
\and V.~Scottez\orcid{0009-0008-3864-940X}\inst{\ref{aff84},\ref{aff121}}
\and M.~Sereno\orcid{0000-0003-0302-0325}\inst{\ref{aff24},\ref{aff27}}
\and M.~Tenti\orcid{0000-0002-4254-5901}\inst{\ref{aff27}}
\and M.~Tucci\inst{\ref{aff55}}
\and M.~Viel\orcid{0000-0002-2642-5707}\inst{\ref{aff12},\ref{aff11},\ref{aff14},\ref{aff13},\ref{aff15}}
\and M.~Wiesmann\orcid{0009-0000-8199-5860}\inst{\ref{aff65}}
\and Y.~Akrami\orcid{0000-0002-2407-7956}\inst{\ref{aff114},\ref{aff122}}
\and I.~T.~Andika\orcid{0000-0001-6102-9526}\inst{\ref{aff61},\ref{aff123}}
\and G.~Angora\orcid{0000-0002-0316-6562}\inst{\ref{aff35},\ref{aff109}}
\and M.~Archidiacono\orcid{0000-0003-4952-9012}\inst{\ref{aff3},\ref{aff20}}
\and F.~Atrio-Barandela\orcid{0000-0002-2130-2513}\inst{\ref{aff124}}
\and E.~Aubourg\orcid{0000-0002-5592-023X}\inst{\ref{aff81},\ref{aff125}}
\and L.~Bazzanini\orcid{0000-0003-0727-0137}\inst{\ref{aff109},\ref{aff24}}
\and J.~Bel\inst{\ref{aff2}}
\and D.~Bertacca\orcid{0000-0002-2490-7139}\inst{\ref{aff101},\ref{aff63},\ref{aff56}}
\and M.~Bethermin\orcid{0000-0002-3915-2015}\inst{\ref{aff126}}
\and F.~Beutler\orcid{0000-0003-0467-5438}\inst{\ref{aff29}}
\and L.~Blot\orcid{0000-0002-9622-7167}\inst{\ref{aff127},\ref{aff76}}
\and M.~Bonici\orcid{0000-0002-8430-126X}\inst{\ref{aff91},\ref{aff40}}
\and S.~Borgani\orcid{0000-0001-6151-6439}\inst{\ref{aff120},\ref{aff12},\ref{aff11},\ref{aff13},\ref{aff15}}
\and M.~L.~Brown\orcid{0000-0002-0370-8077}\inst{\ref{aff128}}
\and S.~Bruton\orcid{0000-0002-6503-5218}\inst{\ref{aff129}}
\and A.~Calabro\orcid{0000-0003-2536-1614}\inst{\ref{aff41}}
\and F.~Caro\inst{\ref{aff41}}
\and C.~S.~Carvalho\inst{\ref{aff103}}
\and F.~Cogato\orcid{0000-0003-4632-6113}\inst{\ref{aff26},\ref{aff24}}
\and S.~Conseil\orcid{0000-0002-3657-4191}\inst{\ref{aff50}}
\and A.~R.~Cooray\orcid{0000-0002-3892-0190}\inst{\ref{aff130}}
\and S.~Davini\orcid{0000-0003-3269-1718}\inst{\ref{aff6}}
\and G.~Desprez\orcid{0000-0001-8325-1742}\inst{\ref{aff131}}
\and A.~D\'iaz-S\'anchez\orcid{0000-0003-0748-4768}\inst{\ref{aff132}}
\and S.~Di~Domizio\orcid{0000-0003-2863-5895}\inst{\ref{aff7},\ref{aff6}}
\and J.~M.~Diego\orcid{0000-0001-9065-3926}\inst{\ref{aff133}}
\and V.~Duret\orcid{0009-0009-0383-4960}\inst{\ref{aff57}}
\and M.~Y.~Elkhashab\orcid{0000-0001-9306-2603}\inst{\ref{aff11},\ref{aff13},\ref{aff120},\ref{aff12}}
\and A.~Enia\orcid{0000-0002-0200-2857}\inst{\ref{aff24}}
\and Y.~Fang\orcid{0000-0002-0334-6950}\inst{\ref{aff64}}
\and A.~G.~Ferrari\orcid{0009-0005-5266-4110}\inst{\ref{aff27}}
\and P.~G.~Ferreira\orcid{0000-0002-3021-2851}\inst{\ref{aff30}}
\and A.~Finoguenov\orcid{0000-0002-4606-5403}\inst{\ref{aff74}}
\and A.~Fontana\orcid{0000-0003-3820-2823}\inst{\ref{aff41}}
\and F.~Fontanot\orcid{0000-0003-4744-0188}\inst{\ref{aff11},\ref{aff12}}
\and A.~Franco\orcid{0000-0002-4761-366X}\inst{\ref{aff134},\ref{aff135},\ref{aff136}}
\and K.~Ganga\orcid{0000-0001-8159-8208}\inst{\ref{aff81}}
\and T.~Gasparetto\orcid{0000-0002-7913-4866}\inst{\ref{aff41}}
\and E.~Gaztanaga\orcid{0000-0001-9632-0815}\inst{\ref{aff9},\ref{aff10},\ref{aff137}}
\and F.~Giacomini\orcid{0000-0002-3129-2814}\inst{\ref{aff27}}
\and F.~Gianotti\orcid{0000-0003-4666-119X}\inst{\ref{aff24}}
\and G.~Gozaliasl\orcid{0000-0002-0236-919X}\inst{\ref{aff138},\ref{aff74}}
\and A.~Gruppuso\orcid{0000-0001-9272-5292}\inst{\ref{aff24},\ref{aff27}}
\and C.~M.~Gutierrez\orcid{0000-0001-7854-783X}\inst{\ref{aff139}}
\and A.~Hall\orcid{0000-0002-3139-8651}\inst{\ref{aff29}}
\and H.~Hildebrandt\orcid{0000-0002-9814-3338}\inst{\ref{aff140}}
\and J.~Hjorth\orcid{0000-0002-4571-2306}\inst{\ref{aff90}}
\and S.~Joudaki\orcid{0000-0001-8820-673X}\inst{\ref{aff43},\ref{aff137}}
\and J.~J.~E.~Kajava\orcid{0000-0002-3010-8333}\inst{\ref{aff141},\ref{aff142}}
\and Y.~Kang\orcid{0009-0000-8588-7250}\inst{\ref{aff55}}
\and V.~Kansal\orcid{0000-0002-4008-6078}\inst{\ref{aff143},\ref{aff144}}
\and D.~Karagiannis\orcid{0000-0002-4927-0816}\inst{\ref{aff109},\ref{aff145}}
\and K.~Kiiveri\inst{\ref{aff73}}
\and J.~Kim\orcid{0000-0003-2776-2761}\inst{\ref{aff30}}
\and C.~C.~Kirkpatrick\inst{\ref{aff73}}
\and S.~Kruk\orcid{0000-0001-8010-8879}\inst{\ref{aff32}}
\and M.~Lattanzi\orcid{0000-0003-1059-2532}\inst{\ref{aff110}}
\and J.~Le~Graet\orcid{0000-0001-6523-7971}\inst{\ref{aff57}}
\and L.~Legrand\orcid{0000-0003-0610-5252}\inst{\ref{aff146},\ref{aff147}}
\and M.~Lembo\orcid{0000-0002-5271-5070}\inst{\ref{aff85},\ref{aff110}}
\and F.~Lepori\orcid{0009-0000-5061-7138}\inst{\ref{aff148}}
\and G.~Leroy\orcid{0009-0004-2523-4425}\inst{\ref{aff149},\ref{aff80}}
\and G.~F.~Lesci\orcid{0000-0002-4607-2830}\inst{\ref{aff26},\ref{aff24}}
\and J.~Lesgourgues\orcid{0000-0001-7627-353X}\inst{\ref{aff150}}
\and T.~I.~Liaudat\orcid{0000-0002-9104-314X}\inst{\ref{aff125}}
\and M.~Magliocchetti\orcid{0000-0001-9158-4838}\inst{\ref{aff58}}
\and A.~Manj\'on-Garc\'ia\orcid{0000-0002-7413-8825}\inst{\ref{aff132}}
\and F.~Mannucci\orcid{0000-0002-4803-2381}\inst{\ref{aff151}}
\and C.~J.~A.~P.~Martins\orcid{0000-0002-4886-9261}\inst{\ref{aff152},\ref{aff153}}
\and L.~Maurin\orcid{0000-0002-8406-0857}\inst{\ref{aff31}}
\and M.~Migliaccio\inst{\ref{aff154},\ref{aff155}}
\and M.~Miluzio\inst{\ref{aff32},\ref{aff156}}
\and A.~Montoro\orcid{0000-0003-4730-8590}\inst{\ref{aff9},\ref{aff10}}
\and G.~Morgante\inst{\ref{aff24}}
\and S.~Nadathur\orcid{0000-0001-9070-3102}\inst{\ref{aff137}}
\and K.~Naidoo\orcid{0000-0002-9182-1802}\inst{\ref{aff137},\ref{aff70}}
\and P.~Natoli\orcid{0000-0003-0126-9100}\inst{\ref{aff109},\ref{aff110}}
\and A.~Navarro-Alsina\orcid{0000-0002-3173-2592}\inst{\ref{aff8}}
\and S.~Nesseris\orcid{0000-0002-0567-0324}\inst{\ref{aff114}}
\and L.~Pagano\orcid{0000-0003-1820-5998}\inst{\ref{aff109},\ref{aff110}}
\and D.~Paoletti\orcid{0000-0003-4761-6147}\inst{\ref{aff24},\ref{aff60}}
\and F.~Passalacqua\orcid{0000-0002-8606-4093}\inst{\ref{aff101},\ref{aff56}}
\and K.~Paterson\orcid{0000-0001-8340-3486}\inst{\ref{aff70}}
\and L.~Patrizii\inst{\ref{aff27}}
\and R.~Paviot\orcid{0009-0002-8108-3460}\inst{\ref{aff4}}
\and A.~Pisani\orcid{0000-0002-6146-4437}\inst{\ref{aff57}}
\and D.~Potter\orcid{0000-0002-0757-5195}\inst{\ref{aff148}}
\and G.~W.~Pratt\inst{\ref{aff4}}
\and S.~Quai\orcid{0000-0002-0449-8163}\inst{\ref{aff26},\ref{aff24}}
\and M.~Radovich\orcid{0000-0002-3585-866X}\inst{\ref{aff63}}
\and K.~Rojas\orcid{0000-0003-1391-6854}\inst{\ref{aff79}}
\and W.~Roster\orcid{0000-0002-9149-6528}\inst{\ref{aff62}}
\and S.~Sacquegna\orcid{0000-0002-8433-6630}\inst{\ref{aff157}}
\and M.~Sahl\'en\orcid{0000-0003-0973-4804}\inst{\ref{aff158}}
\and D.~B.~Sanders\orcid{0000-0002-1233-9998}\inst{\ref{aff46}}
\and A.~Schneider\orcid{0000-0001-7055-8104}\inst{\ref{aff148}}
\and D.~Sciotti\orcid{0009-0008-4519-2620}\inst{\ref{aff41},\ref{aff42}}
\and E.~Sellentin\inst{\ref{aff159},\ref{aff39}}
\and L.~C.~Smith\orcid{0000-0002-3259-2771}\inst{\ref{aff160}}
\and J.~G.~Sorce\orcid{0000-0002-2307-2432}\inst{\ref{aff161},\ref{aff31}}
\and K.~Tanidis\orcid{0000-0001-9843-5130}\inst{\ref{aff30}}
\and C.~Tao\orcid{0000-0001-7961-8177}\inst{\ref{aff57}}
\and F.~Tarsitano\orcid{0000-0002-5919-0238}\inst{\ref{aff162},\ref{aff55}}
\and G.~Testera\inst{\ref{aff6}}
\and R.~Teyssier\orcid{0000-0001-7689-0933}\inst{\ref{aff163}}
\and S.~Tosi\orcid{0000-0002-7275-9193}\inst{\ref{aff7},\ref{aff6},\ref{aff5}}
\and A.~Troja\orcid{0000-0003-0239-4595}\inst{\ref{aff101},\ref{aff56}}
\and A.~Venhola\orcid{0000-0001-6071-4564}\inst{\ref{aff164}}
\and D.~Vergani\orcid{0000-0003-0898-2216}\inst{\ref{aff24}}
\and F.~Vernizzi\orcid{0000-0003-3426-2802}\inst{\ref{aff165}}
\and G.~Verza\orcid{0000-0002-1886-8348}\inst{\ref{aff166},\ref{aff167}}
\and S.~Vinciguerra\orcid{0009-0005-4018-3184}\inst{\ref{aff1}}
\and N.~A.~Walton\orcid{0000-0003-3983-8778}\inst{\ref{aff160}}
\and A.~H.~Wright\orcid{0000-0001-7363-7932}\inst{\ref{aff140}}}
										   
\institute{Aix-Marseille Universit\'e, CNRS, CNES, LAM, Marseille, France\label{aff1}
\and
Aix-Marseille Universit\'e, Universit\'e de Toulon, CNRS, CPT, Marseille, France\label{aff2}
\and
Dipartimento di Fisica "Aldo Pontremoli", Universit\`a degli Studi di Milano, Via Celoria 16, 20133 Milano, Italy\label{aff3}
\and
Universit\'e Paris-Saclay, Universit\'e Paris Cit\'e, CEA, CNRS, AIM, 91191, Gif-sur-Yvette, France\label{aff4}
\and
INAF-Osservatorio Astronomico di Brera, Via Brera 28, 20122 Milano, Italy\label{aff5}
\and
INFN-Sezione di Genova, Via Dodecaneso 33, 16146, Genova, Italy\label{aff6}
\and
Dipartimento di Fisica, Universit\`a di Genova, Via Dodecaneso 33, 16146, Genova, Italy\label{aff7}
\and
Universit\"at Bonn, Argelander-Institut f\"ur Astronomie, Auf dem H\"ugel 71, 53121 Bonn, Germany\label{aff8}
\and
Institute of Space Sciences (ICE, CSIC), Campus UAB, Carrer de Can Magrans, s/n, 08193 Barcelona, Spain\label{aff9}
\and
Institut d'Estudis Espacials de Catalunya (IEEC),  Edifici RDIT, Campus UPC, 08860 Castelldefels, Barcelona, Spain\label{aff10}
\and
INAF-Osservatorio Astronomico di Trieste, Via G. B. Tiepolo 11, 34143 Trieste, Italy\label{aff11}
\and
IFPU, Institute for Fundamental Physics of the Universe, via Beirut 2, 34151 Trieste, Italy\label{aff12}
\and
INFN, Sezione di Trieste, Via Valerio 2, 34127 Trieste TS, Italy\label{aff13}
\and
SISSA, International School for Advanced Studies, Via Bonomea 265, 34136 Trieste TS, Italy\label{aff14}
\and
ICSC - Centro Nazionale di Ricerca in High Performance Computing, Big Data e Quantum Computing, Via Magnanelli 2, Bologna, Italy\label{aff15}
\and
Donostia International Physics Center (DIPC), Paseo Manuel de Lardizabal, 4, 20018, Donostia-San Sebasti\'an, Guipuzkoa, Spain\label{aff16}
\and
IKERBASQUE, Basque Foundation for Science, 48013, Bilbao, Spain\label{aff17}
\and
School of Mathematics, Statistics and Physics, Newcastle University, Herschel Building, Newcastle-upon-Tyne, NE1 7RU, UK\label{aff18}
\and
Fakult\"at f\"ur Physik, Universit\"at Bielefeld, Postfach 100131, 33501 Bielefeld, Germany\label{aff19}
\and
INFN-Sezione di Milano, Via Celoria 16, 20133 Milano, Italy\label{aff20}
\and
Universit\'e de Gen\`eve, D\'epartement de Physique Th\'eorique and Centre for Astroparticle Physics, 24 quai Ernest-Ansermet, CH-1211 Gen\`eve 4, Switzerland\label{aff21}
\and
Technion Israel Institute of Technology, Israel\label{aff22}
\and
Dipartimento di Fisica e Astronomia, Universit\`a di Bologna, Via Gobetti 93/2, 40129 Bologna, Italy\label{aff23}
\and
INAF-Osservatorio di Astrofisica e Scienza dello Spazio di Bologna, Via Piero Gobetti 93/3, 40129 Bologna, Italy\label{aff24}
\and
INFN Gruppo Collegato di Parma, Viale delle Scienze 7/A 43124 Parma, Italy\label{aff25}
\and
Dipartimento di Fisica e Astronomia "Augusto Righi" - Alma Mater Studiorum Universit\`a di Bologna, via Piero Gobetti 93/2, 40129 Bologna, Italy\label{aff26}
\and
INFN-Sezione di Bologna, Viale Berti Pichat 6/2, 40127 Bologna, Italy\label{aff27}
\and
Laboratoire d'Annecy-le-Vieux de Physique Theorique, CNRS \& Universite Savoie Mont Blanc, 9 Chemin de Bellevue, BP 110, Annecy-le-Vieux, 74941 ANNECY Cedex, France\label{aff28}
\and
Institute for Astronomy, University of Edinburgh, Royal Observatory, Blackford Hill, Edinburgh EH9 3HJ, UK\label{aff29}
\and
Department of Physics, Oxford University, Keble Road, Oxford OX1 3RH, UK\label{aff30}
\and
Universit\'e Paris-Saclay, CNRS, Institut d'astrophysique spatiale, 91405, Orsay, France\label{aff31}
\and
ESAC/ESA, Camino Bajo del Castillo, s/n., Urb. Villafranca del Castillo, 28692 Villanueva de la Ca\~nada, Madrid, Spain\label{aff32}
\and
Institut f\"ur Theoretische Physik, University of Heidelberg, Philosophenweg 16, 69120 Heidelberg, Germany\label{aff33}
\and
Department of Physics "E. Pancini", University Federico II, Via Cinthia 6, 80126, Napoli, Italy\label{aff34}
\and
INAF-Osservatorio Astronomico di Capodimonte, Via Moiariello 16, 80131 Napoli, Italy\label{aff35}
\and
Dipartimento di Fisica, Universit\`a degli Studi di Torino, Via P. Giuria 1, 10125 Torino, Italy\label{aff36}
\and
INFN-Sezione di Torino, Via P. Giuria 1, 10125 Torino, Italy\label{aff37}
\and
INAF-Osservatorio Astrofisico di Torino, Via Osservatorio 20, 10025 Pino Torinese (TO), Italy\label{aff38}
\and
Leiden Observatory, Leiden University, Einsteinweg 55, 2333 CC Leiden, The Netherlands\label{aff39}
\and
INAF-IASF Milano, Via Alfonso Corti 12, 20133 Milano, Italy\label{aff40}
\and
INAF-Osservatorio Astronomico di Roma, Via Frascati 33, 00078 Monteporzio Catone, Italy\label{aff41}
\and
INFN-Sezione di Roma, Piazzale Aldo Moro, 2 - c/o Dipartimento di Fisica, Edificio G. Marconi, 00185 Roma, Italy\label{aff42}
\and
Centro de Investigaciones Energ\'eticas, Medioambientales y Tecnol\'ogicas (CIEMAT), Avenida Complutense 40, 28040 Madrid, Spain\label{aff43}
\and
Port d'Informaci\'{o} Cient\'{i}fica, Campus UAB, C. Albareda s/n, 08193 Bellaterra (Barcelona), Spain\label{aff44}
\and
INFN section of Naples, Via Cinthia 6, 80126, Napoli, Italy\label{aff45}
\and
Institute for Astronomy, University of Hawaii, 2680 Woodlawn Drive, Honolulu, HI 96822, USA\label{aff46}
\and
Dipartimento di Fisica e Astronomia "Augusto Righi" - Alma Mater Studiorum Universit\`a di Bologna, Viale Berti Pichat 6/2, 40127 Bologna, Italy\label{aff47}
\and
Instituto de Astrof\'{\i}sica de Canarias, E-38205 La Laguna, Tenerife, Spain\label{aff48}
\and
European Space Agency/ESRIN, Largo Galileo Galilei 1, 00044 Frascati, Roma, Italy\label{aff49}
\and
Universit\'e Claude Bernard Lyon 1, CNRS/IN2P3, IP2I Lyon, UMR 5822, Villeurbanne, F-69100, France\label{aff50}
\and
Institut de Ci\`{e}ncies del Cosmos (ICCUB), Universitat de Barcelona (IEEC-UB), Mart\'{i} i Franqu\`{e}s 1, 08028 Barcelona, Spain\label{aff51}
\and
Instituci\'o Catalana de Recerca i Estudis Avan\c{c}ats (ICREA), Passeig de Llu\'{\i}s Companys 23, 08010 Barcelona, Spain\label{aff52}
\and
Institut de Ciencies de l'Espai (IEEC-CSIC), Campus UAB, Carrer de Can Magrans, s/n Cerdanyola del Vall\'es, 08193 Barcelona, Spain\label{aff53}
\and
UCB Lyon 1, CNRS/IN2P3, IUF, IP2I Lyon, 4 rue Enrico Fermi, 69622 Villeurbanne, France\label{aff54}
\and
Department of Astronomy, University of Geneva, ch. d'Ecogia 16, 1290 Versoix, Switzerland\label{aff55}
\and
INFN-Padova, Via Marzolo 8, 35131 Padova, Italy\label{aff56}
\and
Aix-Marseille Universit\'e, CNRS/IN2P3, CPPM, Marseille, France\label{aff57}
\and
INAF-Istituto di Astrofisica e Planetologia Spaziali, via del Fosso del Cavaliere, 100, 00100 Roma, Italy\label{aff58}
\and
Space Science Data Center, Italian Space Agency, via del Politecnico snc, 00133 Roma, Italy\label{aff59}
\and
INFN-Bologna, Via Irnerio 46, 40126 Bologna, Italy\label{aff60}
\and
University Observatory, LMU Faculty of Physics, Scheinerstr.~1, 81679 Munich, Germany\label{aff61}
\and
Max Planck Institute for Extraterrestrial Physics, Giessenbachstr. 1, 85748 Garching, Germany\label{aff62}
\and
INAF-Osservatorio Astronomico di Padova, Via dell'Osservatorio 5, 35122 Padova, Italy\label{aff63}
\and
Universit\"ats-Sternwarte M\"unchen, Fakult\"at f\"ur Physik, Ludwig-Maximilians-Universit\"at M\"unchen, Scheinerstr.~1, 81679 M\"unchen, Germany\label{aff64}
\and
Institute of Theoretical Astrophysics, University of Oslo, P.O. Box 1029 Blindern, 0315 Oslo, Norway\label{aff65}
\and
Jet Propulsion Laboratory, California Institute of Technology, 4800 Oak Grove Drive, Pasadena, CA, 91109, USA\label{aff66}
\and
Felix Hormuth Engineering, Goethestr. 17, 69181 Leimen, Germany\label{aff67}
\and
Technical University of Denmark, Elektrovej 327, 2800 Kgs. Lyngby, Denmark\label{aff68}
\and
Cosmic Dawn Center (DAWN), Denmark\label{aff69}
\and
Max-Planck-Institut f\"ur Astronomie, K\"onigstuhl 17, 69117 Heidelberg, Germany\label{aff70}
\and
NASA Goddard Space Flight Center, Greenbelt, MD 20771, USA\label{aff71}
\and
Department of Physics and Astronomy, University College London, Gower Street, London WC1E 6BT, UK\label{aff72}
\and
Department of Physics and Helsinki Institute of Physics, Gustaf H\"allstr\"omin katu 2, University of Helsinki, 00014 Helsinki, Finland\label{aff73}
\and
Department of Physics, P.O. Box 64, University of Helsinki, 00014 Helsinki, Finland\label{aff74}
\and
Helsinki Institute of Physics, Gustaf H{\"a}llstr{\"o}min katu 2, University of Helsinki, 00014 Helsinki, Finland\label{aff75}
\and
Laboratoire d'etude de l'Univers et des phenomenes eXtremes, Observatoire de Paris, Universit\'e PSL, Sorbonne Universit\'e, CNRS, 92190 Meudon, France\label{aff76}
\and
SKAO, Jodrell Bank, Lower Withington, Macclesfield SK11 9FT, UK\label{aff77}
\and
Centre de Calcul de l'IN2P3/CNRS, 21 avenue Pierre de Coubertin 69627 Villeurbanne Cedex, France\label{aff78}
\and
University of Applied Sciences and Arts of Northwestern Switzerland, School of Computer Science, 5210 Windisch, Switzerland\label{aff79}
\and
Department of Physics, Institute for Computational Cosmology, Durham University, South Road, Durham, DH1 3LE, UK\label{aff80}
\and
Universit\'e Paris Cit\'e, CNRS, Astroparticule et Cosmologie, 75013 Paris, France\label{aff81}
\and
CNRS-UCB International Research Laboratory, Centre Pierre Bin\'etruy, IRL2007, CPB-IN2P3, Berkeley, USA\label{aff82}
\and
University of Applied Sciences and Arts of Northwestern Switzerland, School of Engineering, 5210 Windisch, Switzerland\label{aff83}
\and
Institut d'Astrophysique de Paris, 98bis Boulevard Arago, 75014, Paris, France\label{aff84}
\and
Institut d'Astrophysique de Paris, UMR 7095, CNRS, and Sorbonne Universit\'e, 98 bis boulevard Arago, 75014 Paris, France\label{aff85}
\and
Institute of Physics, Laboratory of Astrophysics, Ecole Polytechnique F\'ed\'erale de Lausanne (EPFL), Observatoire de Sauverny, 1290 Versoix, Switzerland\label{aff86}
\and
Telespazio UK S.L. for European Space Agency (ESA), Camino bajo del Castillo, s/n, Urbanizacion Villafranca del Castillo, Villanueva de la Ca\~nada, 28692 Madrid, Spain\label{aff87}
\and
Institut de F\'{i}sica d'Altes Energies (IFAE), The Barcelona Institute of Science and Technology, Campus UAB, 08193 Bellaterra (Barcelona), Spain\label{aff88}
\and
European Space Agency/ESTEC, Keplerlaan 1, 2201 AZ Noordwijk, The Netherlands\label{aff89}
\and
DARK, Niels Bohr Institute, University of Copenhagen, Jagtvej 155, 2200 Copenhagen, Denmark\label{aff90}
\and
Waterloo Centre for Astrophysics, University of Waterloo, Waterloo, Ontario N2L 3G1, Canada\label{aff91}
\and
Department of Physics and Astronomy, University of Waterloo, Waterloo, Ontario N2L 3G1, Canada\label{aff92}
\and
Perimeter Institute for Theoretical Physics, Waterloo, Ontario N2L 2Y5, Canada\label{aff93}
\and
Centre National d'Etudes Spatiales -- Centre spatial de Toulouse, 18 avenue Edouard Belin, 31401 Toulouse Cedex 9, France\label{aff94}
\and
Institute of Space Science, Str. Atomistilor, nr. 409 M\u{a}gurele, Ilfov, 077125, Romania\label{aff95}
\and
Institut de Recherche en Astrophysique et Plan\'etologie (IRAP), Universit\'e de Toulouse, CNRS, UPS, CNES, 14 Av. Edouard Belin, 31400 Toulouse, France\label{aff96}
\and
Universit\'e St Joseph; Faculty of Sciences, Beirut, Lebanon\label{aff97}
\and
Departamento de F\'isica, FCFM, Universidad de Chile, Blanco Encalada 2008, Santiago, Chile\label{aff98}
\and
Universit\"at Innsbruck, Institut f\"ur Astro- und Teilchenphysik, Technikerstr. 25/8, 6020 Innsbruck, Austria\label{aff99}
\and
Satlantis, University Science Park, Sede Bld 48940, Leioa-Bilbao, Spain\label{aff100}
\and
Dipartimento di Fisica e Astronomia "G. Galilei", Universit\`a di Padova, Via Marzolo 8, 35131 Padova, Italy\label{aff101}
\and
Departamento de F\'isica, Faculdade de Ci\^encias, Universidade de Lisboa, Edif\'icio C8, Campo Grande, PT1749-016 Lisboa, Portugal\label{aff102}
\and
Instituto de Astrof\'isica e Ci\^encias do Espa\c{c}o, Faculdade de Ci\^encias, Universidade de Lisboa, Tapada da Ajuda, 1349-018 Lisboa, Portugal\label{aff103}
\and
Mullard Space Science Laboratory, University College London, Holmbury St Mary, Dorking, Surrey RH5 6NT, UK\label{aff104}
\and
Cosmic Dawn Center (DAWN)\label{aff105}
\and
Niels Bohr Institute, University of Copenhagen, Jagtvej 128, 2200 Copenhagen, Denmark\label{aff106}
\and
Universidad Polit\'ecnica de Cartagena, Departamento de Electr\'onica y Tecnolog\'ia de Computadoras,  Plaza del Hospital 1, 30202 Cartagena, Spain\label{aff107}
\and
Caltech/IPAC, 1200 E. California Blvd., Pasadena, CA 91125, USA\label{aff108}
\and
Dipartimento di Fisica e Scienze della Terra, Universit\`a degli Studi di Ferrara, Via Giuseppe Saragat 1, 44122 Ferrara, Italy\label{aff109}
\and
Istituto Nazionale di Fisica Nucleare, Sezione di Ferrara, Via Giuseppe Saragat 1, 44122 Ferrara, Italy\label{aff110}
\and
INAF, Istituto di Radioastronomia, Via Piero Gobetti 101, 40129 Bologna, Italy\label{aff111}
\and
Astronomical Observatory of the Autonomous Region of the Aosta Valley (OAVdA), Loc. Lignan 39, I-11020, Nus (Aosta Valley), Italy\label{aff112}
\and
Universit\'e C\^{o}te d'Azur, Observatoire de la C\^{o}te d'Azur, CNRS, Laboratoire Lagrange, Bd de l'Observatoire, CS 34229, 06304 Nice cedex 4, France\label{aff113}
\and
Instituto de F\'isica Te\'orica UAM-CSIC, Campus de Cantoblanco, 28049 Madrid, Spain\label{aff114}
\and
CEA Saclay, DFR/IRFU, Service d'Astrophysique, Bat. 709, 91191 Gif-sur-Yvette, France\label{aff115}
\and
Univ. Grenoble Alpes, CNRS, Grenoble INP, LPSC-IN2P3, 53, Avenue des Martyrs, 38000, Grenoble, France\label{aff116}
\and
Dipartimento di Fisica, Sapienza Universit\`a di Roma, Piazzale Aldo Moro 2, 00185 Roma, Italy\label{aff117}
\and
Aurora Technology for European Space Agency (ESA), Camino bajo del Castillo, s/n, Urbanizacion Villafranca del Castillo, Villanueva de la Ca\~nada, 28692 Madrid, Spain\label{aff118}
\and
Zentrum f\"ur Astronomie, Universit\"at Heidelberg, Philosophenweg 12, 69120 Heidelberg, Germany\label{aff119}
\and
Dipartimento di Fisica - Sezione di Astronomia, Universit\`a di Trieste, Via Tiepolo 11, 34131 Trieste, Italy\label{aff120}
\and
ICL, Junia, Universit\'e Catholique de Lille, LITL, 59000 Lille, France\label{aff121}
\and
CERCA/ISO, Department of Physics, Case Western Reserve University, 10900 Euclid Avenue, Cleveland, OH 44106, USA\label{aff122}
\and
Technical University of Munich, TUM School of Natural Sciences, Physics Department, James-Franck-Str.~1, 85748 Garching, Germany\label{aff123}
\and
Departamento de F{\'\i}sica Fundamental. Universidad de Salamanca. Plaza de la Merced s/n. 37008 Salamanca, Spain\label{aff124}
\and
IRFU, CEA, Universit\'e Paris-Saclay 91191 Gif-sur-Yvette Cedex, France\label{aff125}
\and
Universit\'e de Strasbourg, CNRS, Observatoire astronomique de Strasbourg, UMR 7550, 67000 Strasbourg, France\label{aff126}
\and
Center for Data-Driven Discovery, Kavli IPMU (WPI), UTIAS, The University of Tokyo, Kashiwa, Chiba 277-8583, Japan\label{aff127}
\and
Jodrell Bank Centre for Astrophysics, Department of Physics and Astronomy, University of Manchester, Oxford Road, Manchester M13 9PL, UK\label{aff128}
\and
California Institute of Technology, 1200 E California Blvd, Pasadena, CA 91125, USA\label{aff129}
\and
Department of Physics \& Astronomy, University of California Irvine, Irvine CA 92697, USA\label{aff130}
\and
Kapteyn Astronomical Institute, University of Groningen, PO Box 800, 9700 AV Groningen, The Netherlands\label{aff131}
\and
Departamento F\'isica Aplicada, Universidad Polit\'ecnica de Cartagena, Campus Muralla del Mar, 30202 Cartagena, Murcia, Spain\label{aff132}
\and
Instituto de F\'isica de Cantabria, Edificio Juan Jord\'a, Avenida de los Castros, 39005 Santander, Spain\label{aff133}
\and
INFN, Sezione di Lecce, Via per Arnesano, CP-193, 73100, Lecce, Italy\label{aff134}
\and
Department of Mathematics and Physics E. De Giorgi, University of Salento, Via per Arnesano, CP-I93, 73100, Lecce, Italy\label{aff135}
\and
INAF-Sezione di Lecce, c/o Dipartimento Matematica e Fisica, Via per Arnesano, 73100, Lecce, Italy\label{aff136}
\and
Institute of Cosmology and Gravitation, University of Portsmouth, Portsmouth PO1 3FX, UK\label{aff137}
\and
Department of Computer Science, Aalto University, PO Box 15400, Espoo, FI-00 076, Finland\label{aff138}
\and
 Instituto de Astrof\'{\i}sica de Canarias, E-38205 La Laguna; Universidad de La Laguna, Dpto. Astrof\'\i sica, E-38206 La Laguna, Tenerife, Spain\label{aff139}
\and
Ruhr University Bochum, Faculty of Physics and Astronomy, Astronomical Institute (AIRUB), German Centre for Cosmological Lensing (GCCL), 44780 Bochum, Germany\label{aff140}
\and
Department of Physics and Astronomy, Vesilinnantie 5, University of Turku, 20014 Turku, Finland\label{aff141}
\and
Serco for European Space Agency (ESA), Camino bajo del Castillo, s/n, Urbanizacion Villafranca del Castillo, Villanueva de la Ca\~nada, 28692 Madrid, Spain\label{aff142}
\and
ARC Centre of Excellence for Dark Matter Particle Physics, Melbourne, Australia\label{aff143}
\and
Centre for Astrophysics \& Supercomputing, Swinburne University of Technology,  Hawthorn, Victoria 3122, Australia\label{aff144}
\and
Department of Physics and Astronomy, University of the Western Cape, Bellville, Cape Town, 7535, South Africa\label{aff145}
\and
DAMTP, Centre for Mathematical Sciences, Wilberforce Road, Cambridge CB3 0WA, UK\label{aff146}
\and
Kavli Institute for Cosmology Cambridge, Madingley Road, Cambridge, CB3 0HA, UK\label{aff147}
\and
Department of Astrophysics, University of Zurich, Winterthurerstrasse 190, 8057 Zurich, Switzerland\label{aff148}
\and
Department of Physics, Centre for Extragalactic Astronomy, Durham University, South Road, Durham, DH1 3LE, UK\label{aff149}
\and
Institute for Theoretical Particle Physics and Cosmology (TTK), RWTH Aachen University, 52056 Aachen, Germany\label{aff150}
\and
INAF-Osservatorio Astrofisico di Arcetri, Largo E. Fermi 5, 50125, Firenze, Italy\label{aff151}
\and
Centro de Astrof\'{\i}sica da Universidade do Porto, Rua das Estrelas, 4150-762 Porto, Portugal\label{aff152}
\and
Instituto de Astrof\'isica e Ci\^encias do Espa\c{c}o, Universidade do Porto, CAUP, Rua das Estrelas, PT4150-762 Porto, Portugal\label{aff153}
\and
Dipartimento di Fisica, Universit\`a di Roma Tor Vergata, Via della Ricerca Scientifica 1, Roma, Italy\label{aff154}
\and
INFN, Sezione di Roma 2, Via della Ricerca Scientifica 1, Roma, Italy\label{aff155}
\and
HE Space for European Space Agency (ESA), Camino bajo del Castillo, s/n, Urbanizacion Villafranca del Castillo, Villanueva de la Ca\~nada, 28692 Madrid, Spain\label{aff156}
\and
INAF - Osservatorio Astronomico d'Abruzzo, Via Maggini, 64100, Teramo, Italy\label{aff157}
\and
Theoretical astrophysics, Department of Physics and Astronomy, Uppsala University, Box 516, 751 37 Uppsala, Sweden\label{aff158}
\and
Mathematical Institute, University of Leiden, Einsteinweg 55, 2333 CA Leiden, The Netherlands\label{aff159}
\and
Institute of Astronomy, University of Cambridge, Madingley Road, Cambridge CB3 0HA, UK\label{aff160}
\and
Univ. Lille, CNRS, Centrale Lille, UMR 9189 CRIStAL, 59000 Lille, France\label{aff161}
\and
Institute for Particle Physics and Astrophysics, Dept. of Physics, ETH Zurich, Wolfgang-Pauli-Strasse 27, 8093 Zurich, Switzerland\label{aff162}
\and
Department of Astrophysical Sciences, Peyton Hall, Princeton University, Princeton, NJ 08544, USA\label{aff163}
\and
Space physics and astronomy research unit, University of Oulu, Pentti Kaiteran katu 1, FI-90014 Oulu, Finland\label{aff164}
\and
Institut de Physique Th\'eorique, CEA, CNRS, Universit\'e Paris-Saclay 91191 Gif-sur-Yvette Cedex, France\label{aff165}
\and
International Centre for Theoretical Physics (ICTP), Strada Costiera 11, 34151 Trieste, Italy\label{aff166}
\and
Center for Computational Astrophysics, Flatiron Institute, 162 5th Avenue, 10010, New York, NY, USA\label{aff167}}    

%
%
\abstract{The \Euclid satellite will measure spectroscopic redshifts for tens of millions of emission-line galaxies, allowing for one of the most precise tests of the cosmological model. In the context of Stage-IV surveys such as \Euclid, the 3-dimensional clustering of galaxies plays a key role, both providing geometrical and dynamical cosmological constraints. In this paper, we conduct a comprehensive model-comparison campaign for the multipole moments of the galaxy 2-point correlation function (2PCF) in redshift space. We test state-of-the-art models, in particular the effective field theory of large-scale structure (EFT), one based on the velocity difference generating function (VDG$_{\infty}$), and different variants of Lagrangian perturbation theory (LPT) models, such as convolutional Lagrangian perturbation theory (CLPT) and convolutional Lagrangian effective field theory (CLEFT). We analyse the first three even multipole moments of the 2PCF in the \flagship simulation of emission-line galaxies, which  consists of four snapshots at $z\in\{0.9, 1.2, 1.5, 1.8\}$ covering the redshift range of the \Euclid spectroscopic sample. We study both template-fitting and full-shape approaches and compare the different models in terms of three performance metrics: reduced $\chi^2$, a figure of merit, and a figure of bias. We find that with the template-fitting approach, only the VDG$_{\infty}$ model is able to reach a minimum fitting scale of $\smin=20\Mpc$ at $z=0.9$ without biasing the recovered parameters. Indeed, the EFT model becomes inaccurate already at $\smin=30\Mpc$. Conversely, in the full-shape analysis, the CLEFT and VDG$_{\infty}$ models perform similarly well, but only the CLEFT model can reach $\smin=20\Mpc$ while the VDG$_{\infty}$ model is unbiased down to $\smin=25\Mpc$ at the lowest redshift. Overall, in order to achieve the accuracy required by \Euclid, non-perturbative modelling such as in the VDG$_{\infty}$ or CLEFT models should be considered. At the highest redshift probed by \Euclid, the CLPT model is sufficient to describe the data with high figure of merit. This comparison selects baseline models that perform best in ideal conditions and sets the stage for an optimal analysis of \Euclid data in configuration space.}

%
%
    \keywords{Cosmology: large-scale structure of Universe, theory, cosmological parameters}
%
%
   \titlerunning{2-point correlation function modelling}
   \authorrunning{Euclid Collaboration: M. K\"archer et al.}
   
   \maketitle
%
%
%
%
   
\section{Introduction}
In the last thirty years, large galaxy spectroscopic surveys opened up a window for precise measurements of the clustering of galaxies, essentially dictated by gravity on cosmological scales. The currently accepted theoretical description of our cosmos is the \lcdm model that consists of dark energy in the form of a cosmological constant $\Lambda$ and \cdm. The main probe to extract the clustering information from observations is the galaxy \tpcf or its Fourier counterpart, the galaxy power spectrum. Both statistics depend directly on the parameters of the cosmological model, making them a cosmological probe of prime interest to test our understanding of the Universe. The desire to sample larger and larger volumes of the Universe has led to the development of more comprehensive surveys, which go deeper and cover a larger area of the sky.
Among the most influential spectroscopic surveys are the 2dF Galaxy Redshift Survey \citep[2dFGRS,][]{Colles2001MNRAS_2dFGS}, the 6dF Galaxy Survey \citep[6dFGS,][]{ Jones2009MNRAS_6dFGS}, the Sloan Digital Sky Survey \citep[SDSS,][]{York2000AJ_SDSS}, the VIMOS Public Extragalactic Redshift Survey \citep[VIPERS,][]{Guzzo2014A&A_VIPERS}, the Baryon Oscillation Spectroscopic Survey \citep[BOSS,][]{Dawson2013AJ_BOSS}, and the extended BOSS \citep[eBOSS,][]{Dawson2016AJ_eBOSS}. Currently, Stage-IV surveys such as the Dark Energy Spectroscopic Instrument \citep[DESI,][]{DESI2016} or the \Euclid mission \citep{EuclidSkyOverview} are expected to measure the redshifts of tens of millions of galaxies, producing the largest 3-dimensional galaxy catalogues ever assembled.

The ever increasing precision of measurements has to go hand in hand with improved theoretical models to describe the two-point statistics of galaxies in redshift space. The first challenge is the modelling of the nonlinear clustering of matter. The classical approach for that is the use of Eulerian standard perturbation theory (EPT), often simply referred to as \spt, in which the full matter density contrast is expanded perturbatively up to the desired order, leading to loop corrections to the power spectrum \citep{GoroffEtal1986, Jain1994ApJ_2SPT_Pk}. We refer the interested reader to \citet{Bernardeau2002}, for an extensive review on the subject.
A different approach to \spt  is given by the Lagrangian picture. The latter formalism is based on a displacement field that contains the nonlinear evolution from an initially homogeneous matter distribution \citep{Zeldovich1970}.
\lpt revolves around finding solutions to a perturbative expansion of this displacement field \citep{Buchert1989A&A_LPT, Moutarde1991,Buchert1992MNRAS_1LPT,bouchet1995perturbative}. An advantage of \lpt is the improved modelling of the \bao feature and its damped behaviour, by using a natural resummation scheme \citep{Matsubara2008, Matsubara2008b}. 

Methods to resum terms in \spt were first proposed in the context of renormalised perturbation theory \citep[\glsentryshort{rpt},][]{Crocce2006PhRvD_a_RPT, Crocce2006PhRvD_b_RPT, Crocce2008}, in terms of a propagator expansion that captures the nonlinear evolution from initial densities. This approach leads to a better convergence of the perturbative series in contrast to \spt and has been extended in regularised perturbation theory \citep[RegPT,][]{Bernardeau2008, BernardeauPhysRevD_RegPt, Taruya2012_RegPT}.
Resummation approaches have ultimately converged to a technique known as `infrared-resummation' \citep[IR-resummation]{Seo2008ApJ_ir_resum}, which describes the full power spectrum as a smooth component plus a damped part containing the isolated \bao `wiggly' feature.

In the \eft a radically different approach to solving the problem of convergence in \spt is taken, in addition to relaxing the approximation of an ideal fluid \citep{BaumannEtal2012, Carrasco2012}. Using an explicit split of the density contrast into long- and short-wavelength modes leads to a well-behaved perturbative expansion, where small-scale physical effects are kept non-perturbative and are included via a set of counterterms.
The \eft model led to a major leap in reaching the most nonlinear scales in clustering down to a few times $0.1\kMpc$ in terms of $k$ -- the absolute value of the wavevector -- depending on the considered redshift \citep{Foreman2016JCAP_EFT_simulation_test}.
While originally developed within the Eulerian framework, a similar effective-field approach has also been devised in the Lagrangian picture \citep{Porto2014,Vlah2015}.

The modelling of galaxy 2-point statistics is further complicated by the fact that we are sensitive only to the galaxy density field merely tracing the underlying matter field. The gap between observations and matter predictions is resolved by introducing a `galaxy bias' in the model \citep[see][for a review]{Desjacques2018PhR}. A purely local bias description is given by a power series of the galaxy density in terms of the dark matter density field \citep{Fry1993}.
However, to preserve a zero mean for the galaxy density contrast, the expansion depends on the density variance and renormalised bias parameters have to be defined \citep{McDonald2006PhRvD_Renorm_Bias, McDonald2009JCAP, Assassi2014JCAP}.
Furthermore, for a consistent expansion of the density contrast up to third order, certain non-local bias terms have to be included \citep[\eg][]{Chan2012PhRvD,Assassi2014JCAP}.

Finally, in observations, the apparent galaxy clustering is affected by redshift perturbations due to peculiar velocities, leading to shifted positions called \rsd. \citet{Kaiser1987MNRAS} derived a linear model that accounts for the squashing of the \tpcf on large separations but lacks inclusion of the small-scale Finger of God (FoG) effect \citep{Jackson1972MNRAS}.
\rsd can be accounted for analytically in the streaming model \citep{Peebles1980_LSS_book, Fisher1995ApJ, Scoccimarro2004} or kept perturbatively in a redshift-space analogue of \spt \citep{Scoccimarro1999ApJ}.
Building on the work of \citet{Scoccimarro1999ApJ} and \citet{Scoccimarro2004}, a major breakthrough was achieved with the semi-empirical TNS model \citep{Taruya2010}, which adds corrections to the Kaiser model and analytically approximates the FoG damping term.
A proper treatment of non-local and second-order bias in the correction terms of the TNS model in addition to a non-Gaussian damping factor has been presented in \citet{Sanchez2017}.

The \pdfabbrev of pairwise velocities, needed in the streaming model, can be approximated to be a Gaussian to form the Gaussian-streaming (GS) model \citep{Reid2011MNRAS}. This class of models requires only the first three velocity cumulants to be predicted, which can be done in either \spt or \lpt. Using the Lagrangian approach, the \clpt model emerged  \citep{Carlson2013MNRAS.429.1674C,Wang2014MNRAS.437..588W}.
Extensions of the streaming model beyond a Gaussian \pdfabbrev were studied in different works \citep{Uhlemann2015PhRvD_Edgeworth_streaming,Bianchi2015MNRAS_GS_beyond_a, Bianchi2016MNRAS_GS_beyond_b, Cuesta_Lazara2020MNRAS_GS_beyond, Kuruvilla2018MNRAS_GS_beyond}.
\citet{VlahWhite2019} presented a Fourier analogue of the streaming model, which was compared to the moment expansion approach in the work of \citet{Chen2020JCAP}.
Lastly, the \eft formalism was extended to treat \rsd in \citet{Senator2014arXiv1409_EFT_RSD}, \citet{Lewandowski2018PhRvD_EFT_RSD}, and \citet{Perko2016arXiv161009321P_EFT_bias_RSD}, introducing additional counterterms.

In practical terms, and beyond the nature of the considered model, two main approaches to fitting the data are used: `template fitting' and `full-shape fitting' \citep[but see][]{Brieden2021JCAP_shapefit}.
The former consists of choosing a template linear power spectrum in real space and any shape deviation from this template is captured through the `Alcock--Paczynski' \citep[AP,][]{Alcock1979Natur} parameters.
Amplitude modulations in the template fitting are absorbed in the growth rate of structure, $f$, and linear-theory amplitude of density perturbations in spheres of $8\Mpc$, $\sigeight$. 
Generally, template fitting is close to model-independent. In contrast, the full-shape fitting approach aims at fitting the full-shape multipoles of the \tpcf by varying directly cosmological parameters of the model, including those dictating the shape of the linear real-space power spectrum such as the spectral index.

In this work we aim at a comparative and comprehensive study of state-of-the-art models for the galaxy \tpcf multipoles in redshift space using \Euclid's \flagship simulation \citep{EuclidSkyFlagship}. The goal is to pave the way for a thorough and robust analysis of the \Euclid data, including the use of specific techniques to significantly reduce the computational time and that make it possible to perform a likelihood analysis with full-shape fitting. A specific focus is given on certain performance metrics and on the scales reached by those models, such that a maximum information gain is achieved without biasing the result.
This is crucial to select baseline models for the \tpcf in ideal conditions that can then be further tested for the inclusion of observational systematic effects.
This model comparison is part of a larger effort within the Euclid Collaboration to test models for clustering statistics. 
On the Fourier-space side, this includes the work of \citet{EP-Pezzotta} investigating the bias model on the power spectrum in real space as well as Euclid Collaboration: Camacho et al. (in prep.) comparing redshift-space models of the power spectrum and Euclid Collaboration: Pardede et al. (in prep.) being the analogous work for the bispectrum.
Furthermore, extensions to higher-order statistics in configuration space are studied in \citet{EP-Guidi} in real space with a focus on a combined analysis of the \tpcf and \threepcf and Euclid Collaboration: Pugno et al. (in prep.) with an analogous study of the \tpcf and \threepcf in redshift space.

This article is structured as follows. In Sect.~\ref{sec:galaxy_clustering_redshift_space} we present the modelling of two-point statistics for galaxies in redshift space with a focus on the techniques and models used in this work.
In Sect.~\ref{sec:data} we describe the \flagship simulation used to compare models, and in Sect.~\ref{sec:fitting_procedure} we present the different approaches to fit the data, the priors on the parameters, and introduce the considered performance metrics.
Section~\ref{sec:results} presents the results of our analyses, with a first part focusing on the template fitting approach and a second one on the full-shape fitting approach.
We conclude this work with a discussion of the results in Sect.~\ref{sec:discussion_conclusion}.

\section{Modelling the galaxy 2-point correlation function in redshift space}
\label{sec:galaxy_clustering_redshift_space}
The \tpcf is defined by the cumulant taken over the density contrast at two distinct points in space as
\begin{equation}
    \xi(\rv) = \ev{\delta(\xv) \,\delta(\xv+\rv)}_{\mathrm{c}} \, .
\end{equation}
The density contrast is defined as the fractional overdensity $\delta(\xv) = [\rho(\xv) - \bar{\rho}]\,/\,\bar{\rho}$, where $\bar{\rho}$ denotes the ensemble average of the density field $\rho(\xv)$.
Furthermore, assuming ergodicity replaces the ensemble average with a spatial average.
Under the assumption of statistical homogeneity (invariance under translation) the \tpcf depends only on the difference between the position of two points, the comoving pair-separation vector $\rv$.
The power spectrum $P(\kv)$ is the Fourier transform of the \tpcf and is defined as
\begin{equation}
    (2\pi)^3 \,P(\kv)\,\dirac(\kv+\kv') \equiv \ev{\delta(\kv)\,\delta(\kv')}_{\mathrm{c}} \, ,
\end{equation}
where $\kv$ is the comoving wavevector, $\dirac$ denotes the Dirac delta and $\delta(\kv)$ is the Fourier transform of $\delta(\xv)$.
We denote quantities in Fourier space using $\kv$ or $\kv'$ as arguments and $\xv$, $\rv$, $\sv$, or $\qv$ for configuration space.
In general, the density and velocity fields, as well as their $n$-point correlators, depend on time $t$.
For brevity, where appropriate, we omit this explicit dependence.
We denote the linear density contrast as $\delta_{\mathrm{L}}$, the linear power spectrum as $\Plin$, and the galaxy power spectrum as $\Pgg$, with associated galaxy number density contrast $\deltag$\,.

\subsection{Nonlinear clustering of matter}
In the non-relativistic limit for scales well inside the horizon, matter can be described by a pressureless perfect fluid whose evolution is governed by the continuity and Euler equations, as well as the Poisson equation describing gravitational interactions \citep{Bernardeau2002}.
In the last decades two schools of approaches to solve this system of coupled differential equations have emerged, named Eulerian and Lagrangian perturbation theory.

\subsubsection{Eulerian perturbation theory}
In the Eulerian formulation of perturbation theory, the density contrast is expanded in terms of powers of the initial density contrast, which is assumed to be Gaussian \citep{Bernardeau2002}. The density contrast of the matter field is written order-by-order with a multiplicative split of the time and space dependence, the former being the growth factor.
At first order, the matter density contrast field at any time $t$ can be written as $\deltainit(\xv, t) = D_+(t)\,\deltainit(\xv, t_{\rm ini})$, where $D_+(t)$ is the growing mode of the linear growth factor (normalised to $D_+(t_{\mathrm{ini}})=1$), and $\deltainit(\xv, t_{\rm ini})$ is the initial density contrast.
The Gaussian ansatz for the initial density field leads, as a result of Wick's theorem, to only the moments of even powers of the initial density contrast being non-zero \citep{Bernardeau2002}. This simplifies correlator calculations, and up to one-loop, the matter power spectrum takes the form
\begin{equation}\label{eq:pk_1loop}
    P(\kv,t) = \Plin(\kv,t) + P_{13}(\kv,t) + P_{22}(\kv,t) \, ,
\end{equation}
where $P_{13}$ and $P_{22}$ are integrals over the linear power spectrum $\Plin$ multiplied by a kernel accounting for mode coupling and nonlinear corrections. This can be extended to an arbitrary loop and expressions for the corresponding kernels can be obtained via recursion relations \citep{GoroffEtal1986}.
\spt is inherently flawed by having a poor convergence of the expansion series, that is, two-loop corrections can be larger in amplitude than one-loop corrections at some large $k=|\kv|$.
In addition, loop-corrections can be both positive and negative, hampering further the convergence of the expansion.
Furthermore, loop integrals themselves might exhibit diverging behaviour in the ultraviolet (large $k$) or infrared (small $k$) limit, and it is pure coincidence that $P_{13}$ and $P_{22}$ are finite in amplitude for the cosmology of our Universe.

\subsubsection{Regularised perturbation theory}
To solve the problem of convergence in the \spt approach, RPT performs a different kind of expansion based on propagators, which reorganises loop corrections \citep{Crocce2006PhRvD_a_RPT,  Crocce2006PhRvD_b_RPT, Crocce2008}.
The power spectrum can be constructed by a propagator acting on the initial power spectrum plus mode-coupling terms.
An important property in RPT is that the propagator is exponentially damped at large $k$, while it recovers the linear propagator at small $k$.
The two-point propagators have been generalised to multi-point propagators $\Gamma$ by \citet{Bernardeau2008} in order to connect $n$ initial fields to a final field. This leads to the so-called $\Gamma$-expansion of the power spectrum in terms of integrals over $n$-point propagators.
The mode-coupling terms are now captured by multi-point propagators of three or more points.
The expansion is in powers of $\Plin$ and therefore the first term in the summation contains also the $P_{13}$ contribution of \spt for instance.

A problem in RPT and the propagator expansion comes from a truncated loop expansion of the individual propagator that does not lead to the desired exponential damping. The latter being achieved in the full resummation only. Solving this issue was the main motivation behind RegPT \citep{BernardeauPhysRevD_RegPt, Taruya2012_RegPT} which allows for a continuous interpolation of the low- and high-$k$ behaviour of the propagator at each order in perturbation theory.
This results in a `regularised' multi-point propagator up to the desired loop that exhibits for small $k$ the \spt result as a limit, while at large $k$ it recovers the exponentially damped tree level, as obtained in RPT with a full resummation.

\subsubsection{Effective field theory of large-scale structure}
Eulerian perturbation theory builds on the assumption that the density contrast is small and can be treated perturbatively up until structure formation leads to large $\delta$ where a purely perturbative formalism breaks down.
The \eft formalism \citep{BaumannEtal2012, Carrasco2012} tries to tackle this problem, while also keeping loop integrals well-behaved.
In general, the \eft formalism is based on a split of perturbative quantities, such as $\delta$, into two contributions at long and short wavelengths where the long-wavelength part is obtained via smoothing.
This smoothing is applied to the relevant equations of motion and a stress-tensor contribution is added to the Euler equation accounting for non-ideal fluids.
Importantly, even in the case of perfect fluids, terms with the structure of a stress tensor arise due to the smoothing applied to products of two fields.
This stress tensor describes back-reaction effects of the short modes on the physics of long modes.
For the long-wavelength part of $\delta$, perturbation theory is applicable as the density contrast is small by construction.

The problem of ill-defined loop integrals is solved by introducing a cutoff scale $\Lambda_{\mathrm{cut}}$ up to which those integrals are computed. Cutting the integrals makes the theory dependent on the specific choice of $\Lambda_{\mathrm{cut}}$, which is not desired as there is no real physical reason for the existence of $\Lambda_{\mathrm{cut}}$.
However, the stress tensor adds new contributions to the perturbative expansion of the fields in the form of counterterms that both renormalise the loop integrals and absorb any $\Lambda_{\mathrm{cut}}$ dependence.
In addition, counterterms also include small-scale physics as the stress tensor contains contributions both from long and short wavelengths.
Overall, \eft provides a more theoretically sound theory of matter clustering, while extending the range of models of clustering statistics to higher $k$.
Those gains come at the price of having additional free parameters in the model, which cannot be computed from first principles.

\subsubsection{Lagrangian perturbation theory}
A different approach to \spt is given by the Lagrangian picture.
The Lagrangian formulation of perturbation theory is based on a coordinate transformation from an initial position of the fluid elements $\bm{q}$ to the Eulerian position $\xv$ through a displacement field $\bm{\Psi}$ carrying the time dependence as \citep{Zeldovich1970}
\begin{equation}\label{eq:Lagrangian_mapping}
    \xv(\bm{q}, t) = \bm{q} + \bm{\Psi}(\bm{q}, t) \, .
\end{equation}
Most importantly, the Eulerian density contrast is zero at $\bm{q}$ in the initial position limit.
\lpt revolves around finding solutions to a perturbative expansion of the displacement field $\bm{\Psi}^{(n)}$ that can be described in terms of the initial density field, with its first-order solution $\bm{\Psi}^{(1)}$ known as the `Zeldovich approximation' \citep{Zeldovich1970}.

A major breakthrough was achieved by the formulation of `resummed \lpt' for matter in \citet{Matsubara2008} and biased tracers in \citet{Matsubara2008b}.
In this approach, the galaxy density contrast is expressed with a local Lagrangian bias function $F[\deltainit(\bm{q})]$ such that
\begin{equation}\label{eq:LPT-delta}
    1+\delta_{\mathrm{g}}(\xv,t) = \int \diff^3q \,  F[\deltainit(\bm{q})] \, \delta_{\mathrm{D}}(\xv-\bm{q}-\bm{\Psi}(\bm{q},t)) \, .
\end{equation}
By defining the Fourier transform of the bias function as $F(\delta) = \int \mathrm{e}^{\mathrm{i}\lambda\delta} F[\lambda] \,\diff\lambda/(2\pi)$ and expressing the $\delta_{\mathrm{D}}$-function in Eq. \eqref{eq:LPT-delta} in Fourier space, the \tpcf takes the form \citep{Carlson2013MNRAS.429.1674C}
\begin{equation}\label{eq:xi_LPT}
    \begin{split}
        1+\xi(r) =& 
        \int \diff^3q\int\frac{\diff\lambda_1}{2\pi}\int\frac{\diff\lambda_2}{2\pi}\int \frac{\diff^3k}{(2\pi)^3} \,\mathrm{e}^{\mathrm{i}\kv\cdot(\bm{q}-\bm{r})}\\
        &\times F[\lambda_1]\,F[\lambda_2]\, \ave{\mathrm{e}^{\mathrm{i}(\lambda_1\delta_1+\lambda_2\delta_2 + \kv\cdot\bm{\Delta})}}\, .
    \end{split}
\end{equation}
In this expression, $\delta_i = \deltainit(\qv_i)$, $\qv=\qv_2-\qv_1$, $\bm{\Delta} = \bm{\Psi}(\qv_2, t) - \bm{\Psi}(\qv_1, t)$, which is the displacement-field difference, and $\rv$ is the pair-separation vector. 
The moment containing the exponential depends only on $\qv$ due to statistical invariance under translations.
\citet{Carlson2013MNRAS.429.1674C} showed that with an appropriate expansion, the \tpcf can be computed from a convolution thereby introducing the formalism of \clpt.
An advantage of \lpt in the form of Eq.~\eqref{eq:xi_LPT} is the improved modelling of the \bao feature.
This comes from a natural resummation of contact terms as shown in \citet{Matsubara2008, Matsubara2008b}.
\spt corresponds to a full expansion of the exponential inside the ensemble average, making the IR-resummation technique necessary.

\subsection{Galaxy bias}\label{sec:galaxy_bias}
In galaxy surveys, the target samples are biased tracers of the underlying dark matter field, and the galaxy power spectrum is related to that of matter through galaxy biasing \citep{Desjacques2018PhR}.
Order by order in the expansion of the density contrast, the galaxy density takes contributions of powers of the dark matter density.
However, this picture can be further refined by taking into account non-local contributions to the galaxy density contrast \citep{McDonald2009JCAP, Chan2012PhRvD, Baldauf2012PhRvD_non_local_Lagrangian_bias, Saito2014PhRvD}.
Following the bias expansion of \citet{Assassi2014JCAP}, the galaxy density field can symbolically be written as
\begin{equation}\label{eq:eulerian_bias_expansion}
    \deltag(\xv) = b_1 \delta(\xv) + \frac{b_2}{2} \delta^2(\xv) + \bGtwo\, \Gtwo(\xv) + \bGthree\, \Gthree(\xv)\, ,
\end{equation}
where $b_1$ is the first-order linear bias, $b_2$ is the second-order bias term, and we omitted the time dependence. At this order, the second-order Galileon $\Gtwo$, also known as tidal bias, needs to be included. It is defined as
\begin{equation}\label{eq:Gttwo}
    \Gtwo(\xv) = \left[\nabla_i\nabla_j\,\Phi_{\mathrm{g}}(\xv)\right]^2 - \left[\nabla^2\Phi_{\mathrm{g}}(\xv)\right]^2 \, ,
\end{equation}
acting on the gravitational potential $\Phi_{\mathrm{g}}$ and we  use Einstein's sum convention for repeated indices in the first product.
The second non-local bias operator $\Gthree$ is defined as the difference between two Galileon operators acting on the velocity potential $\Phi_{\mathrm{v}}$ and the gravitational potential.
As described in detail in \cite{Assassi2014JCAP}, the expansion in Eq.~\eqref{eq:eulerian_bias_expansion} uses bare bias parameters and each term needs to be renormalised in order to ensure a vanishing mean of $\delta_{\mathrm{g}}(\xv)$ as well as to remove any cutoff dependence of resulting integrals, for example for the contact term $\delta^2(\xv)$.
Using the Fourier transformation of Eq.~\eqref{eq:eulerian_bias_expansion} and expanding $\langle \deltag(\kv) \,\deltag(\kv')\rangle_c$ up to one-loop leads to separate bias terms for the galaxy power spectrum \citep{Assassi2014JCAP,Simonovic2018JCAP}. The bias terms are, as in matter \spt, integrals over the linear power spectrum with appropriate kernels.

It can be shown that a local bias in the Lagrangian picture creates non-local contributions to the Eulerian density contrast nevertheless \citep{Catelan1998MNRAS_non_local_Lag_bias,Catelan2000MNRAS_non_local_Lag_bias,Baldauf2012PhRvD_non_local_Lagrangian_bias}. By matching coefficients in the expansion, the \LL approximation can be derived such that \citep{Chan2012PhRvD, Saito2014PhRvD}
\begin{equation}\label{eq:LL_bG2_bGam3}
    \bGtwo = -\frac{2}{7}\,(b_1-1)\,, \quad \textrm{and} \quad    \bGthree = \frac{11}{42}\,(b_1-1) \, .
\end{equation}
These relations can be useful if a reduction of the parameter space is desired in the case of insufficient constraining power of the data or strong degeneracies among the parameters.

Similarly to Eulerian bias, the Lagrangian bias function usually takes the form \citep{VlahCastorinaWhite2016}
\begin{equation}
\begin{split}
\label{eq:bias_vlah16}
    F[\deltainit(\qv)] = & 1+\delta_{\mathrm{g}}(\bm{q}) \\
    = & 1 + b_1^{\rm L} \deltainit(\bm{q}) + \frac{1}{2}b_2^{\rm L} \left(\deltainit^2(\bm{q}) - \langle \deltainit^2\rangle\right) \\
    &+ b_{s^2}\left(s^2(\bm{q}) - \langle s^2\rangle\right) + b_{\nabla^2} \frac{\nabla_q^2}{\Lambda_{\mathrm{L}}^2} \deltainit(\bm{q}) \, .
\end{split}
\end{equation}
The superscript `L' refers to Lagrangian biases, as opposed to Eulerian biases defined in Eq.~\eqref{eq:eulerian_bias_expansion}, where the linear bias is related via $b_1 = 1 + b_1^{\rm L}$.
Moreover, in contrast to the Eulerian expansion, here the derivative bias $b_{\nabla^2}$ is included, but this can also be formulated in the Eulerian picture \citep[see \eg][]{McDonald2009JCAP}.
The Lagrangian scale, $\Lambda_{\mathrm{L}}$, can be seen as a distance up to which non-locality is present.
The operator $s^2=s_{ij}s^{ij}$ (using Einstein's summation convention) is 
defined, analogously to the tidal bias in the Eulerian scheme, such that \citep{McDonald2009JCAP}
\begin{equation}
    s_{ij}(\kv) = \left(\frac{k_i k_j}{k^2}-\frac{1}{3}\delta^{\mathrm{K}}_{ij}\right)\deltainit(\kv)
\end{equation}
in Fourier space, where $\delta_{ij}^{\mathrm{K}}$ is the Kronecker delta.
Including those terms in the purely local bias expansion changes Eq.~\eqref{eq:xi_LPT} because of the extended bias function \citep{VlahCastorinaWhite2016}.
Similarly to the \LL approximation for the Eulerian bias parameters, it is possible to express the non-local bias $b_{s^2}$ in terms of $b_1^{\rm L}$ as  \citep{Saito2014PhRvD}
\begin{equation}\label{eq:LL_bs2}
   b_{s^2} = -\frac{4}{7}b_1^{\rm L} \, ,
\end{equation}
under the assumption that the Lagrangian bias expansion is purely local.

\subsection{Redshift-space distortions}
\label{sec:theory_rsd}
In galaxy surveys, the radial distance is measured as a redshift, and therefore the observed position of a galaxy $\bm{s}$ can be written in terms of its comoving position in real space $\xv$ and a velocity-induced Doppler shift as
\begin{equation}
    \label{eq:zspace_position}
    \bm{s} = \xv + \frac{v_{z}(\xv)\,\hat{\bm{z}}}{a\,H} \, ,
\end{equation}
where $H$ is the Hubble parameter and $a$ the scale factor, both depending on time.
Assuming the plane-parallel approximation, the \los is fixed to the $z$-axis and therefore $v_{z}(\xv) = \bm{v}(\xv) \cdot\hat{\bm{z}}$, with $\hat{\bm{z}}$ the unit vector into the $z$-direction.
In Eq.~\eqref{eq:zspace_position} and in the following, we omit the explicit time dependence of all the fields and correlators.
The interested reader is referred to the extensive work by \citet{1998Hamilton_rsd_review} providing, among other aspects, a review of the mathematical and physical description of \rsd and their implications for parameter estimation.

Due to mass conservation between real and redshift space, the density contrast in the respective spaces can be related. This is usually the starting point for any modelling of correlators in redshift space.
The Jacobian corresponding to the mapping into redshift space in Eq.~\eqref{eq:zspace_position} is expanded to make the resulting expressions tractable.
Fourier transforming the formula for mass conservation yields an expression for the density contrast in redshift space \citep{Scoccimarro1999ApJ}.
The power spectrum in redshift space can then be written as \citep[][]{Scoccimarro2004, Taruya2010}
\begin{equation}
\begin{split}\label{eq:Pk_redshiftspace}
    \Ps(\kv) = \int &\diff^3 r\, \mathrm{e}^{\mathrm{i}\kv\cdot\bm{r}} \, \Big\langle \mathrm{e}^{-\mathrm{i}k \mu f \Delta u_z} \\
    &\times \brackets{\delta(\xv)+f\,\nabla_z u_z(\xv)}\,\brackets{\delta(\xv') + f\, \nabla_z u_z(\xv')}\Big\rangle \, ,
\end{split}
\end{equation}
with the velocity difference defined as $\Delta u_z \equiv u_z(\xv)-u_z(\xv')$ and $u_z(\xv) \equiv -v_z(\xv)\,/\,(aHf)$, $\mu$ is defined via $\kv\cdot\hat{\bm{z}} = k\,\mu$, $\rv = \xv-\xv'$, and $f=\diff \ln\delta/\diff \ln a$.
The moment depends only on $\rv$ due to statistical invariance by translation.
A perturbative treatment of $\Ps(\kv)$ works analogously as in real space with an expansion of the density contrast, but the kernels are now in redshift space.
Up to one-loop and including galaxy bias, the galaxy power spectrum in redshift space is given by
\begin{equation}
    \PsSPT(k, \mu) = \Pslin(k,\mu) + \Pstwotwo(k, \mu) + \Psonethree(k, \mu)\,. 
\end{equation}
The exact form of the loop contributions $\Pstwotwo$ and $\Psonethree$ depends on the chosen bias expansion.
Including non-local operators as in Eq.~\eqref{eq:eulerian_bias_expansion} the one-loop expressions can be found in \citet{Assassi2014JCAP}.
The linear order piece constitutes the Kaiser model \citep{Kaiser1987MNRAS} with the characteristic squashing effect on large scales.

\subsubsection{TNS class of models}
\label{sec:theory_TNS}
To improve on the perturbative treatment of \rsd, \citet{Scoccimarro2004} developed a different approach where the full moment in Eq.~\eqref{eq:Pk_redshiftspace} is expressed in terms of cumulants, singling out a common prefactor.
This approach was further developed in \citet{Taruya2010} leading to the TNS model, which empirically assumes either a Lorentzian or Gaussian form for the prefactor that acts as a damping on small scales to mimic the FoG effect.
Keeping only terms up to one-loop order, their final expression of the galaxy power spectrum in redshift space therefore takes the form
\begin{equation}\label{eq:TNS_full_ps}
\begin{split}
    \PsTNS(k, \mu) =& \, D(k, \mu, \sigv)\left[\Pgg(k) + 2\mu^2 f \Pgt(k) + \mu^4 f^2 \Ptt(k) \right. \\
    & \left. + C_A(k, \mu) + C_B(k, \mu)\right] \, ,
\end{split}
\end{equation}
where $\theta\equiv\nabla\cdot\uv$ is the velocity divergence field, while
$\Pgg$, $\Pgt$, and $\Ptt$ are the nonlinear real-space auto galaxy power spectrum, the galaxy-velocity divergence cross power spectrum, and the velocity divergence auto power spectrum, respectively.
The damping function is given by $D(k,\mu, \sigv)$ with $\sigv$ being an effective velocity dispersion.
Traditionally, $\sigv$ is a free parameter to match the specific small-scale dispersion of the galaxy sample under consideration.
For the rest of this work we make the common assumption of a Lorentzian form in $k\mu f\sigv$ for the damping function in the TNS models.
The explicit expressions of the correction terms $C_A(k, \mu)$ and $C_B(k, \mu)$, including linear galaxy bias, can be found in the appendix of \citet{Taruya2010} and \citet{delaTorre2012MNRAS_TNS_testing}.

In this work, we examine different flavours of the TNS model that differ in the way the ingredients entering Eq.~\eqref{eq:TNS_full_ps}, namely the nonlinear power spectra and the perturbative correction term $C_A$, are predicted.
As already discussed, \spt does not yield the best convergence, and here we consider RegPT instead, where expressions for the one-loop and two-loop power spectra are given in \citet{Taruya2012_RegPT}.
As an alternative, we explore a hybrid, simulation-informed model where the nonlinear matter power spectrum is estimated from an emulator.
In this work, we consider the \euclidemulatortwo \citep{Knabenhans-EP9} and \texttt{baccoemu} \citep{Angulo2021MNRAS_BACCO} emulators that learn the nonlinear clustering of matter from a set of $N$-body simulations with the help of machine learning techniques.
They yield much more accurate predictions for small scales where perturbative prescriptions break down.
To obtain the matter-velocity and velocity-velocity spectra from the matter power spectrum, we use the set of fitting functions from \citet{Bel2019}.
The other choice of modelling concerns the $C_A$ correction terms.
The usual procedure is to take the one-loop \spt prescription as in \citet{delaTorre2012MNRAS_TNS_testing} and use those for our hybrid model.
For the RegPT power spectra, we have to be consistent and use the RegPT one-loop or two-loop expansion, where explicit expressions are given in \citet{Taruya2013PhRvD}.
We note that the $C_B$-terms are always computed non-perturbatively taking the product of two nonlinear power spectra as input \citep[Eq.~20 in][]{Taruya2010}. This can therefore result in higher-order $C_B$-terms due the product of, for example, two one-loop power spectra containing contributions with four $\Plin$ in total.

The \vdg model as proposed by \citet{Sanchez2017} is very similar to the TNS model in Eq. \eqref{eq:TNS_full_ps} in the sense of treating the moment in Eq.~\eqref{eq:Pk_redshiftspace} with the same cumulant expansion, although it differs in certain aspects. We consider here an extended version as described in \citet{EggemeierEtal_2023, Eggemeier2025arXiv_COMET_VDG_bispectrum} that includes the counterterms from the \eft formalism.
The \vdg model takes into account non-Gaussianities in the damping function via a non-vanishing kurtosis parametrised by the free parameter $a_{\tm{vir}}$.
The damping function takes the form \citep{Sanchez2017}
\begin{equation}\label{eq:VDG_damping_function}
    D(k, \mu, \avir) = \frac{1}{\sqrt{1-\lambda^2\avir^2}}\exp\left (\frac{\lambda^2\sigv^2}{1-\lambda^2\avir^2}\right) \, ,
\end{equation}
where $\lambda = \mathrm{i}fk\mu$, and $\sigv$ is again the linear velocity dispersion that is predicted in the \vdg model as $\sigv^2=1/3\int \diff^3 k/(2\pi)^3 \, \Plin(k)/k^2$ \citep{EggemeierEtal_2023}.
The functional form can be derived in the large-scale limit (hence the $\infty$ in the name) of the distribution of pairwise-velocity differences exhibiting non-Gaussianities \citep{Scoccimarro2004}.
Another crucial difference to the TNS model concerns the treatment of bias.
While higher-order bias was introduced into the TNS model in the work of \citet{Bautista2021MNRAS_BOSS_clustering}, they did not propagate it into the perturbative expansion of the correction terms (particularly the $C_A$-terms), which contained only $b_1$.
The \vdg model, on the contrary, takes into account contributions from $b_2$ as well as from $\bGtwo$ in the correction terms.

\subsubsection{\eft model}
\label{sec:EFT_redshiftspace}
The \eft model utilises the introduced \eft formalism for renormalised loop contributions and modelling of short-scale physics in redshift space together with a proper treatment of IR-resummation.
We only briefly outline the model, which is presented in more detail in \citet{Ivanov2020JCAP_BOSS_EFT_full_shape} or \citet{2020JCAPDAmico_EFT} as well as \citet{EggemeierEtal_2023}.
In this formalism, the galaxy power spectrum in redshift space can be decomposed into distinct contributions as
\begin{equation}
    \PsEFT(k, \mu) = \Psnlo(k, \mu)  + \Psctr(k, \mu) + \Psstoch(k, \mu) \, ,
\end{equation}
where the first term on the right-hand side is the next-to-leading-order (NLO) IR-resummed galaxy power spectrum in redshift space, and the two remaining contributions are sourced by counterterms and stochastic corrections, respectively.
IR-resummation is necessary in order to improve the modelling of the \bao feature compared to \spt.
In general, the \bao is damped due to the effect of large-scale modes \citep{Eisenstein2007ApJ_BAO_reconstruction}.
\spt corresponds to a full expansion of all modes and IR-resummation tries to regain parts of the resummed expression.
As described in \citet{EggemeierEtal_2023}, the \eft model makes use of a split of the linear power spectrum into wiggly and non-wiggly components as described in \cite{Baldauf-ir-res2015} and \cite{Blas2016a}.
IR-resummation is achieved by expanding only the wiggly component completely as well as the higher-order terms of the non-wiggly component, as these quantities are small by construction.
Due to the resummation of infrared modes, the shortest wavelengths, carrying the information on the sharp \bao peak, are damped leading to the desired effect of a smearing of the \bao feature.
Explicit expressions for the IR-resummed galaxy power spectrum in redshift space are given in \citet{Ivanov2018}.

The contributions to modelling small-scale physics from counterterms that are not captured by the usual perturbative expansion can be included using different parametrisations \citep[see][]{Nishimichi2020}.
The parametrisation used in \citet{EggemeierEtal_2023} takes the form
with three counterterms $c_{0}$, $c_2$, and $c_4$ multiplied by the respective Legendre polynomials as well as an additional counterterm $c_{\rm nlo}$ motivated by a fully perturbative modelling of the small-separation FoG effect \citep[][]{Ivanov2020JCAP_BOSS_EFT_full_shape}.
Lastly, the full set of stochastic contributions \citep[see][]{Desjacques2018PhR, Eggemeier2021PhRvD_bias_joint, EggemeierEtal_2023} includes deviations from a purely Poisson shot noise, as well as scale-dependent shot noise due to, for example, the halo-exclusion effect \citep[][]{Baldauf2013PhRvD_halo_exclusion} and \los effects originating from \rsd \citep[][]{Perko2016arXiv161009321P_EFT_bias_RSD}.
The \eft model can be remapped into the \vdg model by a subtraction of the expansion of the velocity generator and subsequent multiplication by the damping \citep{EggemeierEtal_2023}.
This implies that the \vdg model also contains the three counterterms $c_0$, $c_2$, and $c_4$ and includes IR-resummation.
In order to obtain the prediction for the \tpcf, the power spectra from TNS, \vdg, or \eft are Fourier transformed into configuration space.
In Sect.~\ref{sec:accelerated_models} we comment on the technicalities of this transformation both in physical and numerical senses. 

\subsubsection{The Gaussian streaming class of models}\label{sec:gaussian_streaming_theory}
The streaming model approach is based on a convolution of the real-space \tpcf with the \pdfabbrev $\mathcal{P}$ of pairwise velocities to obtain the redshift-space \tpcf \citep{Peebles1980_LSS_book, Fisher1995ApJ,Scoccimarro2004}.
If the Jacobian of the mapping between real and redshift space is not expanded, tracer density conservation and distant-observer approximation are assumed, the \tpcf in redshift space can be written as \citep{Scoccimarro2004}
\begin{equation}\label{eq:streaming_model}
    1+\xi^s(s_{\parallel}, s_{\perp}) = \int_{-\infty}^{\infty} \diff r_{\parallel} \, \left[1+\xi(r)\right] \,\mathcal{P}(-v_{\parallel}\mathcal{H}^{-1}, \bm{r}) \, ,
\end{equation}
with $s_{\parallel}$ and $s_{\perp}$ are the components of $\sv$ parallel and perpendicular to the \los, respectively. We note also that $s_{\perp}=r_{\perp}$.
The velocity parallel to the \los is given by $v_{\parallel} = -\mathcal{H}(r_{\parallel} - s_{\parallel})$ and $\mathcal{H}=aH$.
The GS model approximates the \pdfabbrev as a Gaussian such that the commonly adopted form is \citep{Reid2011MNRAS, Wang2014MNRAS.437..588W}
\begin{equation}\label{eq:gaussian_streaming}
\begin{split}
    1+\xi_{\mathrm{gg}}^s(s_{\parallel},s_{\perp}) = &\int_{-\infty}^{\infty} \diff r_{\parallel} \, \frac{1+\xi_{\mathrm{gg}}(r)}{\sqrt{2\pi}\, \tilde{\sigma}_{12}(r,\gamma)}  \\
    &\qquad\times\mathrm{exp}\left\{-\frac{[s_{\parallel}-r_{\parallel}-\gamma \,v_{12}(r)]^2}{2\,\tilde{\sigma}^2_{12}(r,\gamma)}\right\} \, ,
\end{split}
\end{equation}
with $r = r_{\parallel}/\gamma$ and \smash{$r=\left(s_{\perp}^2+r_{\parallel}^2\right)^{1/2}$}.
Here, $\gamma$ is defined in real space as $\bm{r}\cdot\hat{\bm{e}} = r\,\gamma$, with $\hat{\bm{e}}$ being a unit vector into the direction of the \los.
In this parametrisation, $v_{12}(r)$ is the absolute value of the mean pairwise velocity directed along the real-space pair-separation vector. Furthermore, $\tilde{\sigma}^2_{12}(r,\gamma)$ is the component of the pairwise velocity dispersion directed along the \los, and we distinguish it from $\sigtwelve$, the variance of the density field in spheres of $12\,\mathrm{Mpc}$, by an argument and a tilde.
The velocity entering $v_{12}(r)$ and $\tilde{\sigma}_{12}(r,\gamma)$ is rescaled by $H^{-1}$ and has hence the unit of a length.
In general, the first three mass-weighted velocity moments, which are the real-space \tpcf $\xi_{\mathrm{gg}}(r)$, mean pairwise velocity $v_{12,n}(r)$, and the pairwise velocity dispersion $\tilde{\sigma}_{12,nm}(r)$ are needed for this model.
The subscripts $n$ and $m$ are hereby referring to components of a vector ($v_{12,n}$) or a matrix ($\tilde{\sigma}_{12,nm}$).
Projecting first $v_{12,n}$ on $\bm r$ via $v_{12} = v_{12,n}\hat r_n$ and then onto the \los gives the factor of $\gamma$ in Eq.~\eqref{eq:gaussian_streaming}.
The full dispersion $\tilde{\sigma}_{12,nm}$ can be split up into components parallel $\tilde{\sigma}_{\parallel}(r)$ and perpendicular $\tilde{\sigma}_{\perp}(r)$ to the pair-separation vector and sum up to
\begin{equation}
    \tilde{\sigma}^2_{12}(r,\gamma) = \gamma^2\,\tilde{\sigma}^2_{\parallel}(r) + (1-\gamma^2)\,\tilde{\sigma}^2_{\perp}(r) \, .
\end{equation}

The ingredients for the GS model as we employ it in this work use the \clpt formalism but can also be computed using EPT as is done in \citet{Reid2011MNRAS} or \citet{Chen2020JCAP}.
As shown in \citet{Wang2014MNRAS.437..588W}, the mass-weighted velocity moments can be obtained by setting up a velocity \mgf similar to the \tpcf in Eq.~\eqref{eq:xi_LPT}.
The resulting moment can be treated with the cumulant expansion theorem and expanding only terms with a vanishing limit $|\bm{q}|\rightarrow\infty$. This keeps terms that are independent of $|\qv|$ exponentiated.
An extended expansion to the one presented in \citet{Wang2014MNRAS.437..588W} is given in the \cleft model developed by \citet{VlahCastorinaWhite2016}. The main idea is to use the \lefttheo formalism and apply it in conjunction with \clpt to the computation of pairwise velocity moments.
The resulting expressions for the real-space \tpcf, the mean pairwise velocity and the pairwise velocity dispersion are very lengthy and we refer the reader to Eqs.~(3.4), (3.7), and (3.10) in \citet{VlahCastorinaWhite2016}.\footnote{We note that there is a typo in their Eq.~(3.7). It is missing a factor of two in front of $g_i \dot \Upsilon_{in}$ and $\dot V^{12}_n$.}
The velocity moments have the set of Lagrangian bias parameters, given by $\{b_1^{\mathrm{L}}, b_2^{\mathrm{L}}, b_{s^2}, b_{\nabla^2}\}$, as free parameters.
In addition, the counterterm contributions are included via several other free parameters, namely $\alpha_{\xi}$ for the \tpcf, $\alpha_v$ and $\alpha_v'$ for the mean pairwise velocity and $\alpha_{\sigma}$ as well as $\beta_{\sigma}$ for the velocity dispersion.
Due to degeneracies in the functional form of the terms up to one-loop involving the derivative bias and the counterterms, only a reduced set of nuisance parameters is actually needed that is $\{b_1, b_2, b_{s^2}, \alpha_{\xi}, \alpha_{v}, \alpha_{\sigma}\}$ (see \citealt{VlahCastorinaWhite2016} and \citealt{Chen2020JCAP}, for more details).
Additionally, \cleft provides a more consistent expansion to the correct order since it keeps only the tree-level piece of $A_{ij}$, the 2-point correlator of the components of $\bm{\Delta}(\qv)$, in the exponential. The latter originates from treating the moment in Eq.~\eqref{eq:xi_LPT} with the cumulant theorem, and analogously for the higher velocity moments. This leads to an expanded $A_{ij}^{\rm{1-loop}}$-term.
For the \cleft model, we use the cumulant in the velocity dispersion of the Gaussian \pdfabbrev instead of the moment.

The classic \clpt model as we use it in this model comparison has three key differences to the more general \cleft model.
First, the full one-loop $A_{ij}$-correlator is kept in the exponential and all the counterterms are set to zero as well as the non-local bias terms $b_{s^2}$ and $b_{\nabla^2}$. 
Second, to be closer to the original presentation of the \clpt model in \citet{Wang2014MNRAS.437..588W} we use the moment version of the velocity dispersion in the Gaussian \pdfabbrev.
Third, differently from their approach, we expand terms only strictly to one-loop, as is done for \cleft. 

\subsection{Multipoles and Alcock--Packzynski distortions}
The effect of \rsd breaks the isotropy of two-point clustering statistics, and the redshift-space power spectrum can be decomposed as
\begin{equation}
\label{eq:pk_multi}
    \Pell\,(k) = \frac{2\ell+1}{2} \int_{-1}^{1} \diff \mu \, \mathcal{L}_\ell(\mu) \, \Ps(k, \mu) \, ,
\end{equation}
with $\mathcal{L}_\ell$ being the Legendre polynomial of order $\ell$.
This expression can then be Fourier transformed to yield correlation function multipoles.
An analogous projection can be done for the anisotropic \tpcf 
\begin{equation}
    \label{eq:xi_multi}
    \xiell\,(s) = \frac{2\ell+1}{2} \int_{-1}^1 \diff \nu \, \mathcal{L}_\ell(\nu) \, \xis(s, \nu) \, ,
\end{equation}
where $\nu$ is defined as $\sv \cdot \hat{\bm{e}} = s\,\nu$ for a redshift-space pair-separation vector $\bm{s}$ and $s =|\sv|$.

Before we compare a model to data we have to incorporate the AP-effect that describes geometric distortions induced by the assumption of a fiducial cosmology, which might differ from the true cosmology of the Universe.
These distortions happen at the conversion of redshifts into comoving distances requiring a cosmology and hence propagate into clustering statistics.
Usually the AP effect is parametrised in terms of two parameters $\qpar$ and $\qperp$ that scale distances parallel and perpendicular to the \los, respectively.
They are defined as
\begin{equation}
    \qpar  := \frac{H'(z)}{H(z)} \quad {\rm and} \quad \qperp  := \frac{d_A(z)}{d_A'(z)} \, ,
\end{equation}
where primed quantities are in the fiducial cosmology from the measurements and $d_A$ is the angular diameter distance.
Leaving free the AP parameters is leveraged in the template-fitting approach to absorb any possible real-space shape modulation.
In Fourier space, the AP effect leads to a rescaling of the power spectrum amplitude as well as its arguments $\mu$ and $k$.
In contrast, in configuration space only the scales (and consequently $\nu$) are affected since the \tpcf is a dimensionless quantity.

\section{Data}
\label{sec:data}
\subsection{Simulated galaxy catalogues}
\begin{table}
    \caption{Fiducial cosmological parameters of the flat \lcdm cosmology assumed to analyse the \flagship simulation. In order, the table shows the value of the dimensionless Hubble parameter $h$, the physical cold dark matter $\omegac$ and baryon density $\omegab$, the total neutrino mass $M_\nu$, the scalar amplitude $\As$, and scalar index $\ns$ of the primordial power spectrum.
  }
  \renewcommand{\arraystretch}{1.5}
  \smallskip
  \label{tab:flagship_cosmology}
  \smallskip
  \begin{tabular}{|c|c|c|c|c|c|}
    \hline
    \rowcolor{blue!5}
    $h$ & $\omegac$ & $\omegab$ & $M_\nu\,[\rm{eV}]$ & $10^9\,\As$ & $\ns$ \\
    \hline
    0.67 & 0.121203 & 0.0219961 & 0 &  2.094273 & 0.97 \\
    \hline
  \end{tabular}
   \tablefoot{\flagship officially claims to have adopted $\ns$ = 0.96, but it was found that using $\ns=0.97$ instead yields a better agreement to the matter power spectrum (see discussion in \citealt{EP-Pezzotta}).}
\end{table}

In this work we make use of mock data built from \flagship, the first $N$-body simulation built by the Euclid Collaboration to reproduce the expected galaxy population and associated properties that will be targeted by \Euclid. The simulation was run with the \pkdgrav code \citep{potter2017pkdgrav}, evolving approximately two trillion dark-matter particles interacting gravitationally in a periodic box of size $L=3780\Mpc$. The reference flat \lcdm cosmology adopted for this run is consistent with \Planck, and the corresponding fiducial values are given in Table~\ref{tab:flagship_cosmology}.
The mass resolution of the simulation, $m_\mathrm{p}\sim2.4\times 10^9\Ms$, allows us to resolve haloes with a typical mass $M_{\rm halo} \sim 10^{12}\Ms$\,, which host the majority of $\mathrm{H}\alpha$ emission line galaxies, the main target of the spectroscopic sample of \Euclid \citep{EuclidSkyFlagship}.  We selected a set of four snapshots covering the redshift range of the spectroscopic sample, $0.9<z<1.8$.
Each snapshot has been populated with galaxies by firstly identifying friends-of-friends haloes with a minimum mass of ten dark matter particles.
Subsequently, haloes have been populated with galaxies using a \hod model \citep{berlind2003hod} that is a good representation of the H$\alpha$ sample that will be observed by \Euclid \citep{EuclidSkyFlagship}.
The \hod has been calibrated to reproduce the mean number density of H$\alpha$ galaxies according to the Model 3 of \citet{pozzetti2016modelling}.
For a more extended description of the \flagship catalogues we refer the reader to \citet{EP-Pezzotta}.
The redshift and number density for the various samples are listed in Table~\ref{tab:hod_samples}.
\begin{table}
  \caption{Specifications of the four galaxy samples used in this paper. The table lists the approximate snapshot redshift and the mean comoving number density $\bar{n}_{\rm gal}$ in a box of size $L=3780\Mpc$.}
   \renewcommand{\arraystretch}{1.5}
    \smallskip
  \label{tab:hod_samples}
    \smallskip
  \begin{tabular}{|c|c|}
    \hline
    \rowcolor{blue!5}
    Redshift  & $\bar{n}_{\rm gal}~\left[\kcMpc\right]$ \\
    \hline
     0.9    & $0.0020$  \\
    \hline
    1.19    & $0.0010$  \\
    \hline
    1.53  & $0.0006$  \\
    \hline
    1.79  & $0.0003$  \\
    \hline
  \end{tabular}
\end{table}

\subsection{Measurements and covariances}
\label{sec:xi_measure}
We measure the binned \tpcf using the natural estimator \citep{peebles1974statistical} with periodic boundary conditions,
\begin{equation}
    \label{eq:PH_estimator}
    \hat{\xi}(s, \nu) = \frac{\NDD\,(s, \nu)}{\NRR\,(s, \nu)} - 1 \, ,
\end{equation}
where the hat symbol denotes an estimator, and the \los is chosen to be along one of the snapshot axes.
We denote with $\NDD$ and $\NRR$ the normalised pair counts in the data and random catalogues, respectively, with the random counts computed analytically, as we can exploit the fact that we measure the \tpcf from a cubic box.
The Legendre multipoles are then computed by performing the integration given in Eq.~\eqref{eq:xi_multi}, which we evaluate for the first three even multipoles $\ell = \{0, 2, 4\}$, capturing the main contribution to \rsd \citep{Kaiser1987MNRAS}.
We measure the \tpcf in the range $0 < r< 200\Mpc$ with a linear bin size of $5\Mpc$.

We produce Gaussian covariances for the \tpcf in redshift space using the approach described in \citet{Grieb2016}.
The covariance between two correlation function multipoles $\xi_{\ell_1}$ and $\xi_{\ell_2}$ evaluated at $s_i$ and $s_j$ is given in their Eqs.~(15) and (18).
This covariance is only valid under the assumption of a Gaussian density field, as otherwise we would have an additional trispectrum contribution to the power spectrum covariance \citep{ScoccimarroZaldarriagaHui1999}.
The three ingredients necessary for the computation of the Gaussian covariance are the survey volume $V_{\rm s}$, in our case that of \flagship, the mean galaxy density $\bar{n}_{\rm g}$, and an anisotropic galaxy power spectrum $\Psgg(k,\mu)$.
We then compute the covariance matrix iteratively, starting from a naive covariance and computing the best-fit $\Psgg$ using the \vdg model, obtained by fitting the \tpcf multipoles with the $\chi^2$-minimisation algorithm \texttt{Minuit} \citep{james1975minuit}. 
We reiterate this procedure five times to converge towards a covariance matrix where the data are well described by the model.
While Camacho et al. (in prep.), performing an analogous analysis in Fourier space, use in addition a rescaled version of the covariance to match the observed volume of \Euclid at \DRthree, we utilise the covariance corresponding to the full volume of \flagship throughout this work.

\section{Fitting procedure}\label{sec:fitting_procedure}
In this section, we describe the fitting procedure, along with samplers, priors, model computation details, and considered free parameters.
Furthermore, we present the various metrics used to quantify the performance of the different models.

Based on the tests provided in Appendix~\ref{sec:appendix_axis_selection_sample_variance}, we chose to analyse the measurements made with the $z$-axis as the \los.
We assume that the data follow a Gaussian likelihood $\mathcal{L}$ (up to a normalisation constant) such that
\begin{equation}
\label{eq:likelihood}
    -2\ln\mathcal{L} = \sum_{i,j}\left[\vec{Q}^{\rm theo}(\bm{\theta}) - \vec{Q}^{\rm data}\right]_i C^{-1}_{ij} \left[\vec{Q}^{\rm theo}(\bm{\theta}) - \vec{Q}^{\rm data}\right]_j \, ,
\end{equation}
where the theory and data vectors, $\vec{Q}^{\rm theo}$ and $\vec{Q}^{\rm data}$, are a concatenation of the monopole, quadrupole, and hexadecapole of the \tpcf, $C_{ij}$ are the components of the covariance matrix, and $\bm{\theta}$ is the parameter vector.
In the likelihood, $i$ and $j$ run from 1 to $3n$ where $n$ is the number of bins in $s$.
The latter depends on the minimum fitting scale, and we have $n=\left[36,\,35,\,34,\,33,\,32\right]$ for $\smin=\left[20,\,25,\,30,\,35,\,40\right]\Mpc$, respectively.
This range of $\smin$ values is motivated by the assumed validity of the covariance matrix but also where we expect to mainly harvest nonlinear information from the 2PCF multipoles.
In addition, the Fourier-space damping used by some of the models (see discussion in Sect.~\ref{sec:accelerated_models}) imposes restrictions on the usage of smaller scales below $20\Mpc$.
An extended discussion with tests can be found in \citet{2022JCAPZhang_2PCF}, also concluding with $\smin=20\Mpc$ as a lower limit.

The posterior distribution of the considered parameter space is explored using the importance-nested sampling algorithm \citep{Feroz2008MNRAS_multimodal_sampling, Feroz2009MNRAS_multinest, Feroz2019OJAp_importance_nested_sampling} implemented in \texttt{PyMultiNest} \citep{Buchner2014A&A_multinest},\footnote{\texttt{PyMultiNest} at \url{https://github.com/JohannesBuchner/PyMultiNest}} a \texttt{Python} wrapper for the original \texttt{MultiNest} written in Fortran.\footnote{\texttt{MultiNest} at \url{https://github.com/JohannesBuchner/MultiNest}}
In the setup of \texttt{MultiNest}, we use \num{3000} live points to make sure that every multimodality is covered in the initial sampling of the parameter space.
Furthermore, as suggested in the documentation, we use an evidence tolerance of $0.5$ and a sampling efficiency of $0.8$, suitable for constraining parameters.
Once converged chains are obtained, those are analysed using the \texttt{getdist} package \citep{Lewis2019arXiv_getdist},\footnote{\texttt{getdist} at \url{https://github.com/cmbant/getdist}} to produce marginalised 2D and 1D posterior distributions.

\subsection{Implementation of models and fast predictions using machine learning}\label{sec:accelerated_models}
To enable faster and more efficient computations of the theoretical models presented in Sect. \ref{sec:theory_rsd}, we make use of several state-of-the-art numerical tools, including FFTlog techniques and emulators.
The RegPT predictions for the TNS model -- utilised for the power spectra and $C_A$ terms from Sect.~\ref{sec:theory_TNS} -- are obtained with the publicly available code \texttt{pyregpt}.\footnote{\texttt{pyregpt} at \url{https://github.com/adematti/pyregpt}}
For the hybrid model as well as the bias terms and $C_B$ terms we use the code \texttt{TNS\_ToolBox},\footnote{\texttt{TNS\_ToolBox} at \url{https://github.com/sdlt/TNS_ToolBox}} which makes use of the FFTlog technique for bias contributions as described in \citet{Simonovic2018JCAP}.
In the hybrid model, the nonlinear matter power spectrum is obtained from the \euclidemulatortwo \citep{Knabenhans-EP9} or \texttt{baccoemu} \citep{Angulo2021MNRAS_BACCO}.
The configuration-space multipoles of the \tpcf are then obtained with the FFTlog transformation via the \texttt{mcfit} package.\footnote{\texttt{mcfit} at \url{https://github.com/eelregit/mcfit}}
We sometimes refer to these flavours of TNS models as the `classic' TNS models as compared to the \vdg model.
For the multipoles of the \eft and \vdg model, both in the template and full-shape fitting, we employ the publicly available emulator code \comet \citep{EggemeierEtal_2023}.\footnote{\comet at \url{https://comet-emu.readthedocs.io/en/latest/index.html}}
These multipoles are then transformed into configuration space using the FFTlog transformation as implemented in the \texttt{hankl} library.\footnote{\texttt{hankl} at \url{https://hankl.readthedocs.io/en/latest/}}

The operation of applying a Fourier transformation to the multipoles for the \vdg, \eft and TNS models to obtain configuration space statistics merits some further discussion.
Theoretically, the power spectrum would be integrated over all $k$, also including modes beyond some $\kmax$, which describes the maximum $k$ where the perturbative model is valid.
The multipoles of the galaxy \tpcf are fairly smooth functions of scale and their main features include the \bao and a rapid change in amplitude towards low $s$.
Therefore, the prediction in Fourier space on scales beyond a $\kmax$ of a few times $0.1\kMpc$ mainly affects the shape of the \tpcf on the smallest $s<20\Mpc$ while the \bao feature is represented by the power spectrum below $\kmax$.
Taking $\smin=20\Mpc$ as the smallest $\smin$ that we consider in this analysis, we are not affected by the precise form of the power spectrum on scales where, for example, the \eft model is not valid any more.
This means that we can also safely apply a Gaussian damping to the predicted power spectrum multipoles of the \vdg and \eft model with a damping scale of $r=0.25\Mpc$, which is necessary to avoid numerical artefacts in the FFTLog algorithm such as ringing.
This damping is particularly important for models that have counterterms scaling as $\propto k^2\Plin$.
For the analysed TNS models we refrain from using a damping as they do not contain counterterms and the power spectrum multipoles approach zero at small scales.

For the \clpt and \cleft models, we build a new emulator based on the infrastructure of \comet to obtain fast predictions, both for the template and full-shape fitting approaches.
\comet is built upon Gaussian processes to emulate the anisotropic power spectrum contributions associated with different bias combinations.
It exploits the evolution mapping approach \citep{Sanchez2020, Sanchez2022} to reduce the dimensionality of the cosmological parameter space, yielding accurate predictions with a relatively compact training set. A full model with \comet is evaluated in around $10\,{\rm ms}$, making it an indispensable asset for the full-shape analysis in this work.
The \lpt models require fast predictions of bias contributions for each velocity statistic before they are fed into the GS integral.
More details about the changes in architecture applied to \comet to emulate \lpt ingredients as well as the validation of the emulator can be found in Appendix~\ref{sec:appendix_emulator}.
It is important to note that the nonlinear process of GS cannot be emulated as it does not factor into individual contributions and hence has to be computed numerically with the \texttt{CLEFT\_GS} code.\footnote{\texttt{CLEFT\_GS} at \url{https://github.com/sdlt/CLEFT_GS}} This code was also used to generate the training set of the emulator.

\subsection{Parameter priors}
The free parameters are split into cosmological and nuisance parameters.
For the cosmological parameters, we set wide uninformative flat priors as given in Table~\ref{tab:priors}.
The priors for the full-shape analysis are basically the limits of \comet.
We keep fixed both the physical energy density parameter for baryons $\omega_{\rm b}$ as well as the spectral index $\ns$, as galaxy clustering measurements on their own do not carry significant information about these parameters.
We also summarise in Table~\ref{tab:priors} the priors for the nuisance parameters, and we list in Table~\ref{tab:model_parameters} the specific parameters required by each of the \rsd model described in Sect. \ref{sec:theory_rsd}.
The FoG parameters are velocity dispersions and hence their priors impose positivity.
We set the stochastic contributions in the \eft and \vdg power spectrum modelling to zero, both for the template fitting and the full-shape analysis, since they are not needed for the \tpcf.
For all counterterms we choose an agnostic approach and use very broad priors although this can lead to projection effects.\footnote{Sometimes called prior volume effects.}
This effect describes an apparent bias in marginalised 1- or 2-dimensional posterior distributions, usually considered in contour plots, originating from a degeneracy in the higher-dimensional parameter space.
It is particularly problematic when this bias occurs in the cosmological parameters.
More nuisance parameters, for example galaxy bias, counterterms, or shot-noise parametrisations, can cause the posterior distribution to become significantly non-Gaussian and thus susceptible to marginalisation effects.
This issue has already been reported in several analyses, including \citet{Ivanov2020JCAP_BOSS_EFT_full_shape}, \citet{ Chudaykin2021PhRvD_EFT_analysis}, \citet{ Philcox2022PhRvD_BOSS_EFT_analysis}, \cite{Simon2023PhRvD_EFT_consistency}, \citet{ Carrilho2023JCAP_BOSS_EFT_analysis}, and \citet{ Hadzhiyska2023OJAp_ProjectionEffects}.
A thorough analysis of projection effects in the context of \Euclid will be found in Euclid Collaboration: Moretti et al. (in prep.).
As a possible extension to the \lcdm model we consider varying the equation of state parameter $w_0$ for dark energy but refrain from using the full Chevallier--Polarski--Linder-parametrisation \citep[hereafter CPL,][]{Chevallier2001IJMPD_CPL_DE, Linder2003PhRvL_CPL_DE} as it requires the simultaneous fitting of snapshots at different redshifts.
Again, we refer to Moretti et al. (in prep.) for an investigation of the full CPL parametrisation.
In this work we always marginalise over the nuisance parameters and therefore only cosmological parameters are compared to the fiducial values from the \flagship simulations.
\begin{table}[t]
\caption{Priors on the cosmological and nuisance parameters of the different \rsd models described in Sect. \ref{sec:theory_rsd}, as employed in the template or full-shape analysis, respectively. The ${\cal U}~[a,b]$ notation represent a uniform prior from $a$ to $b$.}
\renewcommand{\arraystretch}{1.2}
\smallskip
\label{tab:priors}
\smallskip
\begin{tabular}{|c|c|}
\hline
\rowcolor{blue!5}
 Parameter & Priors \\
\hline
\rowcolor{blue!5}
\multicolumn{2}{|c|}{\bf Template analysis} \\
\hline
   $f$ & $\mathcal{U}~[0.5,1.05]$ \\
   $\sigtwelve$ & $ \mathcal{U}~[0.2,1]$ \\
   $\qpar$ & $\mathcal{U}~[0.9,1.1]$ \\
   $\qperp$ & $\mathcal{U}~[0.9,1.1]$ \\
\hline
\rowcolor{blue!5}
\multicolumn{2}{|c|}{\bf Full-shape analysis} \\
\hline
  $h$ & $\mathcal{U}~[0.6, 0.8]$ \\
   $10^9\,A_{\mathrm{s}}$ & $\mathcal{U}~[1.2, 3]$ \\
   $\omega_{\rm c}$ & $\mathcal{U}~[0.085, 0.155]$ \\
   $w_0$ & $\mathcal{U}~[-1.5, -0.5]$ \\
  \hline
  \rowcolor{blue!5}
\multicolumn{2}{|c|}{\bf Galaxy bias parameters} \\
\hline
   $b_1$ & $\mathcal{U}~[0.5,3.5]$ \\
   $b_2$ & $ \mathcal{U}~[-10,10]$ \\
   $\bGtwo$ & $\mathcal{U}~[-20,20]$  or fixed to Eq.~\eqref{eq:LL_bG2_bGam3}\\
   $\bGthree$ & $\mathcal{U}~[-20,20]$  or fixed to Eq.~\eqref{eq:LL_bG2_bGam3}\\
   $b_{s^2}$ & $\mathcal{U}~[-20,20]$ or fixed to Eq.~\eqref{eq:LL_bs2} \\
\hline
\rowcolor{blue!5}
\multicolumn{2}{|c|}{\bf Counterterms} \\
\hline
    $c_0\,[(\Mpc)^2]$ & $\mathcal{U}~[-500,500]$ \\
    $c_2\,[(\Mpc)^2]$ & $\mathcal{U}~[-500,500]$ \\
    $c_4\,[(\Mpc)^2]$ & $\mathcal{U}~[-500,500]$ \\
    $c_{\rm nlo}\,[(\Mpc)^4]$ & $\mathcal{U}~[-800,800]$ \\
    $\alpha_{\xi}\,[(\Mpc)^2]$ & $\mathcal{U}~[-100,200]$ \\
    $\alpha_{v}\,[(\Mpc)]$ & $\mathcal{U}~[-100,200]$ \\
    $\alpha_{\sigma}$ & $\mathcal{U}~[-100,200]$ \\
  \hline
\rowcolor{blue!5}
\multicolumn{2}{|c|}{\bf FoG} \\
\hline
    $\avir\,[(\Mpc)^2]$ & $\mathcal{U}~[0,10]$ \\
    $\sigma^2_{{\rm CLPT},v}\,[(\Mpc)^2]$ & $\mathcal{U}~[0,100]$ \\
    $\sigma_{{\rm TNS},v}\,[\Mpc]$ & $\mathcal{U}~[0,100]$ \\
  \hline
\end{tabular}
\tablefoot{The priors on the linear bias $b_1$ is specified on the Eulerian bias and is transformed into Lagrangian basis internally in the code for \clpt and \cleft. For the nonlinear local bias $b_2$ the same prior is used both in the Lagrangian and Eulerian basis.}
\end{table}
\begin{table}
\caption{Complete list of bias, counterterms, and FoG parameters for each of the \rsd models considered in this work. A tick symbol is used to specify if the considered parameter (row) is present in the considered model (column). The bottom row shows the total number of nuisance parameters for each model.}
  \smallskip
  \renewcommand{\arraystretch}{1.2}
  \label{tab:model_parameters}
    \smallskip
  \begin{tabular}{cccccc}
    \hline
    \rowcolor{blue!5}
    Parameter & \clpt & \cleft & TNS & \vdg & \eft\\
    \hline
    $b_1$  & \checkmark & \checkmark & \checkmark & \checkmark & \checkmark \\
    \hline
    $b_2$ & \checkmark & \checkmark & \checkmark & \checkmark & \checkmark \\
    \hline
    $\bGtwo$ & & & \checkmark & \checkmark & \checkmark \\
    \hline
    $\bGthree$ & & & \checkmark & \checkmark & \checkmark \\
    \hline
    $b_{s^2}$ & & \checkmark & & & \\
    \hline
    $\czero$ & & & & \checkmark & \checkmark \\
    \hline
    $\ctwo$ & & & & \checkmark & \checkmark \\
    \hline
    $\cfour$ & & & & \checkmark & \checkmark \\
    \hline
    $c_{\rm nlo}$ & & & & & \checkmark \\
    \hline
    $\avir$ & & & & \checkmark & \\
    \hline
    $\sigv$ & \checkmark & & \checkmark & & \\
    \hline
    $\alpha_{\xi}$ & & \checkmark & & & \\
    \hline
    $\alpha_{v}$ & & \checkmark & & & \\
    \hline
    $\alpha_{\sigma}$ & & \checkmark & & & \\
    \hline\hline
    tot & 3 & 6 & 5 & 8 & 8 \\
    \hline
  \end{tabular}
\end{table}

\subsection{Performance metrics}\label{sec:performance_metrics}
To conduct a thorough comparison of the different theoretical models, assessing the strengths and weaknesses of each, we make use of three different performance metrics: the goodness of fit, the figure of merit (precision) and figure of bias (accuracy). This approach is very similar to existing performance assessments in the literature \citep{Eggemeier2021PhRvD_bias_joint, Pezzotta2021PhRvD_bias, EP-Pezzotta}.

The goodness-of-fit metric considered in this work is the reduced $\chi^2$, $\chired$, defined via
\begin{equation}
    \chired = \frac{\chi^2}{\rm dof} = -2\frac{\ln \mathcal{L} }{\rm dof}\, ,
\end{equation}
where the \dof are the total number of fitted data points minus the number of free parameters in the model.
The standard deviation of the $\chired$ distribution is given by \smash{$\sigma_{\chi^2}= \left(2/\mathrm{dof}\right)^{1/2}$}.
In general, a good fit to the data is characterised by \smash{$\chired \sim 1$}.
On the one hand, if the model overfits data, then the value of the \smash{$\chired$} is smaller than unity, meaning that the model follows the data too closely, and undesired features (such as noise) are captured by the model as if they were a real signal.
If on the other hand the model underfits the data, meaning a $\chired$ value larger than unity, the model is too rigid to follow the general trend of data, and a more flexible model needs to be devised.
Whenever we present a value for $\chired$, we report the mean value taken over the posterior, as this is the quantity that should be close to unity.

The $\chired$ value alone does not say anything about the accuracy or precision of recovered parameters.
If strong degeneracies between different parameters are present, a $\chired$ value close to unity might be obtained, although the recovered parameters may significantly differ from the fiducial ones.
In our case, we compare three cosmological parameters with fiducial values, for which we can compress the information of accuracy into the \fob, defined as
\begin{equation}\label{eq:FoB}
    \mathrm{FoB} = \sqrt{\sum_{i,j}\left[\pmb{\theta} - \pmb{\theta}_{\rm fid}\right]_i \,\mathcal{S}^{-1}_{ij} \,\left[\pmb{\theta} - \pmb{\theta}_{\rm fid}\right]_j} \, .
\end{equation}
The matrix $\mathcal{S}$ denotes the parameter covariance matrix, which, in our case, is estimated with \texttt{getdist} from the posterior samples obtained with \texttt{MultiNest}.
The \fob can thus be seen as a generalisation of the 1D case, incorporating the full covariance between different parameters.
We are interested in the value of the \fob yielding for instance a 68\% credible region around the fiducial parameter.
It can be shown that this credible region can be obtained by evaluating the percent-point function of the $\chi^2$ distribution with $n$ degrees of freedom at these thresholds \citep{Slotani1964_multivariate_normal}.
The 68\%, 95\%, and 98\% credible regions for the \fob computed over three parameters are then given by a \fob equal to $\sim$1.87, 2.80, and 3.37 respectively.

Lastly, to properly quantify the recovery of parameters from the fitting process we also need a measure of precision. Similarly to the \fob, we need a generalisation to higher dimensional parameter spaces that compresses the information about the precision of the three considered parameters.
To tackle this, \citet{Albrecht2006_DE_task_force} used the inverse of the surface of the 2D posterior for $w_0$--$w_a$ assuming a Gaussian contour and defined it as a \fom. We use the generalisation to higher-dimensional parameter spaces by \citet{Wang2008PhRvD} defined as
\begin{equation}
    \mathrm{FoM} = \frac{1}{\sqrt{\mathrm{det}(\mathcal{S})}} \, .
\end{equation}
For $n$ parameters, the \fom is the inverse of the $n$-dimensional hypervolume spanned by the posterior.
A smaller volume indicates a larger \fom, meaning a better precision on the parameters.

\section{Results}
\label{sec:results}

\subsection{Template-fitting analysis}
In this first section of the results, we present the template-fitting approach, where we assume a fixed linear power spectrum computed with the \flagship cosmology (shown in Table~\ref{tab:flagship_cosmology}), and build different models for $\Psgg$ by varying the nuisance, AP, $f$, and $\sigtwelve$ parameters.
For all the presented results and unless otherwise stated, including also the full-shape fits, the non-local bias parameters for the models are left free to vary in the fit.
In Appendix~\ref{sec:app_TNS_testing} we provide a supplementary comparison within the TNS class of models as described in Sect.~\ref{sec:theory_TNS}.
In this section, we compare the relative performance of the main theory models described in Sect.~\ref{sec:galaxy_clustering_redshift_space} -- namely the \cleft, \clpt, \eft, and \vdg models.
\begin{figure*}
    \centering
    \includegraphics[width=0.9\textwidth]{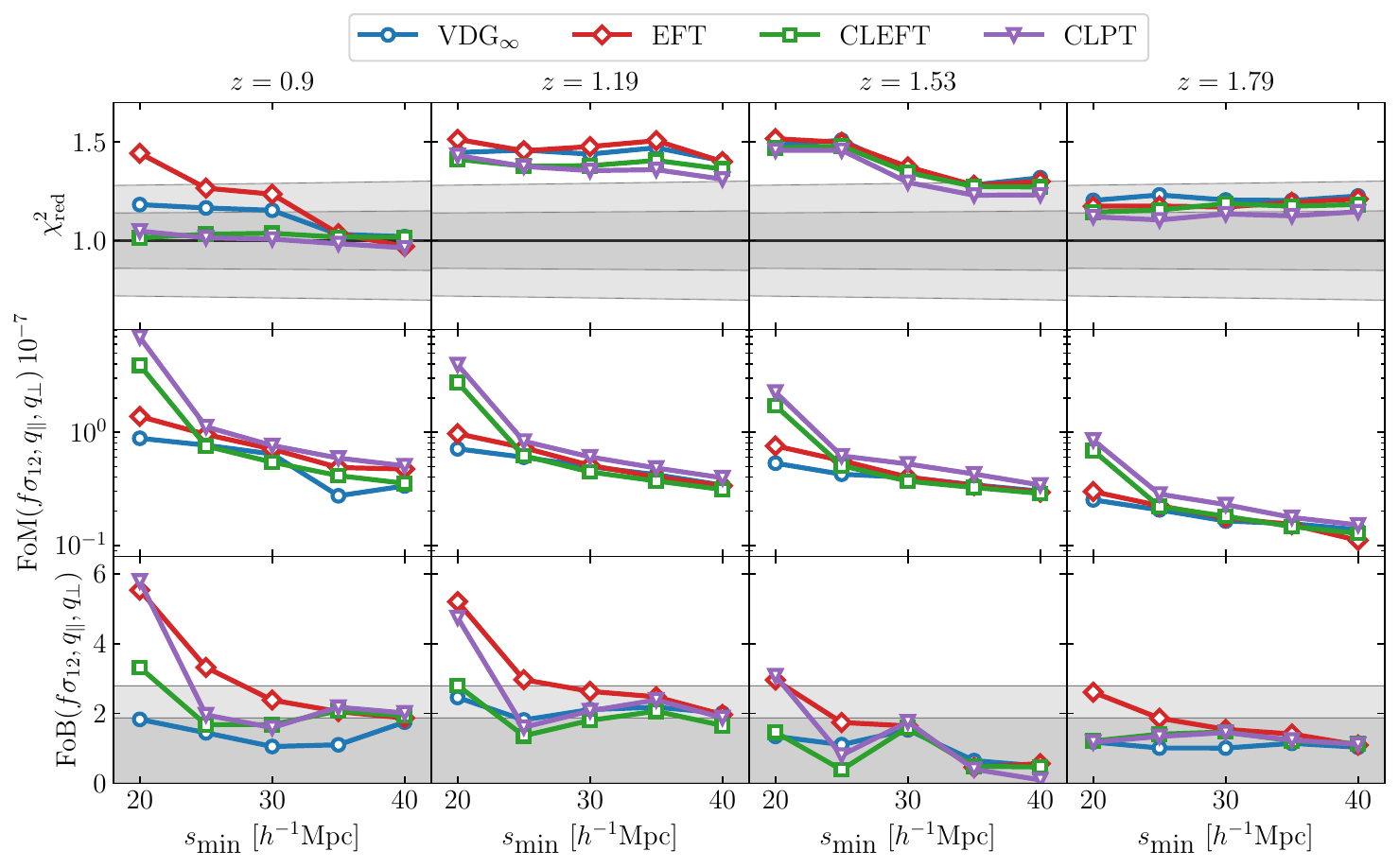}
    \caption{Performance metrics for the fixed-cosmology case of the \vdg, \eft, \cleft, and \clpt model as a function of the minimum fitting scale $\smin$. The shaded regions in the panels displaying the $\chired$ values refer to the standard deviation and twice the standard deviation of the $\chired$ distribution with the model degrees of freedom fixed to seven (\clpt) as a conservative choice. In the \fob panels, the two shaded regions denote the 68-th and 95-th percentiles as described in Sect.~\ref{sec:performance_metrics}.}
    \label{fig:fixed_cosmo_metrics}
\end{figure*}

In Fig.~\ref{fig:fixed_cosmo_metrics}, we show the performance metrics in the fixed-template approach.
As displayed in the upper panels, the $\chired$ values are in general very similar among the different models, particularly at $z\gtrsim1$. We can observe a slight trend of the models getting closer to each other in terms of $\chired$ while going to higher redshifts.
At the lowest redshift, we find an increase of $\chired$ for the \eft model at $\smin = 20\Mpc$, indicating a failure of the model to properly describe strong nonlinear features.
The \fom of all models follows the same trend of continuously decreasing with higher redshifts or $\smin$.
The decrease in terms of redshift is expected as the number density of the galaxy sample gets smaller and therefore the amplitude in the covariance matrix increases and reduces the constraining power.
On the other hand including more data points on small scales (reducing $\smin$) shrinks the uncertainties on the recovered parameters, resulting in an increase in \fom.
The \clpt and \cleft models follow similar trends as the \eft and \vdg models, with the key difference that at $\smin=20\Mpc$ the \fom rises sharply for the former.
This feature is present at all redshifts, although it is less important at higher redshifts.
In general, we obtain the most constraining power with the \clpt model as it has the least free parameters compared with the other models.
We checked the full posterior contours of the \cleft and \clpt model at $z=0.9$ for the strong increase in \fom and found that it is caused by the FoG parameter affecting the velocity dispersion -- that is $\sigv$ for the \clpt model and $\alpha_{\sigma}$ for the \cleft model -- that are much more constrained for $\smin=20\Mpc$ compared to larger $\smin$.
Due to a degeneracy with $\qpar$ the constraints on the FoG parameters shrinks its uncertainty, driving it away from the fiducial value.
The strong increase in the \fom for \cleft and \clpt at $\smin = 20\Mpc$ also explains the consequent increase in \fob, leading to \fob values for both models located outside the 95\% region at $z=0.9$.
\begin{figure*}
    \centering
    \includegraphics[width=0.9\textwidth]{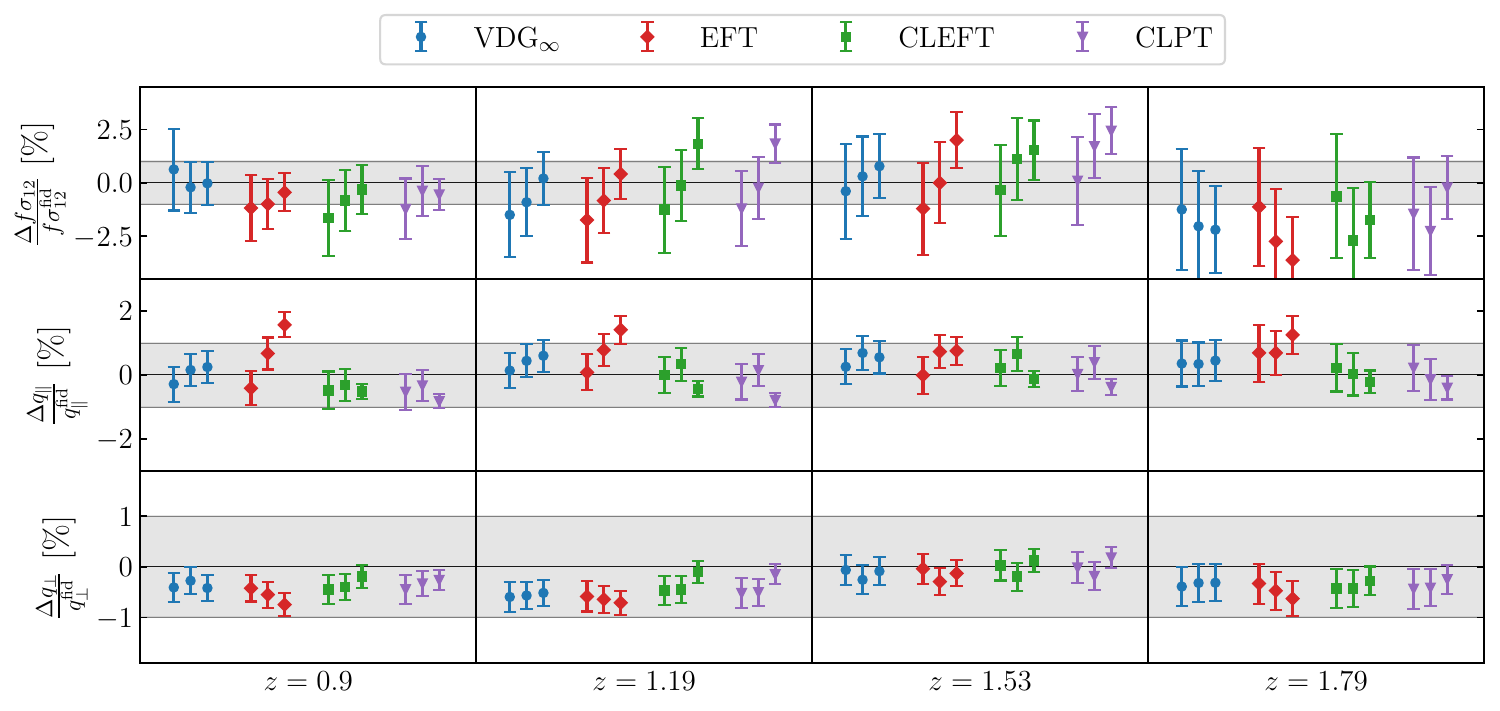}
    \caption{Percent relative difference between the recovered parameters of the template-fitting approach with respect to their fiducial values. Each of the models under consideration (\vdg, \eft, \cleft, \clpt) is represented by three data points for every redshift, corresponding to the recovered values using a progressively smaller value of $\smin \in \{40, 30,20\}\Mpc$ from left to right. The grey shaded area marks an accuracy of 1\% across all parameters.}
    \label{fig:fixed_cosmo_params}
\end{figure*}
A similar trend in terms of \fob can be observed for the \eft model, although it shows a lower \fom on those scales, suggesting an actual failure of the model at $20\Mpc$, as also reinforced by the corresponding $\chired$ value.
However, interpreting the $\chired$ value has to be done with care as we analyse only a single realisation, despite having a large volume, therefore we might be susceptible to sample variance.
This problem is particularly evident when looking at the $\chired$ values for the two intermediate redshifts where all models are outside the $1\sigma$ region of the $\chired$ distribution.
Still, comparing, for the redshift $z=0.9$, the models among each other, shows a significantly higher $\chired$ of the \eft model at $\smin=20\Mpc$ compared to the other models.
We provide further tests concerning sample variance in the full-shape scenario in Appendix\,\ref{sec:appendix_axis_selection_sample_variance}.

In conclusion, our analysis suggests that any model with $\smin=30$--$40\Mpc$  yields consistently good estimates of the \fob, which largely stays within 68\% region and never exceeds the 95\% threshold.
This is particularly remarkable for the \clpt model, being by far the simplest of all models considered here, with only seven free parameters. It performs very similarly to the \cleft model on those scales, suggesting that counterterms for the zeroth and first velocity moments might not be necessary.
We note that the \clpt model has a free parameter to correct the velocity dispersion amplitude -- similar to the \cleft counterterm $\alpha_{\sigma}$ -- but it is scale-independent.
The only model that consistently performs very well in terms of \fob is the \vdg model, even at the lowest redshift and for the smallest minimum fitting scale. In terms of \fom, the models should be compared based on their reach for the \fob, which is the maximum constraining power that can be achieved while providing unbiased parameters.
It is clear that the \fom of the \vdg is then higher compared to the \eft model because the latter can only reach $30\Mpc$ at $z=0.9$.
It even reaches a similar \fom than the \clpt model (at $\smin=30\Mpc$), despite having five more free parameters.
Though the latter is also basically unbiased at $\smin=25\Mpc$, with a \fob just above the 68\% percentile, surpassing the \vdg in terms of \fom.

To further assess the performance on parameter recovery, in Fig.~\ref{fig:fixed_cosmo_params} we present the mean values on the parameters with their respective 68\% uncertainties.
The amplitude parameter $f\sigtwelve$ is recovered within $1\sigma$ for the majority of the models and scale cuts, except at $z=1.79$, where the posterior distribution looks slightly biased towards values smaller than the fiducial one, albeit with larger uncertainties.
When considering the AP parameter $\qpar$, it again becomes clear that the sharp increase in \fom for the \cleft and \clpt models with $\smin = 20\Mpc$ originates from discrepancies in this parameter.
The errorbars visually shrink much more when going from $\smin=30\Mpc$ to $\smin=20\Mpc$ than from $\smin=40\Mpc$ to $\smin=30\Mpc$. The \eft model consistently overestimates $\qpar$, thus explaining the larger \fob seen in Fig.~\ref{fig:fixed_cosmo_metrics}. Apart from these features, $\qpar$ appears to be well recovered within 1\% deviation for most of the considered models and scale cuts. The AP parameter $\qperp$ is also thoroughly recovered with 1\% accuracy, but due to very small errors, the recovered central values are sometimes off by more than $1\sigma$ from the fiducial parameters. This is particularly the case for all the models at $z=0.9$ and $z=1.19$.

\subsection{Full-shape analysis}
\label{sec:results_fullshape}\label{sec:lcdm_cosmology}

In the second part of the results, we perform a full-shape analysis, where a subset of cosmological parameters \{$h,\,\As,\,\omegac$\} is varied in the fit together with the nuisance parameters, effectively recalculating the linear power spectrum at each position in parameter space. The full-shape analysis is extended in Appendix\,\ref{sec:appendix_extended_cosmo} to also incorporate a parametrisation of the dark energy equation of state.

This section presents results from the full-shape analysis when exploring a \lcdm cosmology.
As mentioned before, we leave all non-local bias parameters of the respective models free to vary in the fit, while in Appendix~\ref{sec:appendix_bias_testing} we inspect the validity of \LL approximations and their impact on the performance metrics.

\begin{figure*}
    \centering
    \includegraphics[width=0.9\textwidth]{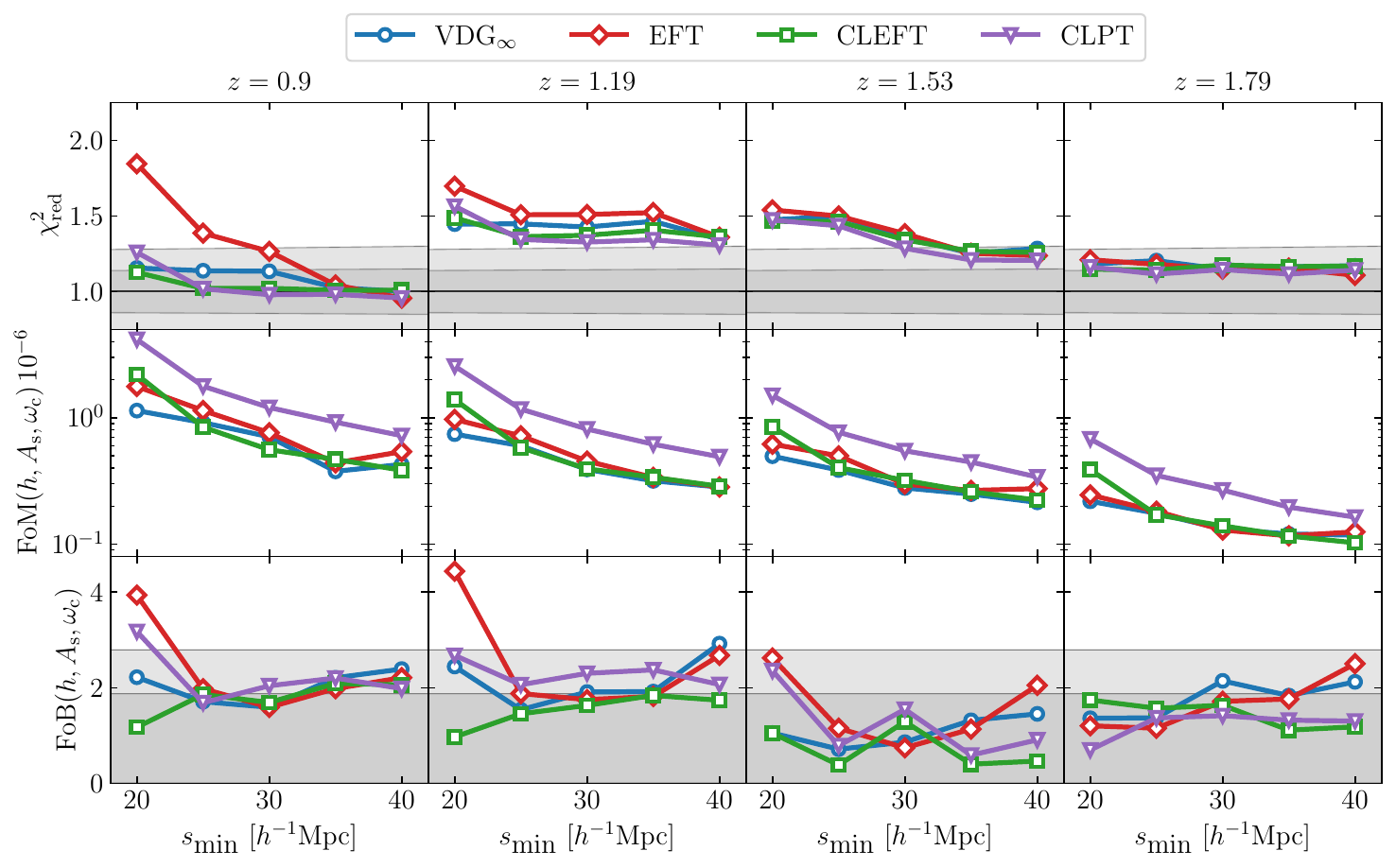}
    \caption{Performance metrics of the \vdg, \eft, \cleft, and \clpt model for the full-shape analysis as a function of the minimum fitting scale $\smin$. The shaded regions in the panels displaying the $\chired$ values refer to the standard deviation and twice the standard deviation of the $\chired$ distribution with the model degrees of freedom fixed to six (\clpt) as a conservative choice. For the \fob, the two shaded regions denote the 68-th and 95-th percentiles as described in Sect.~\ref{sec:performance_metrics}.}
    \label{fig:full_shape_metrics}
\end{figure*}

In Fig.~\ref{fig:full_shape_metrics}, we show the performance metrics of the different \rsd models.
We find that the general trend of the $\chired$ values is visually similar to that of the template-fitting approach.
Though, analogous to the template-fitting result, we might be affected in some cases by sample variance, also in the full-shape analysis.
Still it is useful to assess the trends of the $\chired$ value with scale cuts.
Similarly to the template results, the \eft model fails to describe the data at $z=0.9$ when the minimum fitting scale is $20\Mpc$. However, high $\chired$ values are reported for the two intermediate redshifts for all models, rendering the $\chired$ at $z=0.9$ for some of the models unexpectedly low, indicating sample variance.
To verify this hypothesis we ran full-shape fits to the other \los of the snapshots and found the $\chired$ values at $z=0.9$ to be much higher for all the models, hinting at a sample-variance effect for the smallest redshift.
Details on this test can be found in the Appendix~\ref{sec:appendix_axis_selection_sample_variance}.
Therefore, we should compare the $\chired$ values among models for a given redshift instead of comparing them between different redshifts.
Clearly, at $\smin=20\Mpc$ and $z=0.9$ the \eft model has a significantly higher $\chired$ than all the other models, deviating more than $2\sigma$ from the mean of the $\chired$ distribution.
At $z=1.79$, the $\chired$ values are inside of the $2\sigma$ region and all models seem to fit the data equally well.

Similarly to the template-fitting method, the \fom decreases with higher minimum fitting scale and redshift.
The \clpt model displays the highest \fom because it employs the least number of free parameters, as compared to \eft, \vdg, and \cleft.
The sharp peak at $\smin=20\Mpc$ for \clpt and \cleft reported in the template-fitting analysis is much less pronounced, especially for \cleft, and its \fom follows that of the \eft model.
Consistently with the fixed-template case, the \vdg model exhibits the smallest \fom at $\smin=20\Mpc$.
However, as mentioned before, for a fair comparison, the possible \fom of a model is given by the maximum reach in $\smin$ where the parameters are unbiased.
Therefore, the \vdg model exhibits similar constraining power as the \eft model that can only reach $\smin=25\Mpc$ while the \vdg model is still within the 95\% region at $\smin=20\Mpc$.

Regarding the \fob, the cosmological parameters are generally well recovered with many models and configurations reaching values within the 95\% region, similar to the template-fitting results.
Again, the \eft model displays a high \fob at $\smin=20\Mpc$ and for the two smallest redshifts, recovering biased parameters outside of the 95\% threshold.
The \clpt model is biased more than 1$\sigma$ for almost all $\smin$ at these two redshifts.
At the highest redshift, all models display a \fob within the 68\% region or slightly above, indicating a very good recovery of the fiducial parameters.
Notably, the \eft and \vdg model have a decreasing \fob when going to lower $\smin$ for $z=1.79$ signalling projection effects that are mitigated when more data is added.
Therefore, any slight bias beyond $1\sigma$ could be associated with projection effects rather than a model failure.
A somewhat similar trend can also be observed for the \eft model at $z=1.53$ and the \vdg model at $z=1.19$ where the \fob increases both for low and high $\smin$ with a turn-around in between.
While the increase towards low $\smin$ indicates problems on the theory side, the other end is caused by projection effects.
Also the \cleft model appears to exhibit this effect at the two lowest redshifts.
Throughout all considered redshifts, the \cleft model performs best in terms of \fob, with values below the 68\% limit or slightly above.
Shortly followed by the \vdg model that performs similarly but is biased within the 95\% region for the two lowest redshifts at $\smin=20\Mpc$.

\begin{figure*}
    \centering
    \includegraphics[width=0.9\textwidth]{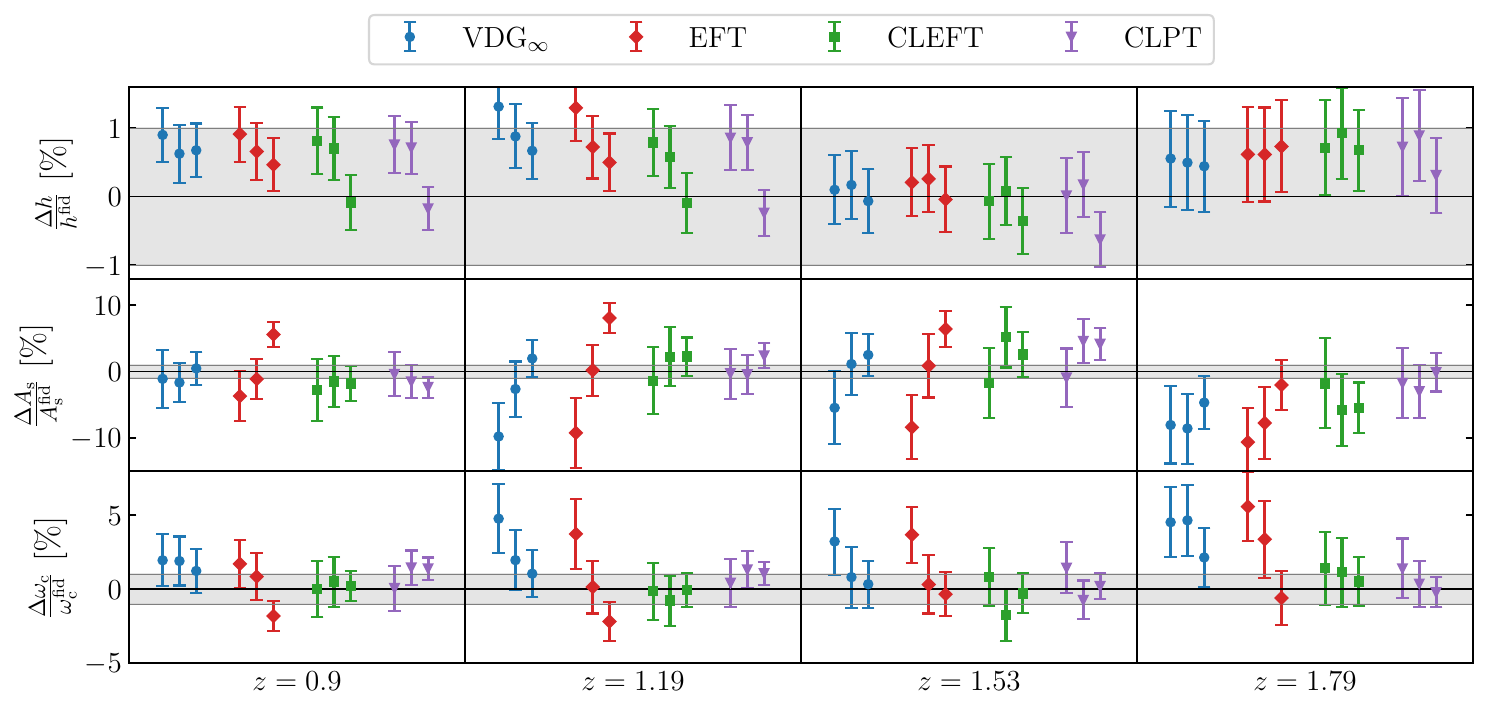}
    \caption{Same as in Fig.~\ref{fig:fixed_cosmo_params}, but for the full-shape analysis constraining the cosmological parameters $h$, $\As$, and $\omegac$.}
    \label{fig:full_shape_params}
\end{figure*}

We present the accuracy on the recovery of the individual cosmological parameters in Fig.~\ref{fig:full_shape_params}.
Similarly to Fig.~\ref{fig:fixed_cosmo_params}, the size of error bars decreases for lower minimum fitting scale and redshift.
The dimensionless Hubble parameter $h$ is well recovered with an accuracy of around 1\% but the central values deviate for many configurations by more than $1\sigma$ at the two lowest redshifts.
In contrast, the amplitude parameter $\As$ exhibits a much stronger scatter together with larger uncertainties.
A similar trend is also observed for the power spectrum in real space \citep{EP-Pezzotta} and also in redshift space (Euclid Collaboration: Camacho et al., in prep.) and is caused by projection effects as $\As$ is strongly degenerate with, for example, the linear bias parameter $b_1$ (see also Appendix\,\ref{sec:appendix_full_contour} for a discussion on the corresponding posterior contours).
This effect is visible on the two intermediate redshifts for the \vdg and \eft model that have two more free parameters than the \cleft model and therefore might be more plagued by this effect.
Indeed, adding information by reducing the $\smin$ from $40\Mpc$ to $30\Mpc$ helps in recovering the fiducial parameter for $\As$ but also $\omegac$.
In general, the central values of $\As$ often lie outside the 1\% region, however, due to the large error bars, they are rarely deviating more than $1\sigma$ at the two lowest considered redshifts.
At $z=0.9$ and $z=1.19$, the minimum fitting scale $\smin=30\Mpc$ seems to yield the best recovery of $\As$ for all models.
In general, the accuracy in the recovery of $\omegac$ lies between that of $h$ and $\As$.
Although systematic effects are noticeably milder than for $\As$, there are still several configurations for which the fiducial parameters are not recovered by the selected models.
Going to the smallest minimum fitting scale helps in reaching the target accuracy of $1\%$ for $\omegac$ that is a clear indication of projection effects (see also Euclid Collaboration: Moretti et al., in prep., for a detailed study).
Interestingly, when assessing the error bars in Fig.~\ref{fig:fixed_cosmo_params} it appears that the gain in uncertainty by including smaller scales is small for certain configuration, for example, the \vdg model at $z=0.9$ for the parameters $h$ and $\omegac$.
However, a more complete picture is given in the \fom in Fig.~\ref{fig:full_shape_metrics}, also measuring the parameter covariance, where we see around a factor of two improvement of the \fom between $\smin=40\Mpc$ and $20\Mpc$.
Of course, the highest absolute gain is achieved with the \clpt model having the least free parameters.

\begin{figure*}
    \centering
    \includegraphics[width=\textwidth]{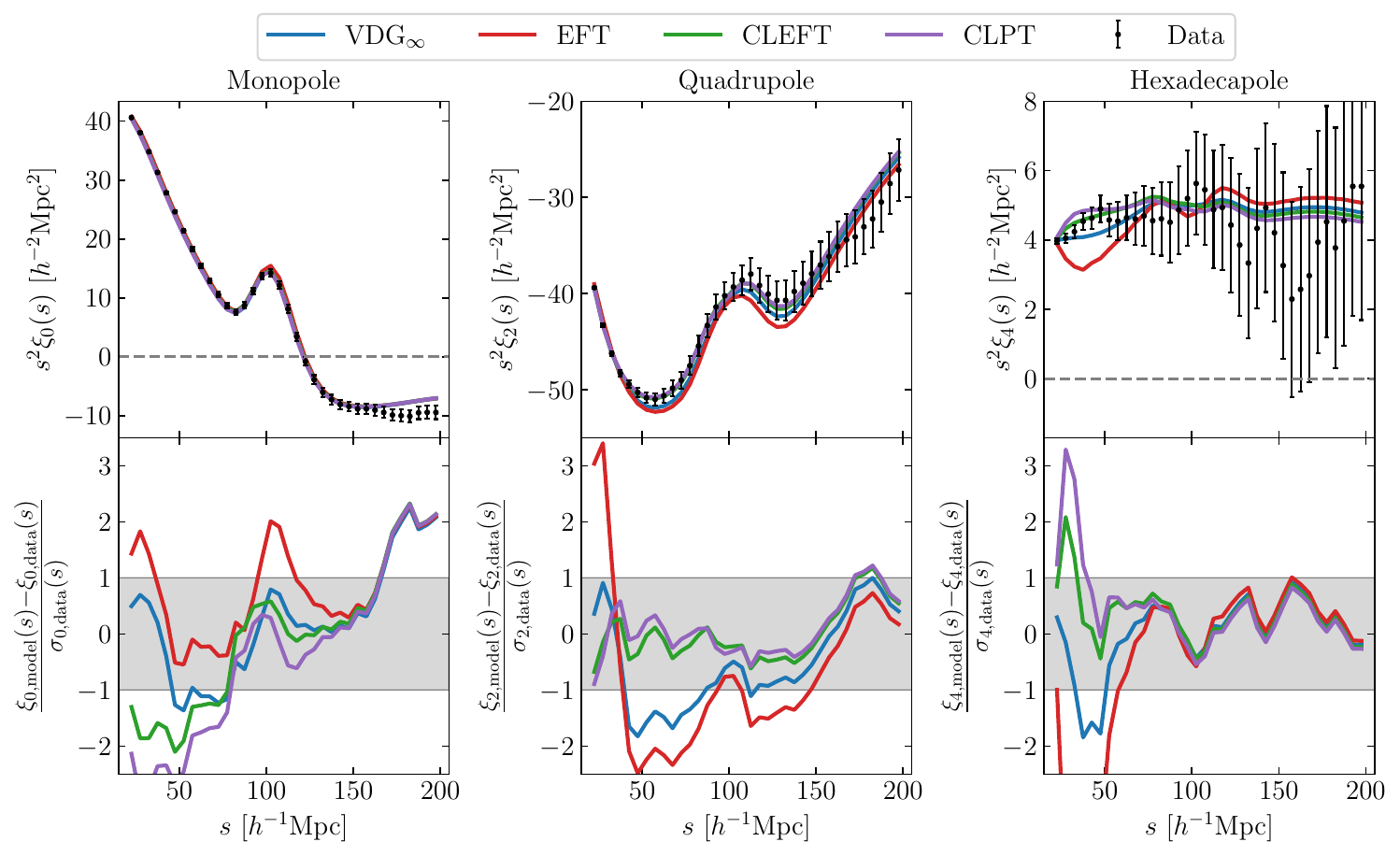}
    \caption{\emph{Top}: Legendre multipoles of the \tpcf as measured from the \flagship snapshot at $z=0.9$ (black errorbars) against the different \rsd model predictions evaluated with the mean values of the corresponding posterior distribution at $\smin=20\Mpc$. \emph{Bottom}: Relative difference between the data points and the different models in units of the diagonal of the covariance matrix (its square root). The grey shaded region refers to a difference of $1\sigma$.}
    \label{fig:xil_best_fit_z0.9}
\end{figure*}

In Fig.~\ref{fig:xil_best_fit_z0.9} we present a comparison between the \tpcf multipoles measured from the H$\alpha$ snapshot at $z=0.9$ against the ones for each of the considered \rsd models predicted using the mean parameters from the posterior samples -- as shown in Fig.~\ref{fig:full_shape_params} -- at $\smin=20\Mpc$.
The \clpt model seems to accurately recover the \bao peak in the monopole, but predicts a wrong amplitude at smaller separations below approximately $ 90\Mpc$, while the \cleft model performs better on those scales, emphasising the necessity to include \eft counterterms.
The \eft model severely overestimates the \bao feature in the monopole, deviating more than $1\sigma$ in terms of the statistical error of the measurements.
However, the fit is much better on these scales for the configuration with $\smin=30\Mpc$ (not shown) where the \eft model recovers also unbiased cosmological parameters with a \fob within the 68\% region (see Fig.~\ref{fig:full_shape_metrics}).
At the largest scales in the monopole -- above 170$\Mpc$ -- we see significant deviations from the data for all considered models indicating underestimated sample variance.
In terms of the quadrupole, both \clpt and \cleft recover the shape and amplitude well, staying within the $1\sigma$ confidence level on almost the whole range of scales considered.
Furthermore, on the quadrupole, the \vdg and \eft models perform worse compared to the \clpt or \cleft models, with the \eft model exhibiting deviations of more than $2\sigma$ on scales smaller than $100\Mpc$.
The behaviour of the models in the hexadecapole differs significantly, with the \eft model displaying particularly large deviations from the data points below $60\Mpc$.
On larger scales, above around $100\Mpc$, the uncertainty on the measurements becomes more significant, making all models consistent with the data.
On those scales also the relative differences among the models is minimal.
Notably, the \vdg and \cleft model seem to agree well with the data on all fitted scales and multipoles and always stay within or just above  $2\sigma$.

\section{Discussion}\label{sec:discussion_conclusion}
In this work we conducted a comprehensive comparison of state-of-the-art models to describe the redshift-space galaxy \tpcf in view of the forthcoming analysis of the main data collected by \Euclid. For this comparison, we made use of mock samples of H$\alpha$ galaxies populated in comoving snapshots of the \flagship simulation at a set of redshifts given by $z\in\{0.9, 1.19, 1.53, 1.79\}$, therefore spanning the range that is currently being explored by \Euclid. To test the model at an exquisite level of precision, we considered mock samples covering the whole simulation box volume, $V=3780\cMpc$.
Since this work was focused on intrinsic modelling systematics regarding \rsd, nonlinear matter clustering, and galaxy bias, the \flagship simulations were deemed adequate and we therefore do not expect our results concerning relative model performance to change significantly on the \flagshiptwo mocks.
The latter are an updated set of mocks on a lightcone with redshift errors and other observational systematics that are used in the forthcoming work of Moretti et al. (in prep.) but would have made this model comparison difficult to interpret due to many different effects at play.

The analysis was divided into two parts. First, we considered the performance of different \rsd models using a template-fitting approach, therefore relying on a template power spectrum at fixed cosmology, and subsequently exploring the parameter space defined by the growth rate $f$, the amplitude of fluctuations $\sigtwelve$, the geometrical distortion parameters $\qpar$ and $\qperp$, and the whole set of nuisance parameters.
Second, we carried out a full-shape analysis, where cosmological parameters are directly sampled in combinations with the nuisance ones, and the linear power spectrum is regenerated at each position in cosmological parameter space.
As the full-shape approach is inherently more demanding in terms of computational resources compared to template fitting, we had to rely on machine learning techniques to speed up the evaluation of multiple models during the exploration of the parameter space. In particular, we made use of the \comet package to produce fast predictions for the \vdg and \eft models, and constructed a new emulator based on the infrastructure of \comet that emulates velocity statistics for the \clpt and \cleft models.

In the template-fitting approach, we found that the \vdg model is the only one whose range of validity extends down to scales of $20\Mpc$ without introducing biases in the recovered parameters, unlike other models which typically break down at approximately $25$--$30\Mpc$, depending on the considered model and redshift. At the highest considered redshift, all models perform comparably well, also down to $20\Mpc$. In general, the parameter set $\{f\sigtwelve,\,\qperp,\,\qpar\}$ is typically recovered with a relative accuracy of $1\%$ or better, especially with the \vdg model. It would be interesting to test the validity of the \vdg formalism to scales smaller than $\smin=20\Mpc$ in future works.
It was not feasible in this analysis mainly due to concerns about the validity of the covariance matrix because of the used high-$k$ damping in the computation of the relevant integral but also because the best-fit \tpcf during the iterative process exhibited sensitivity to scales below and around $20\Mpc$.
In addition, the impact of the Fourier-space damping for the \vdg and \eft models prevented pushing to smaller scales.

When considering constraints from the full-shape analysis, the \clpt and \eft models fail to deliver unbiased measurements of the parameter combination $\{h,\,\As,\,\omegac\}$ on scales of $20\Mpc$. In contrast, the range of validity of the \cleft model reaches this separation while the \fob of the \vdg model is within the 95\% region, which is intriguing, as the two models are based on a completely different theoretical approach. Similarly to the template-fitting method, all four models perform equally well considering higher redshifts or larger minimum separations, and, for the same advocated reasons concerning the covariance matrix, we refrained from testing the performance of the models down to scales smaller than $20\Mpc$. We found that $\As$ could not be recovered as accurately as $h$ and $\omegac$, for which instead we report an accuracy of better than $1\%$ for most of the tested configurations.
Overall, we found that \vdg and \cleft are the best-performing models in terms of recovery of cosmological parameters, although the \cleft model appears to be less prone to projection effect when going to higher $\smin$ while also reaching larger \fom at low $\smin$. 
When focusing on high-redshift snapshots -- where nonlinear evolution is less important -- a simpler model such as \clpt can yield higher \fom values by a factor of $1.2$--$2$ while still recovering unbiased cosmological parameters.
\citet{Ramirez2024arXiv_DESI_ModellingChallenge_ConfigSpace} also reported a good performance of the \cleft model in terms of recovering the fiducial parameters under different configurations, additionally comparing the latter against the \lpt-based model described in \citet{ChenVlahCastorinaWhite2020, Chen2020JCAP} in a full-shape analysis, and finding good agreement.
This leaves open space in the future to explore how this model performs on the same galaxy population explored in this work, and how it ultimately compares to the \vdg model.

We supplemented this analysis with an extended appendix where we provide, among supporting material, additional tests concerning extended cosmologies and bias relations.
We tested the validity of the \LL approximations to express the non-local bias parameters in terms of the linear bias $b_1$. We found no compelling evidence for significant deterioration of the goodness of fit and the \fob when the parameter space is reduced, although we do observe a noticeable gain in the \fom of a factor of 1.4 and more when all non-local bias parameters are fixed in the \LL approximation.
Leaving the non-local bias parameters free yields a more theoretically sound expansion up to one-loop. Indeed  hints for deviations from the local approximations were found in \citet{Lazeyras_2016} and \citet{Abidi_HaloBias}. Furthermore, there are compelling reasons why the non-local bias parameter $\bGthree$ has to be kept free when the derivative bias is not fitted for \citep{Sanchez2017}.
However, derivative bias contributions are included in the counterterms if present in the model.
It is nevertheless encouraging to find the same performance of the model, even improvement in terms of \fom, when the parameter space of the bias is simplified with physical motivation.
This result is particularly important in light of projection effects and advocates considerations of model simplification when applied to real data even though purely theoretical reasons tell otherwise.
Of course, our bias tests should be repeated with different \hod catalogues and a \Euclid \DRone covariance matrix where projection effects might be stronger in order to validate realistic gains.

We also extended the analysis to explore cosmologies beyond \lcdm in the form of the \wcdm model, with the equation of state parameter $w_0$ free to vary.
We refrained from using the full CPL parametrisation as such dynamical dark energy requires the simultaneous fitting of several redshift bins.
We refer the interested reader to Euclid Collaboration: Moretti et al. (in prep.) where a very thorough analysis of projection effects in extended cosmologies and possible mitigation strategies is presented.
In the \wcdm model, we find that the combination $\{h, \As, \omegac, w_0\}$ is more biased, that is, exhibits a larger \fob, with respect to the one already tested for the \lcdm analysis, which only includes $\{h, \As, \omegac\}$.
The response of the different models to the additional parameter was very diverse over the $\smin$ and redshifts. The \cleft and \clpt models reacted the least, except on $\smin=20\Mpc$ where a significant increase in \fob was detected.
This bias might be due to new degeneracies with $w_0$ that worsen the constraining power and can possibly bias the results.
A clear boost of projection effects for the \vdg and \eft models was observed via an increase of the \fob at higher redshifts and larger $\smin$.

This work evidenced the importance of \eft counterterms in the modelling of the nonlinear clustering of matter in order to go beyond the assumption of a pressure-less fluid and also include higher-derivative and velocity bias as well as small-scale \rsd effects.
Care has to be taken, however, concerning the standard \eft model transformed into configuration space as it does not appear to be able to reach the smallest nonlinear scales of $20\Mpc$.
Instead, improved modelling of the FoG damping has to be included, as is done in the \vdg model that clearly outperforms the \eft model.
This result is in line with the Fourier space analyses undertaken in Euclid Collaboration: Camacho et al. (in prep.), where the \eft model breaks earlier than, for example, the \vdg model.
Nevertheless, in full-shape analysis the approach of using GS to map from real to redshift space achieves a similar performance to the \vdg model, making these promising models to be applied to real \Euclid data.
These prescriptions form a set of baseline models that performed best in the ideal conditions of the mocks used in this work.
The final choice that will be used to analyse \Euclid data, however, will be influenced by a model's ability to additionally include observational systematic effects.
Therefore, this work will be extended in the future to include redshift interlopers \citep{EP-Risso} or systematic effects of lightcones as arising from lensing that have to be incorporated in the modelling of clustering statistics \citep{jelic-cizmek2021importance, breton2022impact, EP-JelicCizmek}.

\begin{acknowledgements}
This work received support from the French government under the France 2030 investment plan, as part of the Excellence Initiative of Aix Marseille University – amidex (AMX-19-IET-008 – IPhU).
This research has made use of computing facilities operated by Centre de donn\'ees Astrophysiques de Marseille (CeSAM) at Laboratoire Astrophysique de Marseille (LAM), France.
MAB acknowledges support from project “Advanced Technologies for the exploration of the Universe”, part of Complementary Plan ASTROHEP, funded by the European Union - Next Generation (MCIU/PRTR-C17.I1).
This research made use of \texttt{matplotlib}, a \texttt{Python} library for publication quality graphics \citep{Hunter:2007_matplotlib}.
\AckEC  
\end{acknowledgements}

\bibliography{my, Euclid}

\begin{appendix}

\section{Axis selection and sample variance}\label{sec:appendix_axis_selection_sample_variance}

In this work, we performed a model comparison between various \rsd models using comoving snapshots of the \flagship simulation as a testing ground. In this appendix, we assess the relative performance of the models considering different spatial \los directions, which are the $x$- and $y$-axis, as compared to the $z$-axis used in the main analysis. One possible approach could have been to average the measurements over the three main axes of the simulation box, leading to a suppressed variance. However, this would complicate both the computation of the covariance matrix and the interpretation of the averaged $\chired$ statistics, especially in configuration space \citep{smith2021reducing}.
This is the reason why we only considered measurements along a single \los in the main text. 

\begin{figure}
    \centering
    \includegraphics[width=\columnwidth]{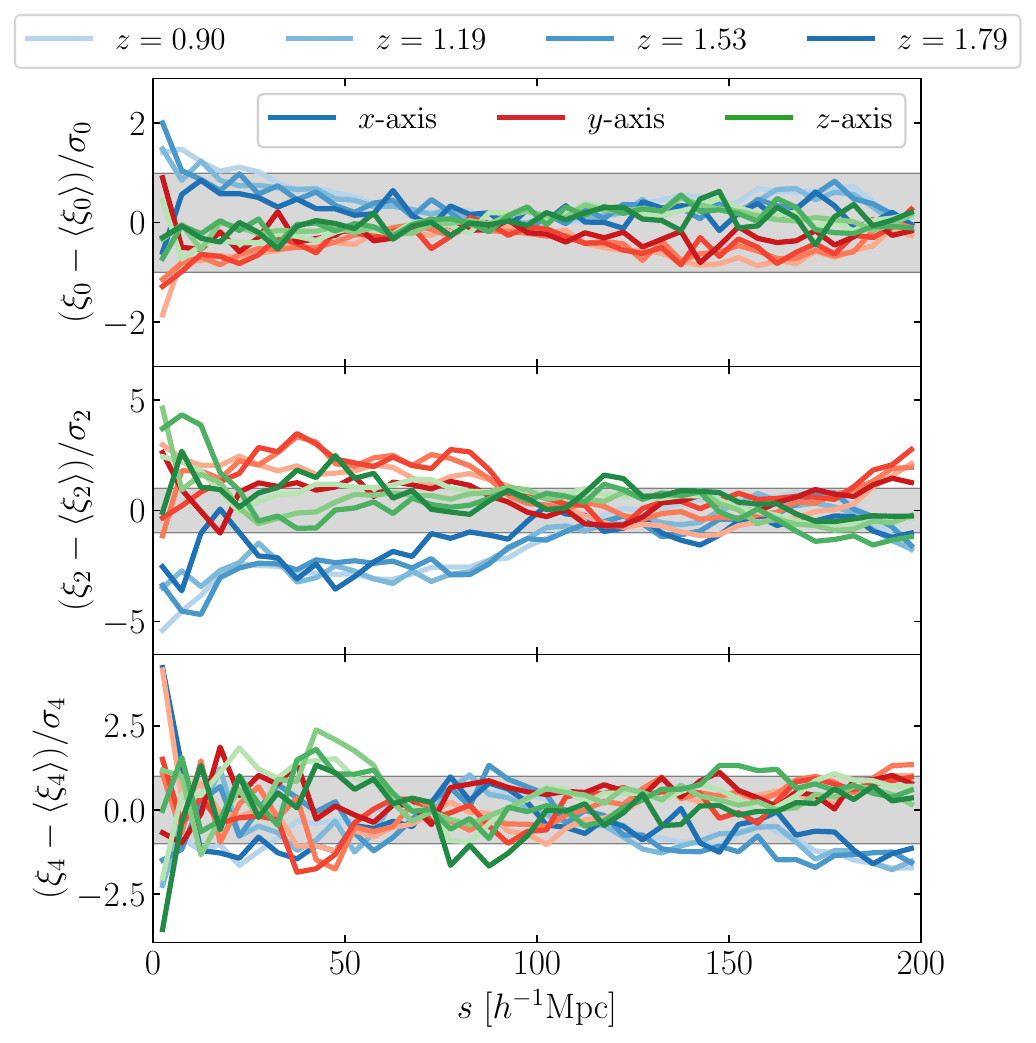}
    \caption{Difference between the \tpcf multipoles measured assuming different \los directions, as indicated in the legend, and the ones obtained by averaging the different \los, normalised by the square root of the diagonal entries of the covariance matrix, for which we use the average of the three \los.
    The colour tone from light to dark indicates the redshift in increasing order. 
    From top to bottom, the panels display the monopole, quadrupole, and hexadecapole.
    Differences at the $1\sigma$ level are marked by the grey shaded area.}
    \label{fig:appendix_LOS_chi2}
\end{figure}

In Fig.~\ref{fig:appendix_LOS_chi2}, we show residuals between measurements of the \tpcf multipoles along different \los and their mean, expressed in units of their dispersion, which we computed using the average of the three \los.
By inspection, it emerges that choosing the $x$-axis as \los, or to some extent the $y$-axis, leads to significant differences in the quadrupole on scales up to $100\Mpc$.
A similar trend can be observed in the monopole on small scales for the $x$-axis, whereas the hexadecapole seems to be very similar between the different \los measurements, although the hexadecapole exhibits the largest uncertainty.
Based on these observations, we chose the $z$-axis as our primary \los direction for the analysis, as it most closely matches the averaged measurement and is thus expected to be the least biased.
To assess this selection more quantitatively, we computed the $\chi^2$ obtained by taking the difference between the individual \los and the averaged one on scales above $20\Mpc$, corresponding to the minimum separation explored in this work. By repeating this calculation for the four different redshifts and summing up the corresponding $\chi^2$ values, the $z$-axis exhibits the smallest overall $\chi^2$ compared to the other two axes.

\begin{figure*}
    \centering
    \includegraphics[width=0.9\linewidth]{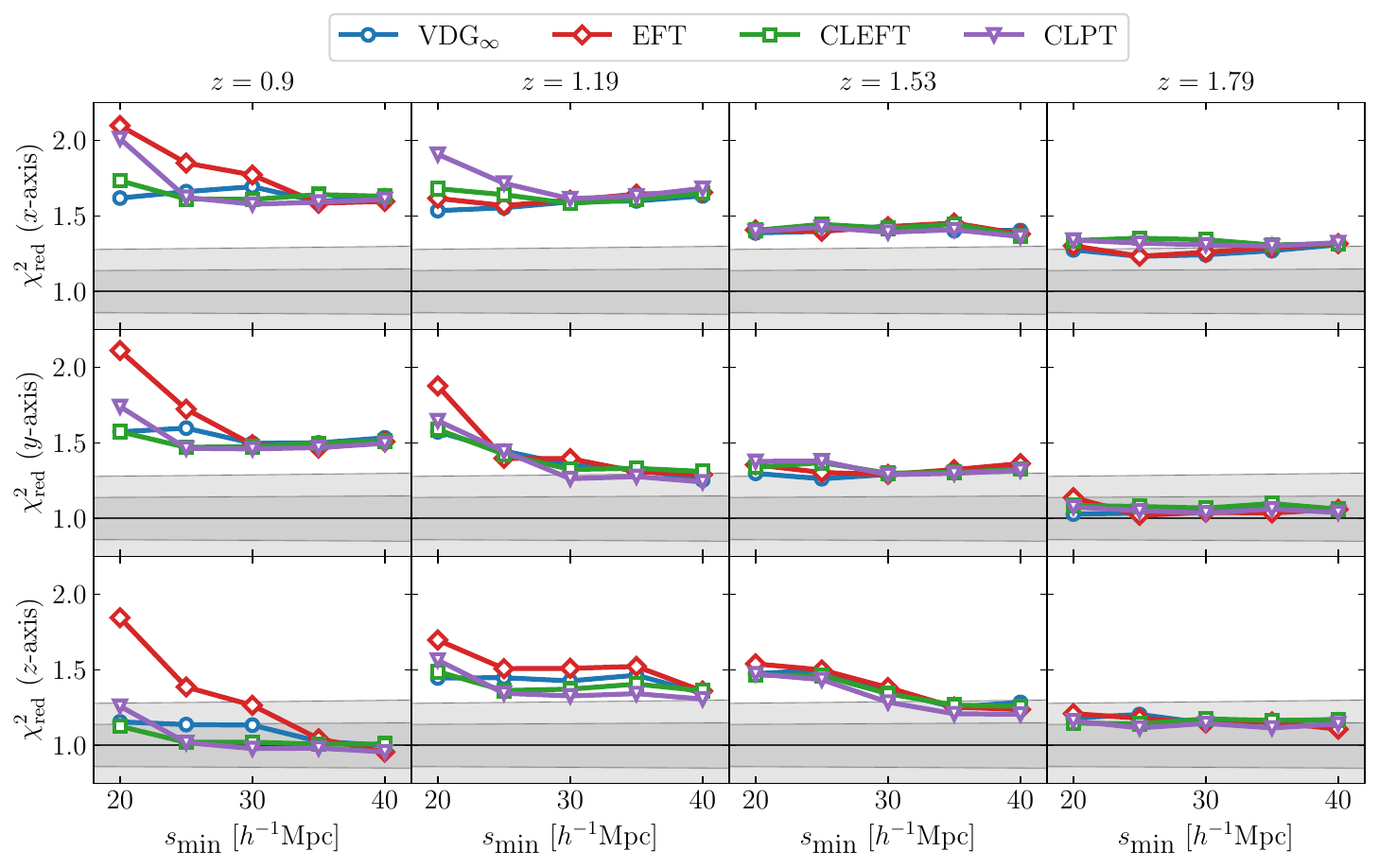}
    \caption{Mean $\chired$ values obtained from the full-shape analysis assuming a \los directed along the $x$-, $y$-, and $z$-axis. The symmetric shaded regions mark the standard deviation and twice the standard deviation of the $\chired$ distribution using six (\clpt) free parameters as a conservative choice.}
    \label{fig:chi_squ_red_x_y_z_axis}
\end{figure*}

A drawback of using this approach, even when considering a snapshot covering an outstanding volume such as that of \flagship, is that the analysed data vectors are partially influenced by sample variance, especially on the largest scales considered.
For this reason, we additionally ran the full-shape analysis presented in Sect. \ref{sec:results_fullshape} also on the $x$- and $y$-axis to check potential differences in the recovered $\chired$ value.
In Fig.~\ref{fig:chi_squ_red_x_y_z_axis} we show the outcome of this test when employing either of the three Cartesian axes as the \los in the three different rows, respectively.
The largest discrepancy can be seen at $z=0.9$, where the $x$- and $y$-axis show deviations from the mean beyond two times the dispersion of the corresponding $\chired$ distribution, which is not present for the $z$-axis (except partially for the \eft model).
For these \los the $\chired$ values progressively become smaller moving to higher redshift, as expected due to less relevant nonlinear effects and hence a better fit to the data.
In addition, although rather difficult to see in Fig.~\ref{fig:appendix_LOS_chi2}, at $z=1.79$ the multipoles of the different \los are better in agreement with each other.
This hints at the low $\chired$ values that we obtained when fitting the $z$-axis at $z=0.9$ are indeed partially due to sample variance, but at the same time also the higher $\chired$ values at the two intermediate redshifts seem to be affected.
It is evident when looking at the $x$-axis that $z=1.19$ and $z=1.53$ have almost no dependence on $\smin$, sometimes even slightly decreasing towards $20\Mpc$, which is very counter-intuitive.
All of this assumes the snapshots observed from different \los to be independent, which is not strictly true.
Interestingly, the feature of the \eft model to exhibit significantly higher $\chired$ when fitting down to $\smin=20\Mpc$ at $z=0.9$ does persist for the other choices of \los, indicating a true failure of the model.
Therefore, we advocate a comparison of general trends and the $\chired$ values among different models for a given redshift rather than precise values between different snapshots.

\section{Internal consistency among the TNS class of models}\label{sec:app_TNS_testing}
In this appendix we provide a consistency test between the results obtained with the different models of the TNS class. As described in Sect. \ref{sec:theory_TNS}, all these models differ only in the nature of their building blocks -- perturbative or from simulations -- but for the \vdg model, which additionally includes \eft counterterms and has a different modelling of the velocity difference generator (the damping function). It has to be noted that $\sigeight$ is kept fixed in the classic TNS models as the code we used did not allow us to vary it freely.
To facilitate a fair comparison among the TNS models we keep $\sigtwelve$ fixed in the \vdg model.

\begin{figure*}
    \centering
    \includegraphics[width=0.9\textwidth]{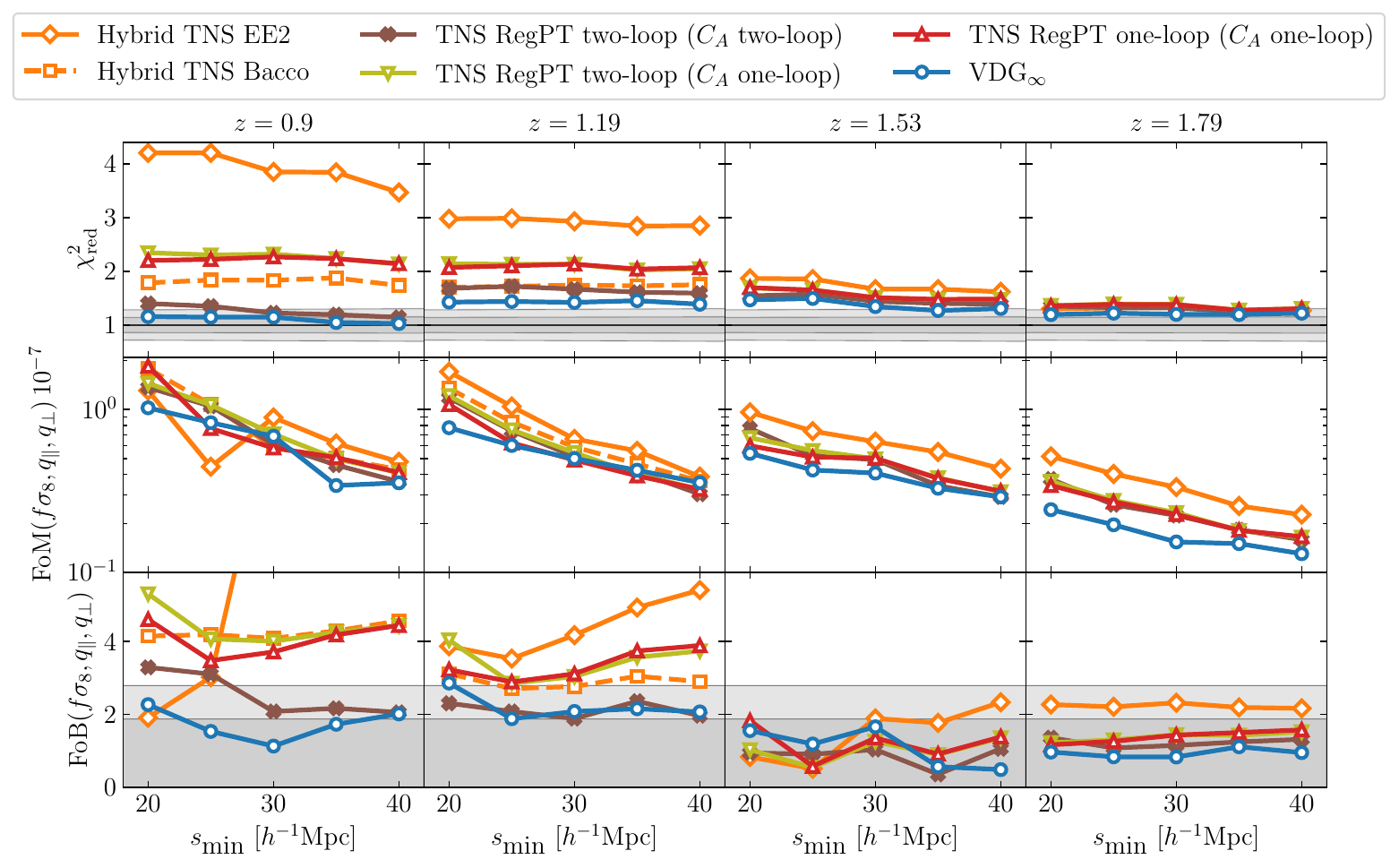}
    \caption[Performance metrics for TNS class of models]{Performance metrics for the TNS class of models as a function of the minimum fitting scale $\smin$. The symmetric shaded regions in the panels displaying the $\chired$ values refer to the standard deviation and twice the standard deviation of the $\chired$ distribution with the model degrees of freedom fixed to eight (all the classic TNS models) as a conservative choice. For the \fob, the two shaded regions denote 68-th and 95-th percentiles  as described in Sect.~\ref{sec:performance_metrics}. To be precise for the \vdg model we actually show the \fom and \fob for the parameters $\{f\sigtwelve, \qpar, \qperp\}$.}
    \label{fig:TNS_family_metrics}
\end{figure*}

The performance metrics of these models are presented in Fig.~\ref{fig:TNS_family_metrics}. We find that, at high redshift, the models perform similarly across all performance metrics, with no significant differences. This is expected since, at these redshifts, nonlinear features have not had enough time to grow significantly; hence more sophisticated modelling is unnecessary and does not improve the fits and the recovery of parameters. However, significant differences emerge in the two lowest redshift bins. The hybrid TNS model using \euclidemulatortwo exhibits a very high $\chired$ and a biased parameter recovery as seen from the \fob. We found that this is mostly due to the \euclidemulatortwo producing relatively noisy estimates of the nonlinear matter power spectrum. When transformed into configuration space, this introduces significant numerical artefacts that increase the $\chired$ value.
A similar result was found by \cite{Chen2025SCPMA_CSST_emu}, where in their Fig.~7 they have residuals in the correlation function.
If instead we use the \texttt{baccoemu} \citep{Angulo2021MNRAS_BACCO} power spectrum emulator, which only spans the two first redshift bins, we can significantly reduce the $\chired$ values and approach unity. This shows the importance of having noise-free predictions of the power spectrum to ensure the stability of the Fourier transform and obtain accurate \tpcf predictions.

An improvement with respect to the hybrid TNS model in terms of $\chired$ is achieved when the nonlinear power spectra are predicted in a fully analytic way using RegPT (at either one- or two-loop).
Interestingly, the two cases are almost indistinguishable in their \fob and \fom suggesting that a slightly higher reach in $k$ for the nonlinear power spectra does not seem to yield significant improvements. On the contrary, if the correction term $C_A$ is computed up to two-loop along with nonlinear power spectra at two-loop (brown lines) a significant improvement in the $\chired$ values and \fob can be observed while the \fom stays about the same.
This model achieves a similar $\chired$ than the \vdg model (blue line), although the latter shows a worse \fom at the lower minimum fitting scale.
It has to be noted that this comparison is somewhat unfair as we compare predictions at two-loop (TNS) to a one-loop model (\vdg).
The decrease in \fom is somehow expected as the \vdg model has three more degrees of freedom (the counterterms). However, the \vdg model also has much better \fob at $z=0.9$.
This makes the \vdg model overall the best model in this comparison, giving a consistently unbiased recovery of the parameters down to $z=0.9$ while only being just above the 68\% region for $z=1.19$ and exhibiting a slight lack of constraining power towards lower $\smin$. The higher potential of the \vdg model with respect to the other tested TNS models might come from the addition of the short-scale \eft physics and a more consistent treatment of bias terms.
In addition, the more theoretically sound form of the damping function might add to the improved performance on small scales. Overall, even if the foundation in the TNS class of model is the same, the details in the predictions of the different ingredients and terms have a non-negligible impact on the performance of the models.

\section{Bias relations in full-shape fitting}\label{sec:appendix_bias_testing}
In the baseline analysis carried out in the main text of this work, non-local bias parameters were left free to vary in the fit.
However, as described in Sect.~\ref{sec:galaxy_bias}, by assuming a purely local bias expansion in Lagrangian space, non-local bias parameters at later times can be expressed in terms of the linear bias $b_1$.
This can be used to reduce the size of the parameter space, thus leading to tighter constraints on the remaining parameters.
This is particularly relevant when the data are not able to constrain the whole set of nuisance parameters, as it may potentially occur in the \DRone having the smallest volume compared to later releases.

\begin{figure*}
    \centering
    \includegraphics[width=0.9\textwidth]{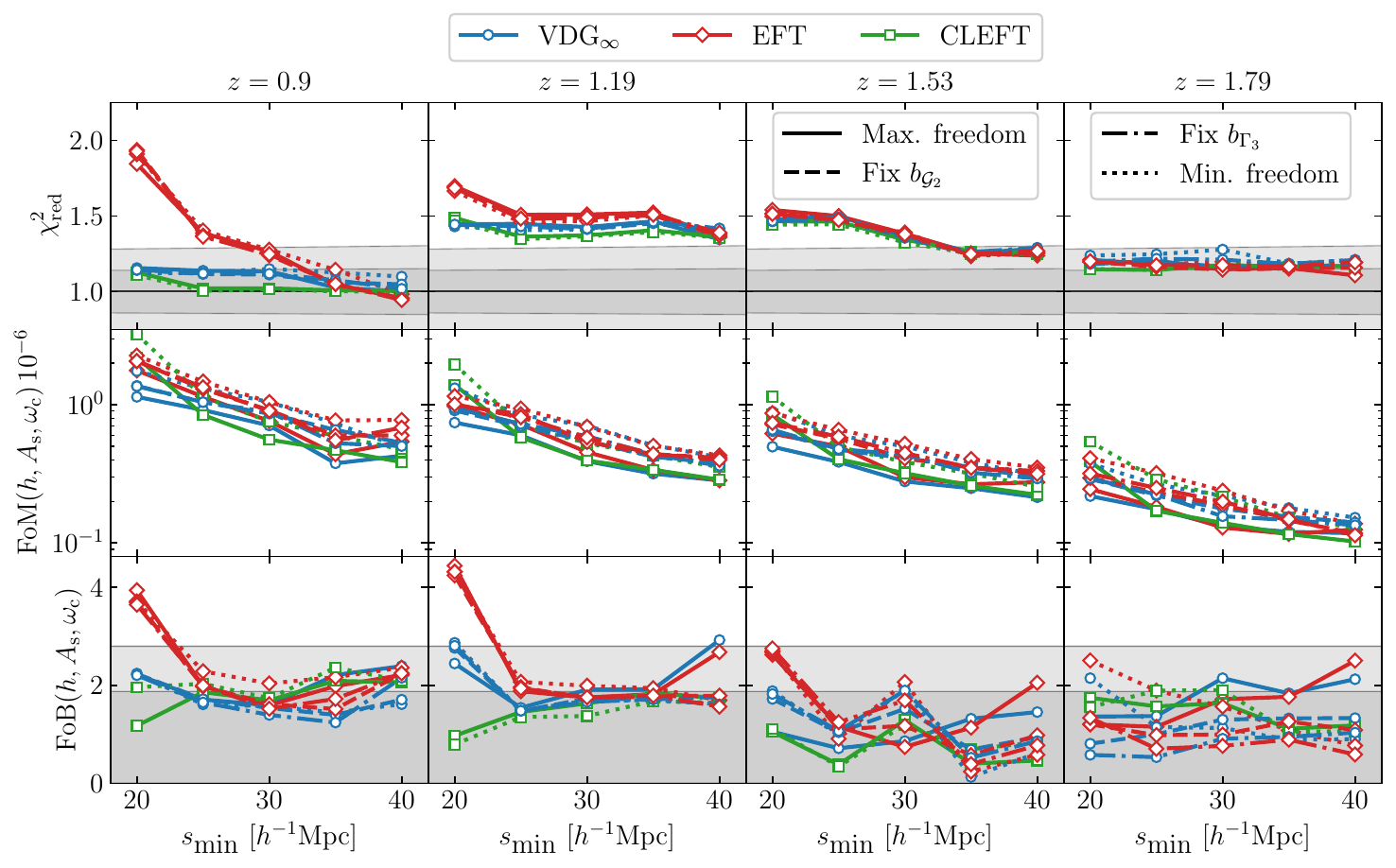}
    \caption[Summary of bias relation analysis]{Comparison between the performance metrics of the \vdg, \eft, and \cleft models with different degrees of freedom in terms of the non-local bias parameters -- $\bGtwo$ and $\bGthree$ for the \vdg and \eft models, or $b_{s^2}$ for the \cleft model. Specifically, in addition to the configuration already explored in the main text, we test configurations where either of the non-local biases are fixed to the \LL relation, as indicated in the legend. We denote with `maximum' and `minimum' freedom the configurations where none or all non-local bias parameters are expressed in terms of the \LL approximation.
    The shaded regions in the \fob panels represent the 68\% and 95\% credible regions as described in Sect.~\ref{sec:performance_metrics}. The symmetric shaded areas in the $\chired$ panels represent the standard deviation and twice the standard deviation of a $\chired$ distribution with eight degrees of freedom -- as in the \cleft model with the minimum freedom setup -- as a conservative approach.}
    \label{fig:full_shape_bias_comparison}
\end{figure*}

In Fig.~\ref{fig:full_shape_bias_comparison}, we show the performance metrics for the \vdg, \eft, and \cleft models -- the \clpt model is not considered here since it does not include any non-local bias contribution -- for different configurations where either of the non-local bias parameters from the set $\{\bGtwo,\bGthree,b_{s^2}\}$ are expressed in terms of $b_1$ using Eqs.~\eqref{eq:LL_bG2_bGam3} or \eqref{eq:LL_bs2}.
As a general observation, these parameters appear to have essentially no impact on the $\chired$ values, regardless of the configuration that is selected. This is expected since a fraction of the nuisance parameters of these models is degenerate with each other -- most notably, the two non-local bias parameters and $b_2$ -- thus resulting in a similar goodness of fit also when the \LL approximation is used.
The \fom shows the expected behaviour of an improved precision on the cosmological parameters when the degrees of freedom are reduced, sometimes up to a factor of 1.4 and more when all bias parameters are expressed in terms of $b_1$.
Regardless of the configuration, the constraining power increases with smaller $\smin$ and lower redshifts, which is the same behaviour as observed in the template and full-shape fitting approaches in Figs.~\ref{fig:fixed_cosmo_metrics} and\,\ref{fig:full_shape_metrics}.
For the \fob, the situation is more diverse. For some configurations, fixing non-local bias parameters can lead to a less biased recovery of the parameters -- for instance, for the \eft model at $z=1.79$ --  a clear indication of a mitigation of projection effects.
The opposite is also found, for instance, for the \vdg model at $z=1.53$.

Overall, the changes are only significant for specific configurations in terms of redshift and $\smin$.
In general, all the models perform well, modulo the flaws discussed in Sect.~\ref{sec:lcdm_cosmology}, regardless of the choice to fix or leave free the non-local bias parameters.
Therefore, we can conclude that while leaving all bias parameters free represents a more theoretically sound prescription, using the \LL approximations does not worsen the overall model performance with the advantage of more constraining power and possibly less projection effects.

\section{Full contours for \lcdm model}\label{sec:appendix_full_contour}
\begin{figure}
    \centering
    \includegraphics[width=\columnwidth]{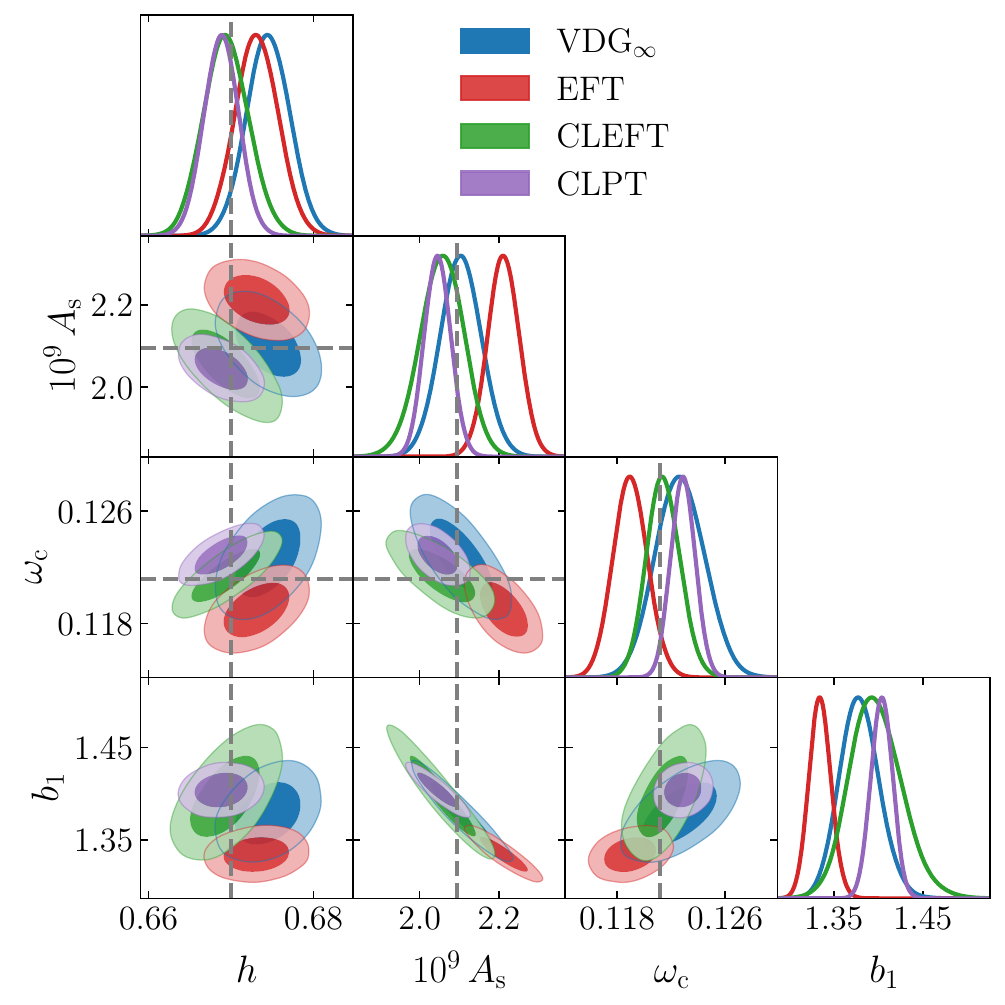}
    \caption{Marginalised posterior distribution of the parameters $h$, $\As$, $\omegac$, and $b_1$ at $z=0.9$ assuming a \lcdm cosmology and a scale cut at $\smin=20\Mpc$. Different \rsd models are indicated with different colours, as shown in the legend.
    Dashed grey lines mark the fiducial values of the parameters.
    The 2D contours show the 68\% and 95\% credible intervals. We note that the Lagrangian linear bias is converted into Eulerian bias to facilitate a comparison with the \vdg and \eft model.}
    \label{fig:contour_LCDM}
\end{figure}

To illustrate the degeneracy among cosmological parameters and the linear bias $b_1$, in Fig.~\ref{fig:contour_LCDM} we present the marginalised posteriors of $\{h, \As, \omegac, b_1\}$ for all the main \rsd models considered.
The minimum fitting scale is set to $\smin = 20\Mpc$ and $z=0.9$. 
As already shown in Figs. \ref{fig:full_shape_metrics} and particularly \ref{fig:full_shape_params}, the \cleft model recovers all cosmological parameters fairly well while the \vdg model recovers too large a value for $h$.
In contrast, the \eft model shows significant bias both in $\As$ and $\omegac$, and similarly for the \clpt model, although less strongly.
The strong degeneracy of $\As$ and the linear bias $b_1$, both affecting the overall amplitude of the \tpcf, 
can lead to projection effects and hampers an unbiased and accurate recovery of $\As$.

\section{Constraints on \wcdm cosmology}\label{sec:appendix_extended_cosmo}
In this section we revisit the analysis presented in Sect. \ref{sec:lcdm_cosmology} extending the \lcdm model to include a varying dark energy equation of state. The most straightforward approach is to vary the parameter $w_0$ while keeping its time-evolution fixed, in the so-called \wcdm model. It is instructive to study the possible loss in constraining power induced by opening the parameter space, and investigate possible artificial gains in the $\chired$ value as \flagship has a fixed $w_0=-1$.

\begin{figure*}
    \centering
    \includegraphics[width=0.9\textwidth]{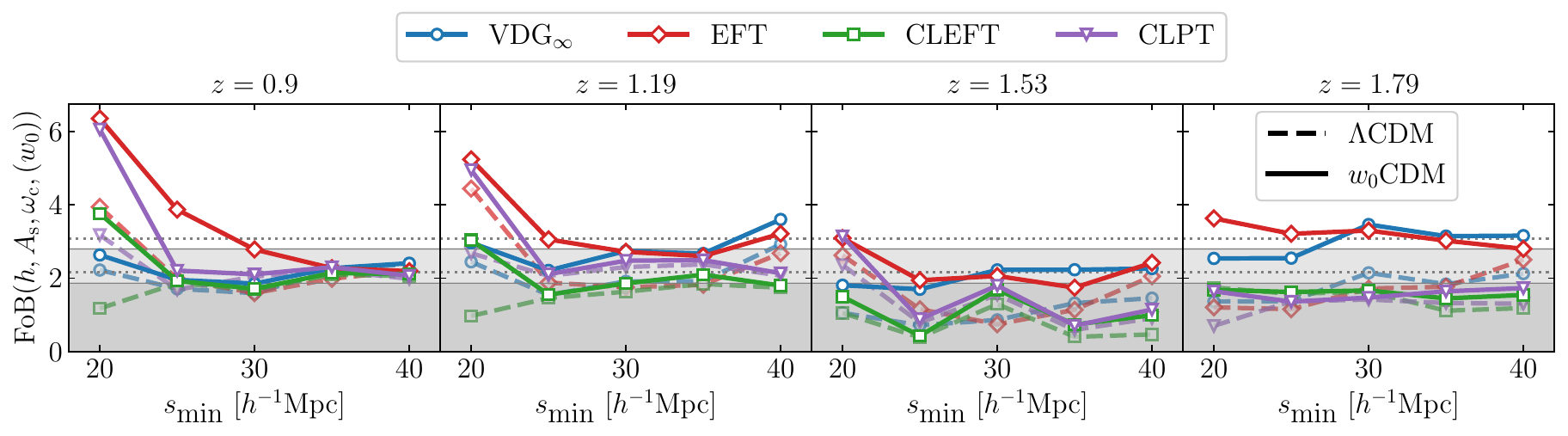}
    \caption[Constraints for the \wcdm cosmology]{\fob of the extended parameter space of the \wcdm model (solid lines) compared to the one already presented in Fig. \ref{fig:full_shape_metrics} for the \lcdm model (dashed lines, less opaque).  In the former case the \fob is calculated from the combination of the cosmological parameters sampled in the \lcdm case that is $\{h,\,\As,\,\omegac\}$ plus the dark energy parameter $w_0$. The 68\% and 95\% credible intervals for the \fob distribution for three parameters are denoted with solid grey regions while for four parameters they are shown with dotted grey lines. The latter are given by 2.17 and 3.08 for 68\% and 95\%, respectively.}
    \label{fig:full_shape_FoB_w0_comparison}
\end{figure*}

In Fig.~\ref{fig:full_shape_FoB_w0_comparison} we present the \fob obtained when considering either a \lcdm or a \wcdm cosmology, where in the second case the \fob is computed also considering the additional free parameter $w_0$.
In general, it appears that the recovery of the four cosmological parameters $\{h, \As, \omegac, w_0\}$ in the \wcdm cosmology is more biased with respect to the set of parameters of the \lcdm cosmology.
The \cleft and \clpt models are only marginally affected by the specific value of $\smin$ above $20\Mpc$; for this specific scale cut the recovered parameters are strongly biased, exceeding the $95\%$ credible region at $z=0.9$ and $z=1.19$.
The \vdg model exhibits a homogeneous response over almost all $\smin$ to the extension of the parameter space, which is the strongest at the highest redshift.
In fact, all values of the \fob for redshifts $z=1.19$ and $z=1.79$ are outside the $68\%$ credible region, hereby signalling biased parameters.
The increase of \fob towards higher $\smin$ and redshift suggests an enhancement of projection effects with respect to the \lcdm case because we expect the \wcdm model to perform equally as good or better.
Leaving free $w_0$ opens up new degeneracies with existing parameters that will possibly boost projection effects and in turn diminish the \fob.
At the same time, we also observe biased parameters at $z=0.9$ and $\smin$ below $30\Mpc$ for the \eft and \clpt model signalling an even earlier breaking of the model when $w_0$ is left free.
Moreover, at $z=1.79$, the \eft model has a \fob outside the $68\%$ credible region for all $\smin$, similarly to the \vdg model. 
Overall, this shows that opening up the parameter space with a possibly strongly degenerate parameter can lead to a severely biased recovery of the cosmological parameters.
Although we do not show it here, we investigated the impact on the $\chired$ value and found that it improves mainly for the two lowest redshifts and at the smallest considered $\smin$.
This underlines the presence of projection effects at larger $\smin$ where the $\chired$ value stays virtually the same but we recover more biased parameters in many configurations.

\begin{figure}
    \centering
    \includegraphics[width=\columnwidth]{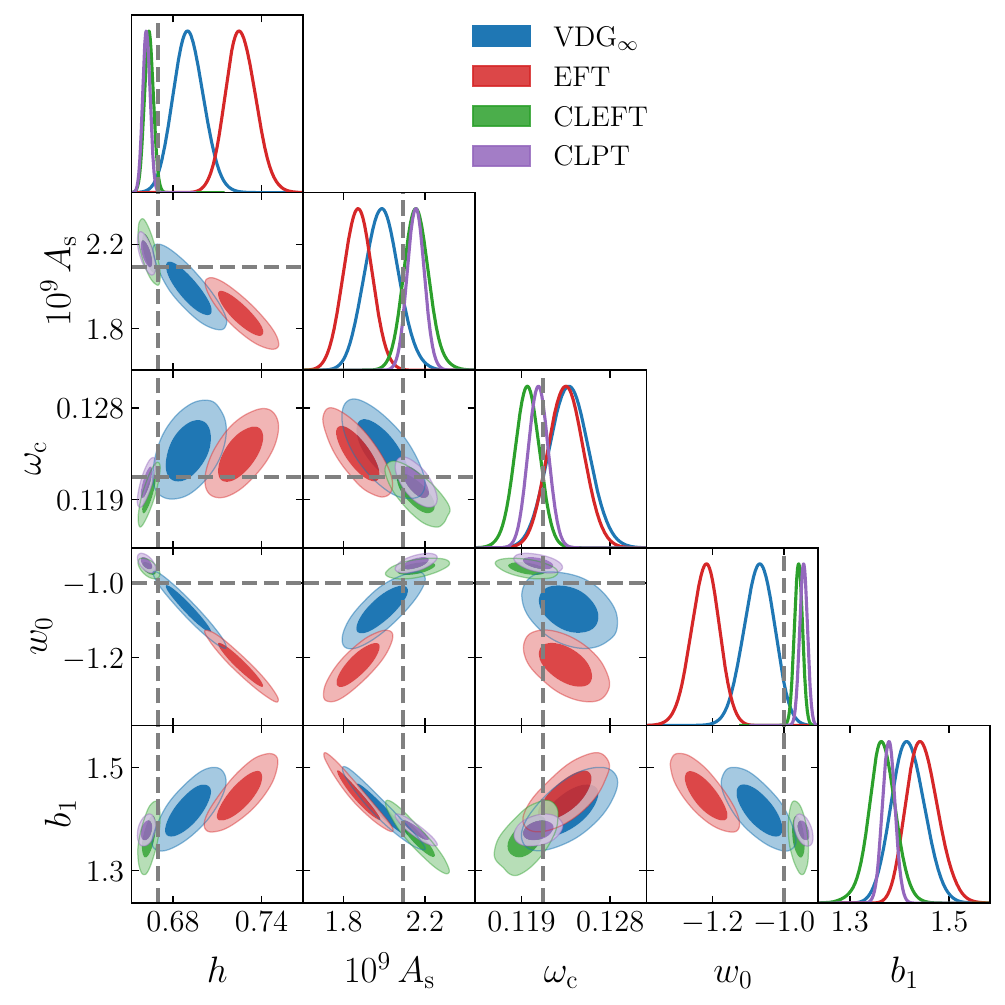}
    \caption{Marginalised posterior distribution of the cosmological parameters of the \wcdm model, referring to the snapshot at $z=0.9$ and the configuration with $\smin=20\Mpc$. Different \rsd models are represented by different colours, as indicated in the legend, while grey dashed lines mark the fiducial values of the parameters. The 2D contours show the 68\% and 95\% credible intervals.}
    \label{fig:contour_wCDM}
\end{figure}

The effect of opening up the parameter space can also be nicely appreciated when looking directly at the contours as shown in Fig.~\ref{fig:contour_wCDM}. They exhibit quite different behaviours from what is seen in Fig.~\ref{fig:contour_LCDM}.
First of all, we observe a very strong degeneracy in the $w_0$--$h$ plane for the \eft and \vdg model that hampers constraints on the individual parameters.
While an unbiased recovery of the three parameters $\{h, \As, \omegac\}$ can be achieved for some of the considered \rsd model in the \lcdm case, the additional degree of freedom of the \wcdm model leads to substantial shifts in terms of $h$.
This is particularly clear for the \eft model, which deviates clearly by more than 1$\sigma$ on all parameters, except $\omegac$, although having a rather large uncertainty.
The \vdg model deviates less from the fiducial parameters while exhibiting similar uncertainties except for $\omegac$, where the 1-dimensional posterior distribution is overlapping with the \eft model.
We checked the same figure but for $\smin=40\Mpc$ and as the \fob in Fig.~\ref{fig:full_shape_FoB_w0_comparison} suggests the 1-dimensional contours for the \vdg and \eft models are more consistent with the fiducial value of $h$, $w_0$, and $\As$.
Rather than signalling projection effects, which should be stronger at larger $\smin$, there seems to be a compensating effect pushing both $w_0$ and $h$ away from their fiducial values when going to smaller $\smin$.
In contrast, the \clpt and \cleft models have much smaller uncertainties in Fig.~\ref{fig:contour_wCDM} but similar absolute deviations from fiducial parameters compared to the \vdg model.

Overall, we find that leaving $w_0$ free to vary the fit introduces significant shifts in the marginalised posterior distribution of the cosmological parameters. Additionally, for some of the considered \rsd models -- \eft and \clpt -- the premature breaking of the model with more aggressive scale cuts is enhanced with respect to the \lcdm case.

\section{Details about the emulator}\label{sec:appendix_emulator}
In this section, we describe the emulator for the GS models that we built upon the infrastructure of \comet.
It is based on a \GP, which assumes a prior multivariate Gaussian distribution from which functions can be sampled.
This prior distribution takes a mean and a covariance matrix (also known as kernel function) as input, where the latter determines the shape of the functions that this prior describes and has to be chosen in the beginning when setting up the \GP.
More precisely, a kernel measures the covariance between data points in terms of their Euclidean distance.
Once the Gaussian prior is conditioned on the observed data, this results in a Gaussian posterior whose mean can be used to predict observations at new points.
For more details on a \GP, we refer to the seminal book by \citet{Rasmussen_GP_book} as well as to \citet{EggemeierEtal_2023}, which describes the setup of the \GP used in \comet.
In the case of \comet, the observed data are the contributions to the power spectrum multipoles at sets of cosmological parameters.
What makes \comet powerful is its implementation of the evolution mapping approach \citep{Sanchez2020, Sanchez2022} that reduces the dimensionality of the cosmological parameter space, yielding accurate predictions while keeping the training set reasonably small.
This is achieved by splitting up the cosmological parameters into those that affect solely the shape of the linear matter power spectrum and those that affect its amplitude.
The latter are degenerate with the redshift and therefore are called evolution parameters.
Due to this degeneracy, it is in fact not necessary to train the emulator explicitly on all of them; instead, only a single parameter is enough.
In the case of \comet this parameter is $\sigtwelve$, which is very similar in amplitude to $\sigeight$ but does not depend on $h$. 
Hence, the only parameters \comet is trained on are given by $\{\omega_{\tm{b}}, \omega_{\tm{c}}, \ns, \sigtwelve, f\}$.
The inclusion of $\sigtwelve$ does also include the dependence on the redshift, as they are degenerate at the linear level.

In Table~\ref{tab:emulated_quantities}, we list the bias combinations that need to be emulated for the \clpt and \cleft model, respectively.
For completeness, we consider also an expansion only up to linear order that we denote as Zeldovich approximation or `ZA' model.
Therefore, only the linear bias $b_1^{\mathrm{L}}$ is needed in this model (compare Table~\ref{tab:emulated_quantities}).
\citet{white2014zeldovich} showed that using the Zeldovich approximation for $\bm{\Psi}$ in conjunction with the GS model provided a better fit to the quadrupole of the \tpcf than using only the Zeldovich approximation for the real-space clustering \emph{and} mapping from real to redshift space.
However, he expanded beyond linear order, that is, including terms such as $\xi_{\mathrm{L}}^2$ -- the linear \tpcf squared -- that we neglect here; hence his model deviates significantly in the theoretical descriptions from our simpler implementation.
We use the moment version of the velocity dispersion in the Gaussian \pdfabbrev for the ZA model -- as also done in the \clpt model described in Sect.~\ref{sec:gaussian_streaming_theory}.

The structure of the emulator for \lpt models is somewhat simpler compared to \comet because we do not need to emulate ratios to the linear power spectrum.
Hence, it requires only the separate emulator for $\sigtwelve$ in order to obtain the correct $\sigtwelve$ for the native parameter space in case of a full-shape analysis.
Furthermore, since the growth factor $f$ in the \lpt models factors out of the velocity correlators, we do not need to emulate over it, simplifying the native parameter space to $\{\omegab, \omegac, \ns, \sigtwelve\}$.
For the training of the emulator, we generated \num{2000} training samples in a Latin hypercube spanning the same limits in terms of $\omega_{\tm{c}}$, $\omega_{\tm{b}}$, $\ns$, and $\sigtwelve$ as in the \comet code.
\citet{EggemeierEtal_2023} showed that the chosen limit on $\sigtwelve$ covers redshifts up to around $z=3$.
For the dedicated emulator of $\sigtwelve$ in terms of the shape parameters, 750 training samples were used, the same number as taken in \comet.
The performance is assessed by producing a validation set of 1500 cosmologies randomly selected in the range of the emulator.
For each cosmology of the validation test, we first compute the first three even multipoles ($\ell = 0, 2, 4$) of the \tpcf using the exact \lpt ingredients from \texttt{CLEFT\_GS}.
The bias parameters are fixed to results from a template fitting such that reasonable values are used.
For the ZA model, we simply use the value of $b_1$ and $\sigv$ that we obtained with the \clpt model.
The growth rate $f$ can be computed from cosmological parameters and is necessary for the GS model.
The AP-parameters $\qpar$ and $\qperp$ are set to unity.
The derivative bias $b_{\nabla^2}$ as well as counterterms $\alpha'_v$ and $\beta_{\sigma}$ are set to zero in this analysis, since these are also set to zero in the fitting, both in the case of the template and full-shape approaches. Although not included in the validation, the emulator does in principle yield predictions also for those contributions.
The numerically computed multipoles are then compared with the predictions from the emulator.

Using this training and validation setup we optimised the choice of internal data transformation as well as kernel function.
We changed the internal transformation of the data in the code to a min-max transformation in order to yield more accurate emulation of the \tpcf.
In this transformation first the minimum is subtracted followed by a division by the maximum resulting in data that are constrained to be between zero and one.
We also considered the so-called `power transformer' as implemented in \texttt{scikit-learn} \citep{2011Pedregosa_scikit_learn},\footnote{\texttt{scikit-learn} at \url{https://scikit-learn.org/stable/index.html}} which applies the Yeo--Johnson algorithm \citep{Yeo2000_PowerTransform} to make data follow closer a Gaussian distribution, but did not find satisfactory results in the end.
In addition, we investigated the optimal choice of kernel function that might be different from the choice for \comet for Fourier space emulation.
This is motivated by configuration-space statistics such as the \tpcf usually varying continuously in linear space but not in logarithmic space as Fourier statistics do.
We found the linear combination of a Matern$(\nu=5/2)$ and radial basis function (RBF) kernel to be the most performant \citep[explicit expressions can be found in Eqs.~2.16 and 4.17 in][]{Rasmussen_GP_book}.
This is in contrast to \comet, which replaces the $\text{Matern}(\nu=5/2)$ with a $\text{Matern}(\nu=3/2)$ kernel.
In fact, the latter combination did lead to only marginally worse results.

\begin{table}[]
    \caption{Emulated galaxy bias contributions to the different velocity statistics in the three different \lpt-based models.}
    \smallskip
    \label{tab:emulated_quantities}
    \smallskip
    \begin{tabular}{ccccc}
    \hline
    \rowcolor{blue!5}
        & & & & \\[-8pt]
    \rowcolor{blue!5}
      Moment & Bias term & ZA  & \clpt & \cleft \\
       \hline
       \multirow{13}{*}{$\xi(r)$} & 1 & \checkmark & \checkmark & \checkmark \\
        & $b_1^{\mathrm{L}}$ & \checkmark & \checkmark & \checkmark\\
       \noalign{\vskip 1pt}
        & $b_2^{\mathrm{L}}$ &  & \checkmark & \checkmark\\
       \noalign{\vskip 1pt}
       &$b_1^{\mathrm{L}} b_2^{\mathrm{L}}$ &  & \checkmark & \checkmark\\
       \noalign{\vskip 1pt}
       &$\left(b_1^{\mathrm{L}}\right)^2$ & \checkmark & \checkmark & \checkmark\\
       \noalign{\vskip 1pt}
       \noalign{\vskip 1pt}
       &$\left(b_2^{\mathrm{L}}\right)^2$ &  & \checkmark & \checkmark\\
       &$b_{s^2}$ &  & & \checkmark\\
       &$b_{s^2}^2$ &  & & \checkmark\\
       &$b_1^{\mathrm{L}} b_{s^2}$ &  & & \checkmark\\
       \noalign{\vskip 1pt}
       &$b_2^{\mathrm{L}} b_{s^2}$ &  & & \checkmark\\
       \noalign{\vskip 1pt}
       &$b_{\nabla^2}$ &  & & \checkmark\\
       &$b_1^{\mathrm{L}} b_{\nabla^2}$ &  & & \checkmark\\
       &$\alpha_{\xi}$ & & & \checkmark\\
       \hline
       \multirow{10}{*}{$v_{12}(r)$}&$1$ & \checkmark & \checkmark & \checkmark\\
       &$b_1^{\mathrm{L}}$ & \checkmark & \checkmark & \checkmark\\
       \noalign{\vskip 1pt}
       &$b_2^{\mathrm{L}}$ & & \checkmark & \checkmark\\
       \noalign{\vskip 1pt}
       &$b_1^{\mathrm{L}} b_2^{\mathrm{L}}$ &  & \checkmark & \checkmark\\
       \noalign{\vskip 1pt}
       &$\left(b_1^{\mathrm{L}}\right)^2$ & & \checkmark & \checkmark\\
       \noalign{\vskip 1pt}
       &$\left(b_2^{\mathrm{L}}\right)^2$ &  & \checkmark & \checkmark\\
       &$b_{s^2}$ &  & & \checkmark\\
       &$b_1^{\mathrm{L}} b_{s^2}$ &  & & \checkmark\\
       &$\alpha_{v}$ & & & \checkmark\\
       &$\alpha_{v}'$ & & & \checkmark\\
       \hline
       \multirow{7}{*}{$\tilde{\sigma}_{\parallel, \perp}(r)$}&$1$ & \checkmark & \checkmark & \checkmark\\
       &$b_1^{\mathrm{L}}$ & & \checkmark & \checkmark\\
       \noalign{\vskip 1pt}
       &$b_2^{\mathrm{L}}$ & & \checkmark & \checkmark\\
       \noalign{\vskip 1pt}
       &$\left(b_1^{\mathrm{L}}\right)^2$ &  & \checkmark & \checkmark\\
       &$b_{s^2}$ &  & & \checkmark\\
       &$\alpha_{\sigma}$ & & & \checkmark\\
       &$\beta_{\sigma}$ & & & \checkmark\\
       \hline
    \end{tabular}
    
\end{table}

\subsection{Emulator performance}
\begin{figure*}
    \centering
    \includegraphics[width=0.9\textwidth]{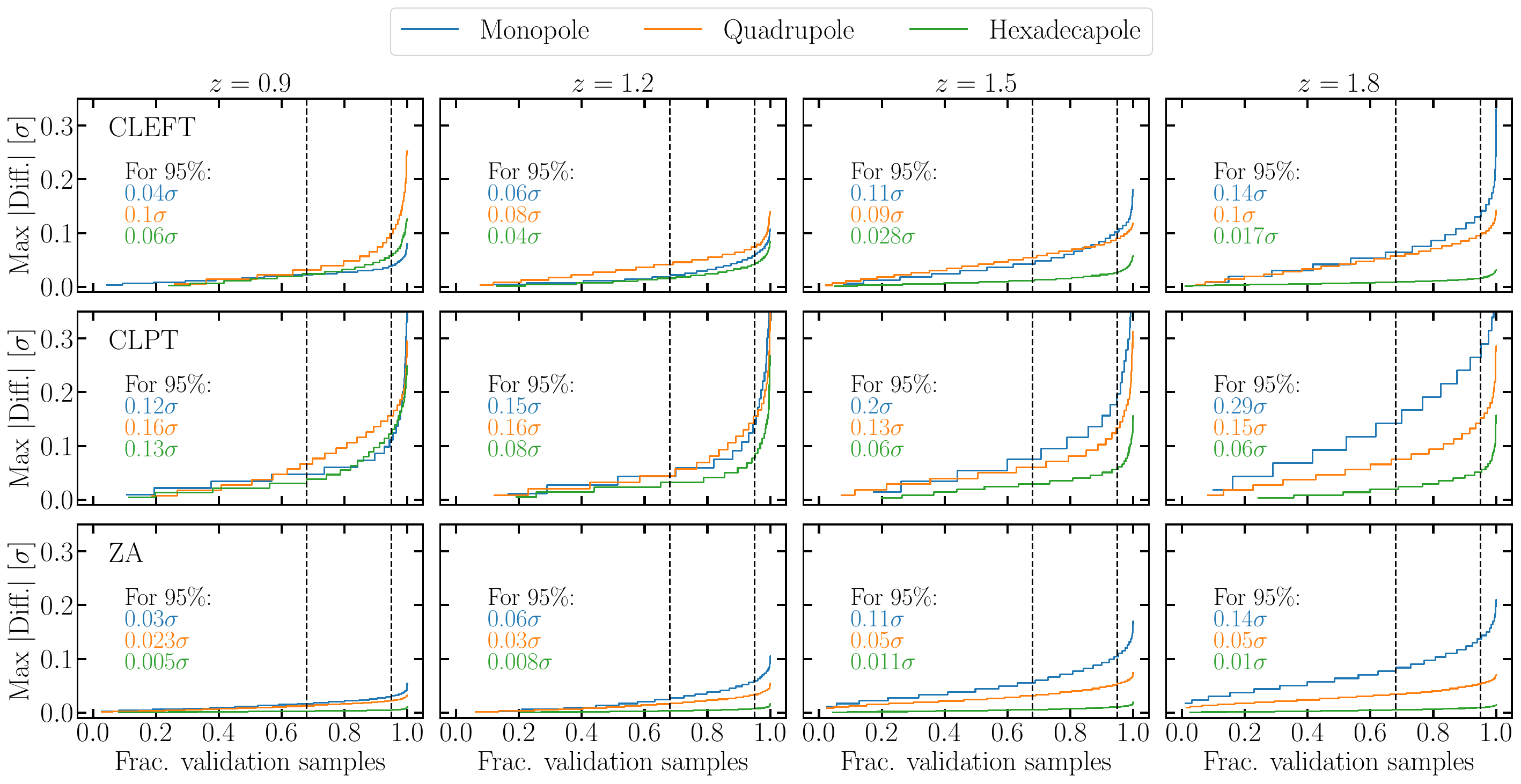}
    \caption{Emulator performance for the \cleft, \clpt, and ZA model. The four columns represent the four validation redshifts. Shown is always the maximum difference in standard deviations as a function of the number of samples of the validation set. The standard deviation is hereby the square root of the diagonal of the \flagship covariance matrix as described in Sect.~\ref{sec:data}. The different colours refer to the three multipoles. The vertical dashed lines indicate a relative fraction of 68\% and 95\% of the total number of 1500 validation samples, respectively. The annotated values give the maximum difference in standard deviations for 95\% of the samples.}
    \label{fig:emulator_Emulator_performance}
\end{figure*}

Equipped with this kernel and optimal choice of data transformation, we can study the emulator performance in more detail.
In Fig.~\ref{fig:emulator_Emulator_performance}, we present the difference on the correlation function multipoles between the numerical code and the emulated ingredients for the \clpt, \cleft, and ZA model.
To ease comparison with \comet, as the \lpt emulator has a very similar architecture, this figure follows precisely the layout of Fig.~C1 in \citet{EggemeierEtal_2023}.
Therefore, Fig.~\ref{fig:emulator_Emulator_performance} shows the differences between the emulator and exact code divided by the square root of the diagonal of the covariance matrix as described in Sect.~\ref{sec:data}.
This is the covariance assigned to the measurements of the \flagship simulation on which we used the emulator.
If the absolute value of this ratio is well below unity, the uncertainty coming from the emulation is negligible compared to the uncertainty of the data and can be neglected in the fitting process.
It has to be noted that the covariance that we use in the validation is not exactly for redshifts $z\in\{0.9, 1.2, 1.5, 1.8\}$, rather for the redshifts of the snapshots, which differ slightly.
However, the predicted correlation functions by the numerical code and emulator are for the approximated redshifts. Since the latter differ by around 0.5--2\% compared to the snapshot redshifts, using the \flagship covariances should only marginally bias the result.
To be more precise, in Fig.~\ref{fig:emulator_Emulator_performance} we show the maximum difference in standard deviations as a function of the fraction of validation samples.
Therefore, for all three models, the recovery of the monopole is the worst, but we stay equal or below $0.2\sigma$ for all models and redshifts for 95\% of validation samples, except for the \clpt model at $z=1.8$ where this value reaches $0.29\sigma$.
The quadrupole is better recovered with a difference of around $0.1\sigma$ to $0.16\sigma$ for the \cleft and \clpt model at 95\% of validation samples.
The ZA reaches a much better performance with a difference in the quadrupole below or equal to $0.05\sigma$. This is intuitively expected since the ZA model requires much fewer terms in the velocity statistics to emulate.
Finally, the hexadecapole is well recovered for all models and redshifts below $0.1\sigma$ except with the \clpt model at $z=0.9$ reaching $0.13\sigma$ at 95\% of validation samples.

The better performance of the \cleft emulator as compared to \clpt in terms of $\sigma$-deviations might be attributed to the GS that uses the cumulant version of the velocity dispersion in the case of \cleft.
In addition, also the expansion of the exponential in the velocity correlators is done differently, as specified in Sect.~\ref{sec:gaussian_streaming_theory}.
The GS might also explain why the emulator performance for the monopole gets worse the higher the redshift, although the values on the diagonal of the covariance matrix increase with redshift.
A validation of the velocity statistics might directly shine some light on this feature but is difficult to do in practice, as we do not have a covariance for those readily available.
Concerning the specific number of \num{2000} training samples, we did a training of the \cleft emulator at $z=0.9$ by randomly depleting the training sample and found the results in the upper leftmost panel of Fig.~\ref{fig:emulator_Emulator_performance} to have converged.
The expected gain in performance by increasing this number is therefore only marginal and the cosmologies appear to be well sampled.

\end{appendix}

\end{document}